\documentclass[twocolumn]{emulateapj}

\usepackage{makecell}
\usepackage{amsmath}
\usepackage[utf8]{inputenc}
\usepackage{graphicx}
\usepackage{adjustbox}
\usepackage[figuresright]{rotating}
\usepackage{color}
\usepackage[colorlinks]{hyperref}
\hypersetup{
    colorlinks,	
    citecolor=blue,
}

\newcommand{\be}{\begin{eqnarray}}
\newcommand{\ee}{\end{eqnarray}}

\newcommand{\nustar}{\textit{NuSTAR}}
\newcommand{\sw}{\textit{Swift}}

\shorttitle{High-density reflection in XRBs}
\shortauthors{Liu et al.}

\begin{document}

\title{High-density reflection spectroscopy of black hole X-ray binaries in the hard state}

\author{Honghui Liu\altaffilmark{1}, Jiachen Jiang\altaffilmark{2}, Zuobin Zhang\altaffilmark{1}, Cosimo Bambi\altaffilmark{1, \dag}, Andrew C. Fabian\altaffilmark{2}, Javier A. Garc{\'\i}a\altaffilmark{3,4}, Adam Ingram\altaffilmark{5}, Erin Kara\altaffilmark{6}, James F, Steiner\altaffilmark{7}, John A. Tomsick\altaffilmark{8}, Dominic J. Walton\altaffilmark{2,9}, Andrew J. Young\altaffilmark{10}}

\altaffiltext{1}{Center for Field Theory and Particle Physics and Department of Physics, 
Fudan University, 200438 Shanghai, China. \email[\dag E-mail: ]{bambi@fudan.edu.cn}} 
\altaffiltext{2}{Institute of Astronomy, University of Cambridge, Madingley Road, Cambridge CB3 0HA, UK}
\altaffiltext{3}{Cahill Center for Astronomy and Astrophysics, California Institute of Technology, 1216 E. California Boulevard, Pasadena, CA 91125, USA}
\altaffiltext{4}{Dr. Karl Remeis-Observatory and Erlangen Centre for Astroparticle Physics, Sternwartstr. 7, D-96049 Bamberg, Germany}
\altaffiltext{5}{School of Mathematics, Statistics, and Physics, Newcastle University, Newcastle upon Tyne NE1 7RU, UK}
\altaffiltext{6}{MIT Kavli Institute for Astrophysics and Space Research, MIT, 77 Massachusetts Avenue, Cambridge, MA 02139, USA}
\altaffiltext{7}{Harvard-Smithsonian Center for Astrophysics, 60 Garden Street, Cambridge, MA 02138, USA}
\altaffiltext{8}{Space Sciences Laboratory, University of California Berkeley, CA 94720}
\altaffiltext{9}{Centre for Astrophysics Research, University of Hertfordshire, College Lane, Hatfield AL10 9AB, UK}
\altaffiltext{10}{H.H. Wills Physics Laboratory, Tyndall Avenue, Bristol BS8 1TL, UK}

\begin{abstract}
We present a high-density relativistic reflection analysis of 21 spectra of six black hole X-ray binaries in the hard state with data from \textit{NuSTAR} and \textit{Swift}. We find that 76\% of the observations in our sample require a disk density higher than the 10$^{15}$~cm$^{-3}$ assumed in the previous reflection analysis. Compared with the measurements from active galactic nuclei, stellar mass black holes have higher disk densities. Our fits indicate that the inner disk radius is close to the innermost stable circular orbit in the luminous hard state. The coronal temperatures are significantly lower than the prediction of a purely thermal plasma, which can be explained with a hybrid plasma model. If the disk density is fixed at 10$^{15}$~cm$^{-3}$, the disk ionization parameter would be overestimated while the inner disk radius is unaffected.
\end{abstract}



\section{Introduction} 
\label{intro}


Accreting black holes are efficient at converting gravitational energy into electromagnetic radiation \citep[e.g.][]{Shakura1973, Thorne1974}. The accretion process is believed to be responsible for many high energy phenomena in the Universe, i.e., active galactic nuclei (AGNs) and X-ray binaries (XRBs). Moreover, accretion is an important ingredient in understanding the growth of supermassive black holes (SMBHs) and galaxies \citep[e.g.][]{King2003ApJ...596L..27K, Alexander2012NewAR..56...93A, Fabian2012ARA&A..50..455F}. The X-ray emission is a powerful probe to study the innermost regions of accreting black holes \citep[e.g.][]{Mushotzky1993,McHardy2006}.

One of the primary components in the X-ray spectra of black holes is the power-law continuum with a high-energy cutoff. This component is thought to originate from inverse Compton scattering of seed photons by a hot plasma (the corona) near the black hole \citep[e.g.][]{Shapiro1976ApJ...204..187S, Haardt1993ApJ...413..507H}. This coronal emission can illuminate the optically thick accretion disk and produce the reflected emission that is featured with fluorescent emission, photoelectric absorption and the Compton scattering hump \citep[e.g.][]{George1991, Garcia2010}. Due to strong relativistic effects near the black hole, the observed reflected features are significantly skewed \citep[e.g.][]{Fabian1989, Dauser2010, Bambi2017bhlt.book.....B} and contain rich information about the spacetime \citep[e.g.][]{Reynolds2014, Bambi2017, Bambi2021, Reynolds2021} and disk-corona geometry \citep[e.g.][]{Fabian2009Natur.459..540F,Fabian2012,Wilkins2012MNRAS.424.1284W}. This technique has been successfully applied to many black hole XRBs and AGNs \citep[e.g.][]{Garcia2015, Walton2016, Jiang2018, Liu2019}

Previous reflection models assume the disk electron density to be $\log(n_{\rm e}/{\rm cm}^{-3})=15$. This value is appropriate for very massive (e.g. $M_{\rm BH}>10^8~M_{\odot}$) black holes in AGNs \citep[e.g.][]{Jiang2019MNRAS.483.2958J} but the standard disk model \citep{Shakura1973} predicts a higher density for less massive black holes. For AGNs, reflection-based analysis has shown that a disk density larger than $\log(n_{\rm e}/{\rm cm}^{-3})=15$ is sometimes required for SMBHs with $M_{\rm BH}<10^7~M_{\odot}$ and the measured densities are consistent with the prediction of a radiation pressure-dominated disk \citep[e.g.][]{Jiang2019AGN,Mallick2022}. The high-density reflection model is able to explain the ``soft excess'' feature found in many Seyfert galaxies \citep[e.g.][]{Arnaud1985} without including an additional component \citep[e.g.][]{Garcia2019Mrk509}. The soft X-ray reverberation signature also supports the reflection origin of the soft excess \citep[e.g.][]{DeMarco2013, Kara2013MNRAS.428.2795K}. 

Moreover, a higher disk density can also relieve the puzzle of super-solar iron abundance in some reflection spectral modelling \citep[e.g.][]{Tomsick2018, Jiang2019gx339}. Since the standard disk model predicts an anti-correlation between the disk density and black hole mass times the square of the Eddington-scaled accretion rate ($M_{\rm BH}\dot{m}^2$) for a radiation pressure-dominated disk \citep[e.g.][and Eq.~\ref{ne_ra}]{Svensson1994}, we expect an even higher density for the accretion disk of XRBs (especially in the hard state). This has been confirmed with an analysis of individual sources but the measured disk densities are lower than theoretical predictions of a radiation pressure or gas pressure-dominated disk \citep[see Fig.~8 of][]{Jiang2019AGN}.

Previous high-density reflection modelling of black hole XRBs has only been conducted for a few sources \citep[e.g.][]{Reis2008, Reis2009, Reis2011, Steiner2011,Reis2012ApJ...751...34R, Steiner2012MNRAS.427.2552S,Walton2012MNRAS.422.2510W,Chiang2012MNRAS.425.2436C, Tomsick2018, Jiang2019gx339,Jiang2020MNRAS.492.1947J, Connors2021, Chakraborty2021}. Moreover, on the $\log(n_{\rm e})$--$\log(M_{\rm BH}\dot{m}^2)$ diagram, there is still an obvious gap between the XRBs and AGNs sample \citep[see Fig.~8 of][]{Jiang2019AGN}. In this work, we conduct a systematic analysis on the broadband X-ray spectra of six black hole XRBs in the hard state that have not been studied with high-density reflection models. The data selection and reduction are described in Sec.~\ref{obs_red}. Sec.~\ref{s-ana} explains the spectral analysis. In Sec.~\ref{dis} we discuss the results.

\section{Observations and data reduction}\label{obs_red}

We select six black hole XRBs from the BlackCAT catalog\footnote{\url{https://www.astro.puc.cl/BlackCAT/transients.php}} \citep{Corral-Santana2016}. All the sources have measurements of their distances, enabling the estimation of their bolometric luminosities through broadband spectral fitting. We also require the sources to have been observed by \nustar{} so the Compton hump is captured. Tab.~\ref{source} shows information about the selected sources. For those sources that do not have measurements of their black hole mass, we assume a value of $10\pm1~M_{\odot}$ throughout this work to calculate the Eddington-scaled luminosity. For each source, we select the \nustar{} observations in the hard state that show relativistic reflection features. One of the main features of the reflection by high-density accretion disk is the enhanced soft X-ray emission \citep{Jiang2019AGN}. Therefore, we also include contemporaneous (on the same day) \sw{} data if possible. Details of the selected observations are listed in Tab.~\ref{obs}. Note that a few other sources may also match our selection criteria but they have already been studied with reflection models of variable electron density (e.g. GX 339--4, \citealt{Jiang2019gx339}; GRS~1716--249, \citealt{Jiang2020MNRAS.492.1947J}; 4U~1630--47, \citealt{King2014ApJ...784L...2K}, \citealt{Connors2021}; MAXI~J1348--630, \citealt{Chakraborty2021}). We do not include those sources in the analysis of this work. MAXI~J1820+070 also matches the selection criteria, but there is still debate on its accretion geometry \citep[e.g.][]{Buisson2019,Zdziarski2022}. We will explore this source in a future publication.

\begin{table*}
    \centering
    \caption{Selected sources in this work}
    \label{source}
    \renewcommand\arraystretch{2.0}
    \begin{tabular}{lcccc}
        \hline\hline
        Source           & Distance (kpc)       & Mass ($M_{\odot}$) & Inclination ($^{\circ}$) & Ref \\
        \hline
        MAXI~J1535--571  & $4.1_{-0.5}^{+0.6}$  & - & - & 1  \\
        \hline
        GRS~1739--278    & 6--8.5 & - & -  & 2  \\
        \hline
        GS~1354--64      & 25--61 & - & $<79$ (binary)  & 3  \\
        \hline
        IGR~J17091--3624 & 11--17 & 8.7--15.6 & - & 4,5 \\
        \hline
        H~1743--322      & $8.5\pm0.8$ & - & $75\pm3$ (jet) & 6 \\
        \hline
        V404~Cyg         & $2.39\pm0.14$ & $9.0_{-0.6}^{+0.2}$ & $67_{-1}^{+3}$ (binary) & 7,8 \\
        \hline
    \end{tabular} \\

    \textit{Note.} Selected sources and their properties. For references about the measurements of the distance, mass and inclination angle: (1) \cite{Chauhan2019}; (2) \cite{Greiner1996}; (3) \cite{Casares2009ApJS..181..238C}; (4) \cite{Rodriguez2011}; (5) \cite{Iyer2015}; (6) \cite{Steiner2012}; (7) \cite{Miller-Jones2009}; (8) \cite{Khargharia2010}. Note that there is still debate on the distance of GS~1354--64 (also known as BW~Cir). A smaller value has been reported by \cite{Gandhi2019MNRAS.485.2642G}.
\end{table*}


\subsection{\textit{NuSTAR}}

We produce cleaned event files for both FPMA and FPMB with the tool \texttt{nupipeline} v0.4.9 and the calibration version 20220301. The source spectra are then extracted from circular regions centered on the sources using the \texttt{nuproducts} task. The background spectra are extracted from source-free areas with polygon regions created with \texttt{ds9}. Note that for V404~Cyg, five flux-resolved spectra are extracted from the two observations (see details in Sec.~\ref{v404}).

\subsection{\textit{Swift}}

The \sw{}/XRT data are all in the Window Timing (WT) mode and are free from pile-up effects. The cleaned events files are produced using \texttt{xrtpipeline} v0.13.7 and the last calibration files as of September 2021. We extract the source spectra from circular regions centered on the source with a radius of 100 arcsec. The background regions are chosen to be annuli with an inner radius of 110 arcsec and an outer radius of 200 arcsec. We only include events with grade 0.



\section{Spectral analysis}\label{s-ana}

Spectral fittings are conducted with \texttt{XSPEC} v12.12.1 \citep{xspec}. The \nustar{} data are used in the 3--79 keV band and the \sw{} data in the 1--10 keV band. All data are grouped to ensure a minimum of 30 counts per bin. We implement element abundances of \cite{Wilms2000} and cross-sections of \cite{Verner1996}. $\chi^2$ statistics is used to find the best-fit values and uncertainties (at 90\% confidence level unless otherwise specified) of the model parameters.

\begin{figure*}
    \centering
    \includegraphics[width=0.33\linewidth]{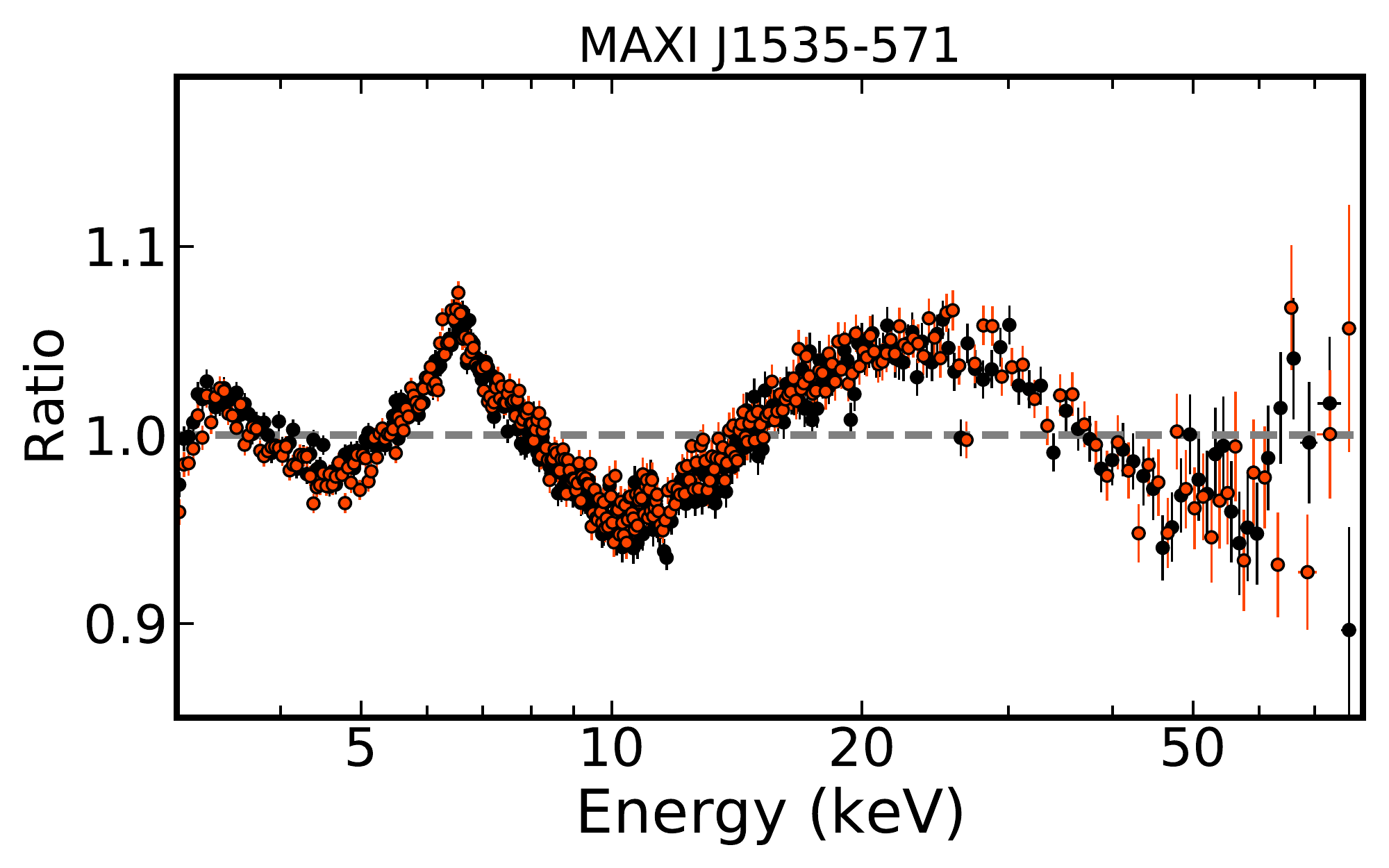}
    \includegraphics[width=0.33\linewidth]{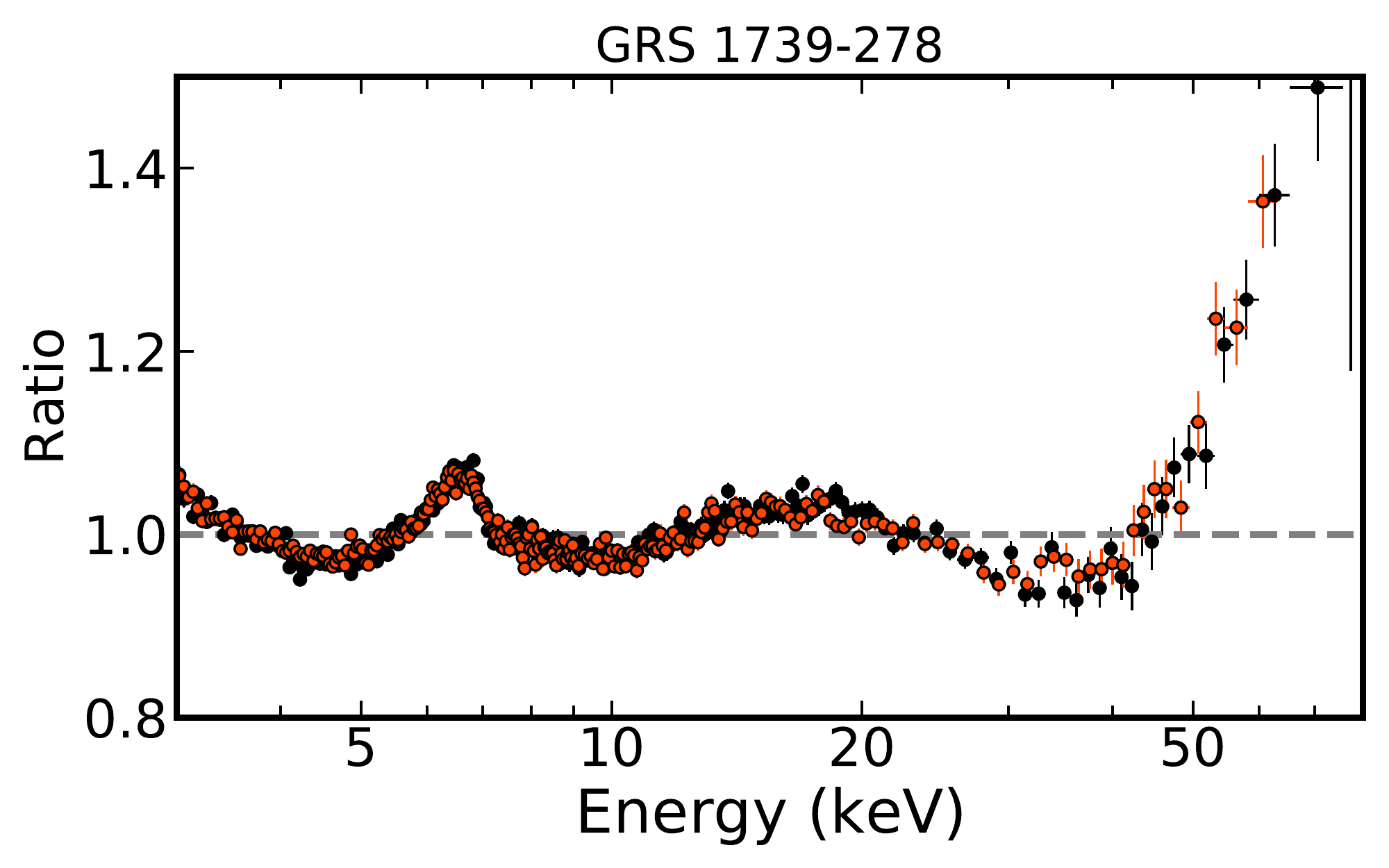}
    \includegraphics[width=0.33\linewidth]{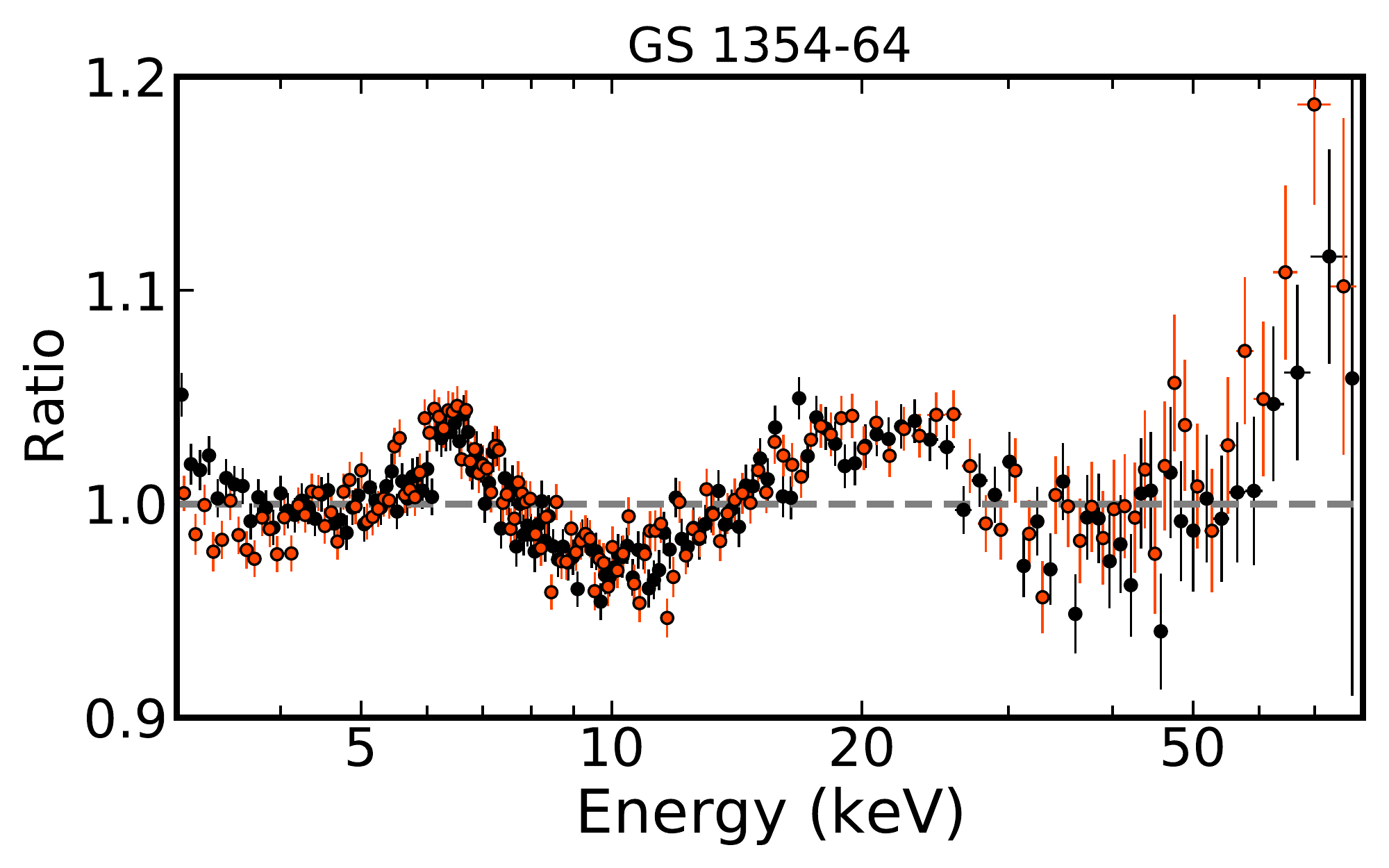}\\
    \includegraphics[width=0.33\linewidth]{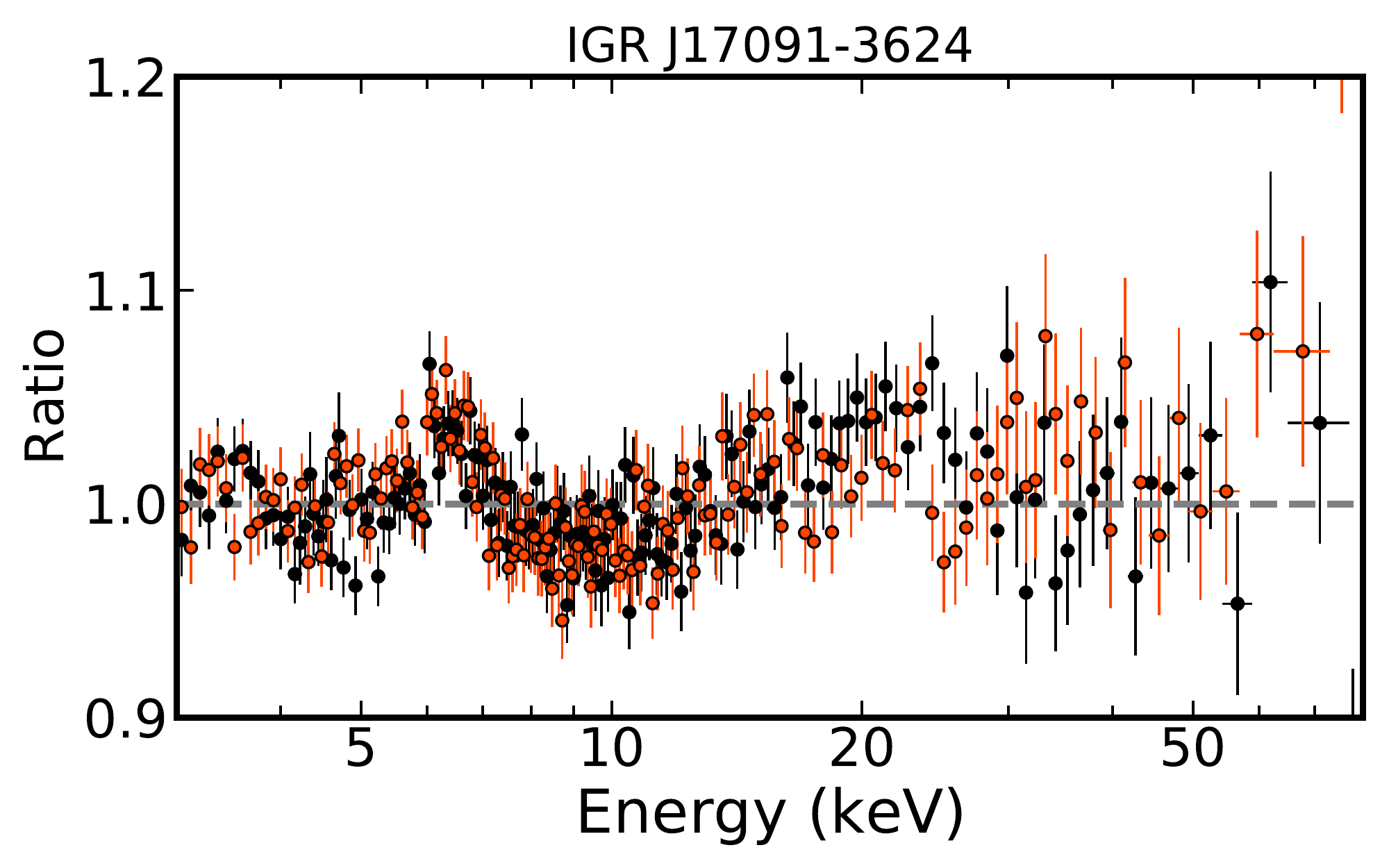}
    \includegraphics[width=0.33\linewidth]{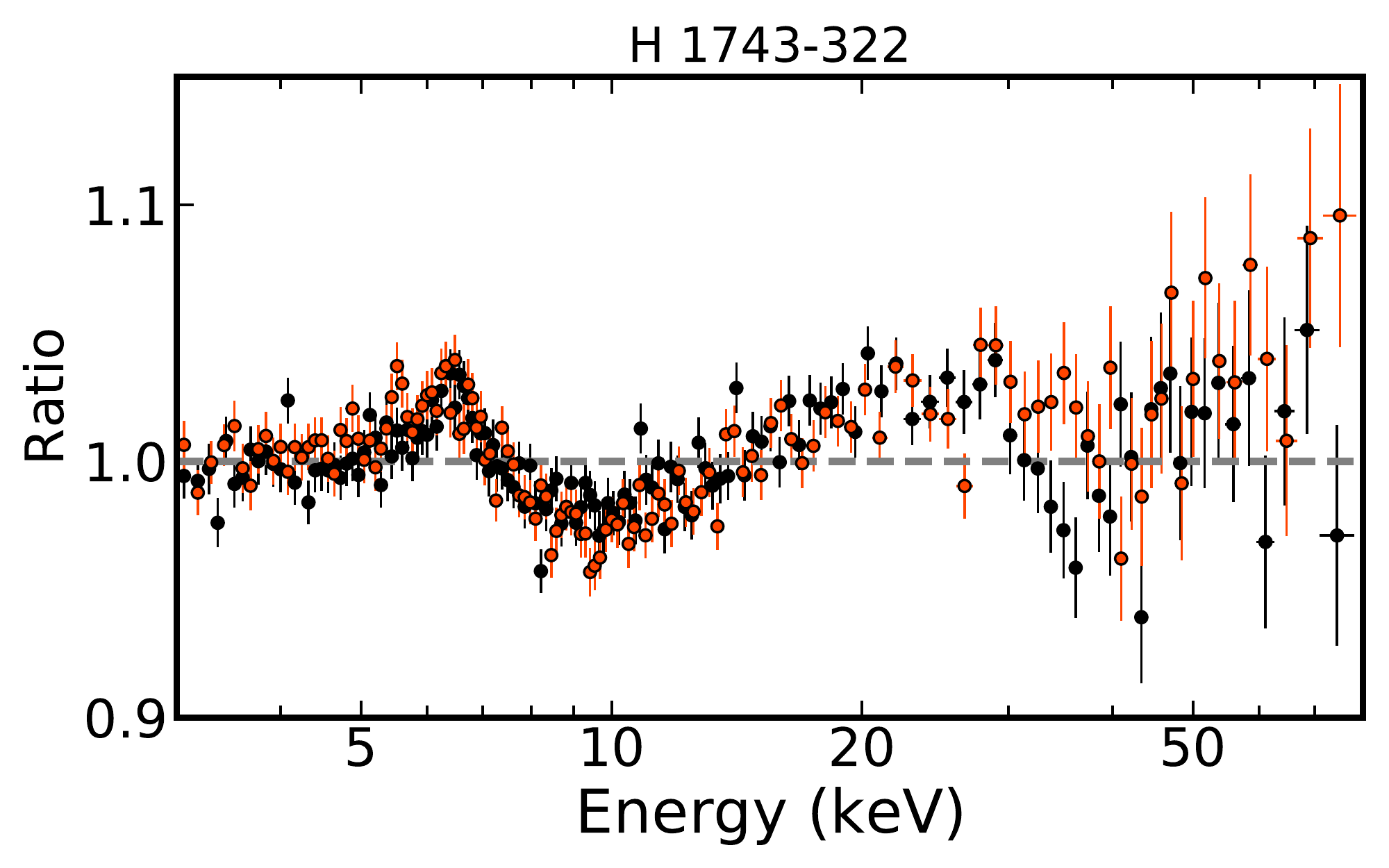}
    \includegraphics[width=0.33\linewidth]{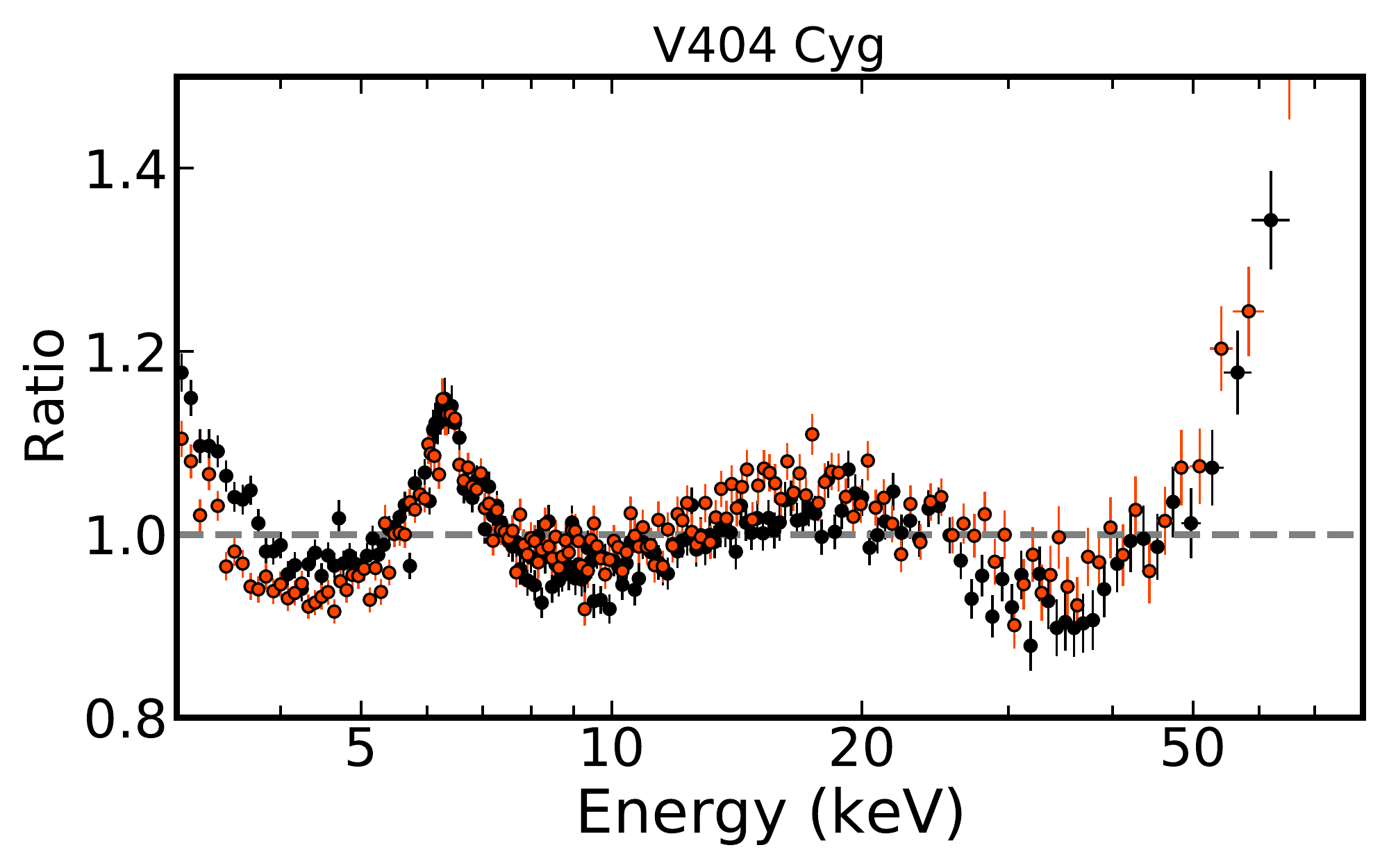}
    \caption{Data to model ratio when fitting with a simple absorbed continuum model: \texttt{constant*tbabs*(diskbb+nthcomp)}. Only one observation is shown for each source. The black and red colors represent data from FPMA and FPMB respectively. Data are rebinned for visual clarity.}
    \label{ironlines}
\end{figure*}

We first fit the broadband spectra from each source with a simple absorbed continuum model: \texttt{constant * tbabs * ( diskbb + nthcomp )}. The \texttt{constant} model is to fit the cross-normalization between instruments. The absorption by Galactic inter-stellar medium is modelled by \texttt{tbabs} \citep{Wilms2000}. The \texttt{diskbb} \citep{Mitsuda1984} component is to fit the thermal emission from the multicolor accretion disk. The Comptonization model \texttt{nthcomp} \citep{Zdziarski1996, Zycki1999} fits the coronal emission. The data to model ratios for this test are shown in Fig.~\ref{ironlines}. Common features for the plots in Fig.~\ref{ironlines} are a broad line feature around 6--7 keV and a hump above 20 keV. These features are known as indications of a relativistic reflection component in the spectra \citep[e.g.][]{Ross1999}. In some cases, e.g. GRS~1739--278, we also see a strong tail above 50~keV. This high energy excess may result from extra emission of the jets. Another explanation could be that the corona is a hybrid plasma with both thermal and non-thermal particles \citep[e.g.][Jiang et al. submitted]{Parker2015}. The non-thermal component can reduce the coronal temperature and produce more hard X-ray photons compared to the pure-thermal model \citep[e.g.][]{Coppi1999, Fabian2017}.

Then we model the reflection features with relativistic reflection models: \texttt{constant * tbabs * ( cflux * nthcomp + cflux * relconv * reflionx\_HD)}.
In this model, \texttt{reflionx\_HD}\footnote{The model is free to download from \url{https://www.michaelparker.space/reflionx-models}. The fits file used here is \texttt{reflionx\_HD\_nthcomp\_v2.fits}.} is a rest-frame disk reflection model calculated with the \texttt{reflionx} code \citep{Ross2005}. The incident spectrum to the accretion disk is assumed to be described by \texttt{nthcomp}. Therefore, we link the photon index ($\Gamma$) and the coronal temperature ($kT_{\rm e}$) parameters between \texttt{nthcomp} and \texttt{reflionx\_HD}. Other free parameters of the reflection model include the ionization parameter $\xi=L/nr^2$ (where $L$ is the ionizing luminosity from the primary source, $n$ is the density and $r$ is the distance from the ionizing source), the iron abundance ($A_{\rm Fe}$) and the electron density that can vary between $\log(n_{\rm e}/{\rm cm}^{-3})=$ 15-22. The convolution kernel \texttt{relconv} \citep{Dauser2010, Dauser2013} is required to include the relativistic effects. In this way, the model calculates the angle-averaged spectrum with only some minor bias (smaller than the statistical errors) in the estimates of some parameters for low disk viewing angles \citep{Tripathi2020MNRAS.498.3565T}. This model has parameters like the black hole spin ($a_{*}$), the disk inclination angle ($i$), the inner disk radius ($R_{\rm in}$) and the emissivity profile. In this work, we fix the spin parameter in order to constrain the $R_{\rm in}$. The spin is fixed at $a_*=0.2$ for H~1743-322 \citep{Steiner2012}, $a_*=0$ for IGR~J17091--3624 (see Sec.~\ref{igrj17091}) and at the maximum value (0.998) for the other sources (see Sec.~\ref{detail} for details). The inclination angle and the inner disk radius are left free during the fit. For the emissivity profile, we implement a broken power-law profile (i.e., $\epsilon\propto 1/r^{q_{\rm in}}$ for $R_{\rm in}<r<R_{\rm br}$ and $\epsilon\propto 1/r^{q_{\rm out}}$ for $R_{\rm br}<r<R_{\rm out}$ where $R_{\rm br}$ is the breaking radius). In cases that $q_{\rm out}$ and $R_{\rm br}$ are not constrained, a power-law emissivity ($q_{\rm in}=q_{\rm out}$) is used instead. We include a \texttt{cflux} model on each additive component to calculate the flux in the 0.1--100~keV band. Note that in some cases, a distant reflection component, a disk thermal emission component or additional absorption components are required to fit the data (see Fig.~\ref{eemod_de} and Sec.~\ref{detail} for details). 

To illustrate the impact on the measurements of spectral parameters after including a variable electron density, we also fit each spectrum with $\log(n_{\rm e})$ setting to 15, which is a value adopted by traditional reflection models. The best-fit parameters of the two sets of fittings are shown in Tab.~\ref{para_reflionx} and Tab.~\ref{para_n15}. Moreover, we also test the \texttt{relxill} model \citep{Garcia2014} with $\log(n_{\rm e})=15$ (see Sec.~\ref{relxillcp} and Tab.~\ref{para_relxill}).

\begin{figure}
    \centering
    \includegraphics[width=0.95\linewidth]{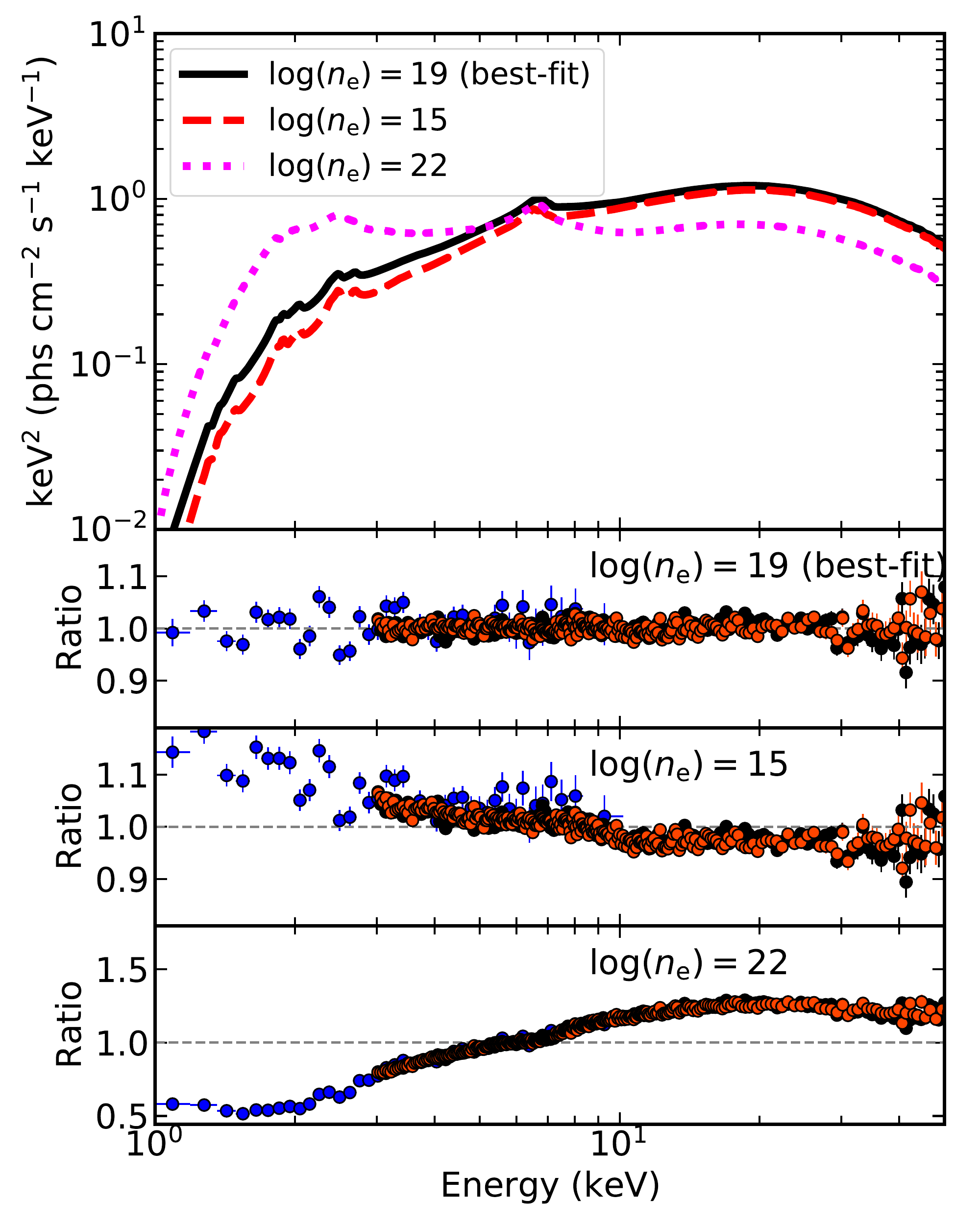}
    \caption{Upper: The black solid line shows the best-fit reflection component for GRS~1739--278 with a disk density of 10$^{19}$~cm$^{-3}$. The red dashed and magenta dotted lines represent the cases when the density is set to 10$^{15}$~cm$^{-3}$ and 10$^{22}$~cm$^{-3}$ respectively without changing the other parameters. Lower three panels: The data to model ratios for the three cases in the upper panel. The blue, black and red data represent \sw{}-XRT, \nustar{}-FPMA and \nustar{}-FPMB respectively.}
    \label{example}
\end{figure}

Fig.~\ref{example} shows an example of high-density reflection of GRS~1739--278. The best-fit electron density for this source is $\log(n_{\rm e}/{\rm cm}^{-3})=19.0\pm0.4$. Allowing the electron density to be a free parameter improves the $\chi^2$ by 60 with one more degree of freedom (see Tab.~\ref{para_reflionx} and Tab.~\ref{para_n15}). If fixing the density at $\log(n_{\rm e}/{\rm cm}^{-3})=15$ (without refitting the spectrum), the reflection component would be depressed in the soft X-ray band, and there appears to be an excess in the residuals below 2~keV (see the third panel of Fig.~\ref{example}). On the contrary, if increasing the density to $\log(n_{\rm e}/{\rm cm}^{-3})=22$, the soft X-ray emission is apparently strengthened. This is a result of stronger free-free process on the disk when the density gets higher, which could also increase the disk surface temperature (see left panel of Fig.~\ref{pressure_balance}).

\begin{table*}
    \centering
    \caption{Summary of observations analyzed in this work}
    \label{obs}
    \renewcommand\arraystretch{1.8}
    \begin{tabular}{lcccccccc}
        \hline\hline
        Source  & Obs ID & Exposure & Start date & Obs ID & Exp. (ks) \\
          & (NuSTAR) & (ks) & yyyy-mm-dd &  (Swift) & (Swift)  \\
        \hline
        MAXI~J1535--571 & 90301013002 & 10.3 & 2017-09-07 & &  \\
        \hline
        GRS~1739--278 & 80002018002 & 29.7 & 2014-03-26 & 00033203003 & 1.9   \\
        \hline
        GS~1354--64 & 90101006002 & 24.0 & 2015-06-13 & 00033811005 & 2.0   \\
                   & 90101006004 & 29.7 & 2015-07-11 & 00033811017 & 0.2  \\
                   & 90101006006 & 35.3 & 2015-08-06 & &  \\
        \hline
        IGR~J17091--3624 & 80001041002 & 43.3 & 2016-03-07 & 00031921099 & 1.5  \\
                        & 80202014002 & 20.2 & 2016-03-12 & 00031921104 & 1.9  \\
                        & 80202014004 & 20.7 & 2016-03-14 & 00031921106 & 1.0 \\
        \hline
        H~1743--322      & 80001044002 & 50.4 & 2014-09-18 & &  \\
                        & 80001044004 & 61.1 & 2014-09-23 & &  \\
                        & 80001044006 & 25.7 & 2014-10-09 & &  \\
                        & 80002040002 & 28.3 & 2015-07-03 & &  \\
                        & 80202012002 & 65.9 & 2016-03-13 & &  \\
                        & 80202012004 & 65.7 & 2016-03-15 & &  \\
                        & 90401335002 & 38.4 & 2018-09-19 & &  \\
                        & 80202012006 & 65.7 & 2018-09-26 & &  \\
        \hline
        V404~Cyg & 90102007002 & 17.7 & 2015-06-24 & &   \\
                 & 90102007003 & 6.2 & 2015-06-25 & &   \\
        \hline
    \end{tabular} \\
\end{table*}

\section{Results and Discussion}\label{dis}

\subsection{Comparisons with the low density model}
\label{dis-compare}

\begin{figure*}
    \centering
    \includegraphics[width=0.33\linewidth]{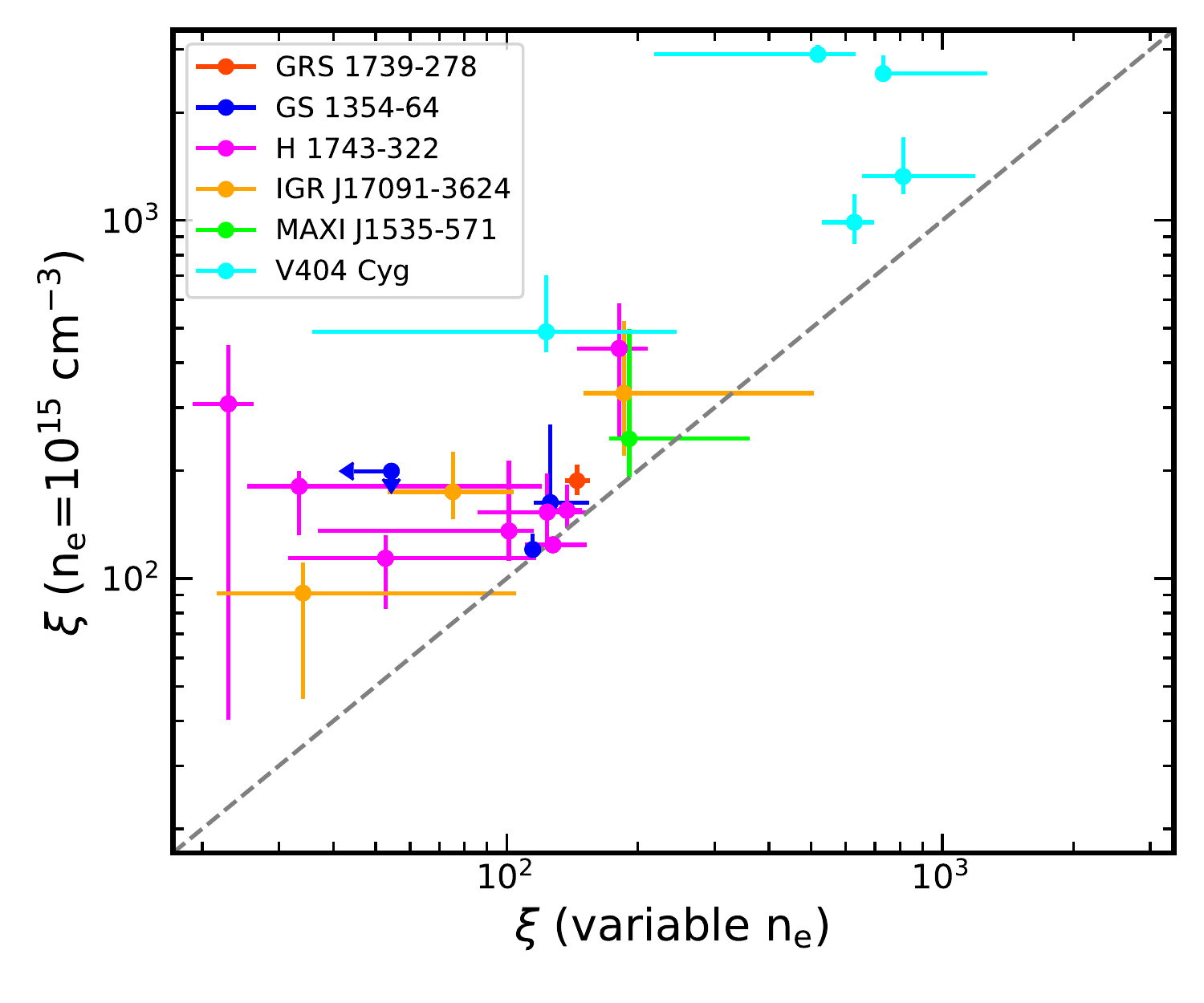}
    \includegraphics[width=0.33\linewidth]{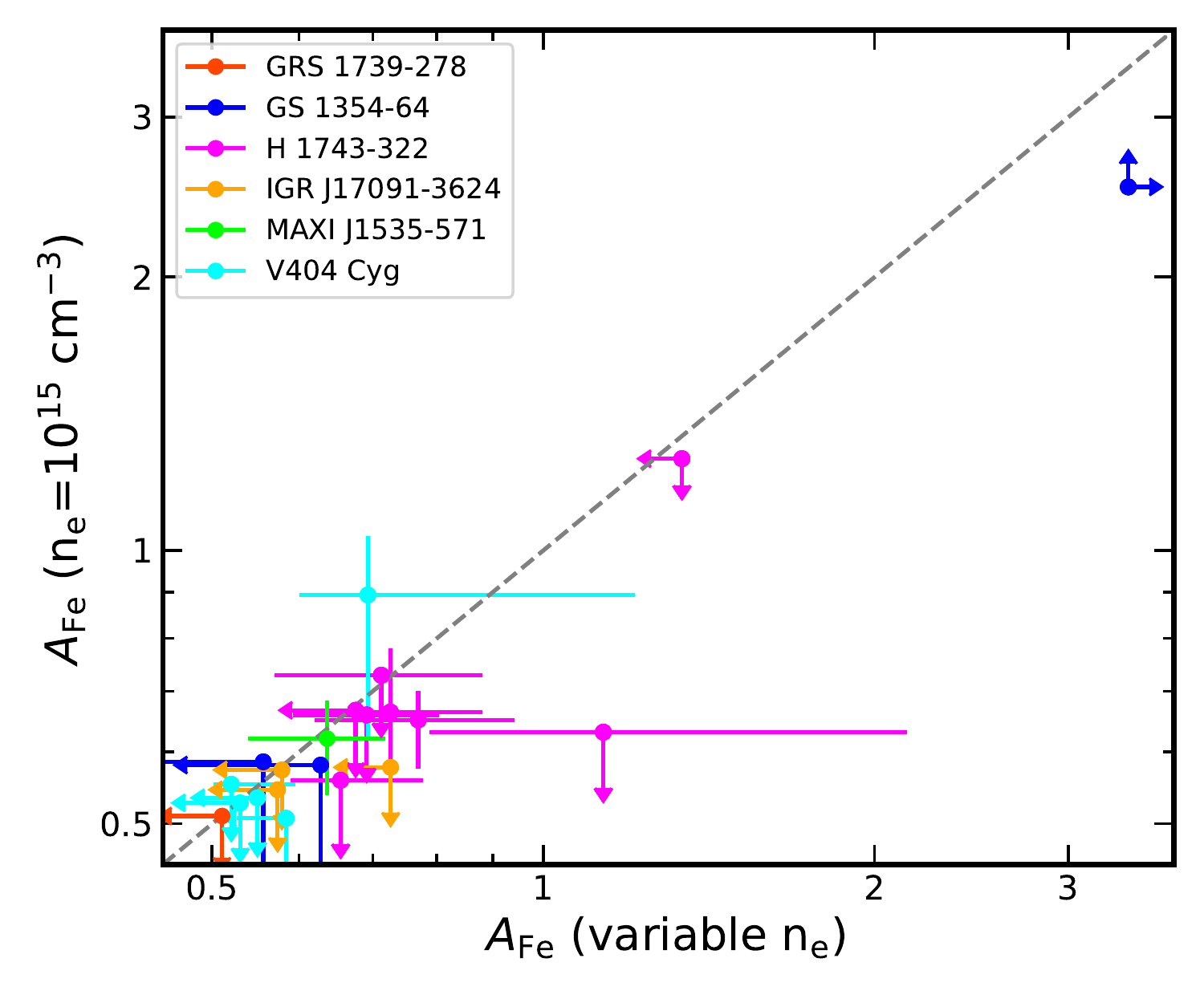}
    \includegraphics[width=0.33\linewidth]{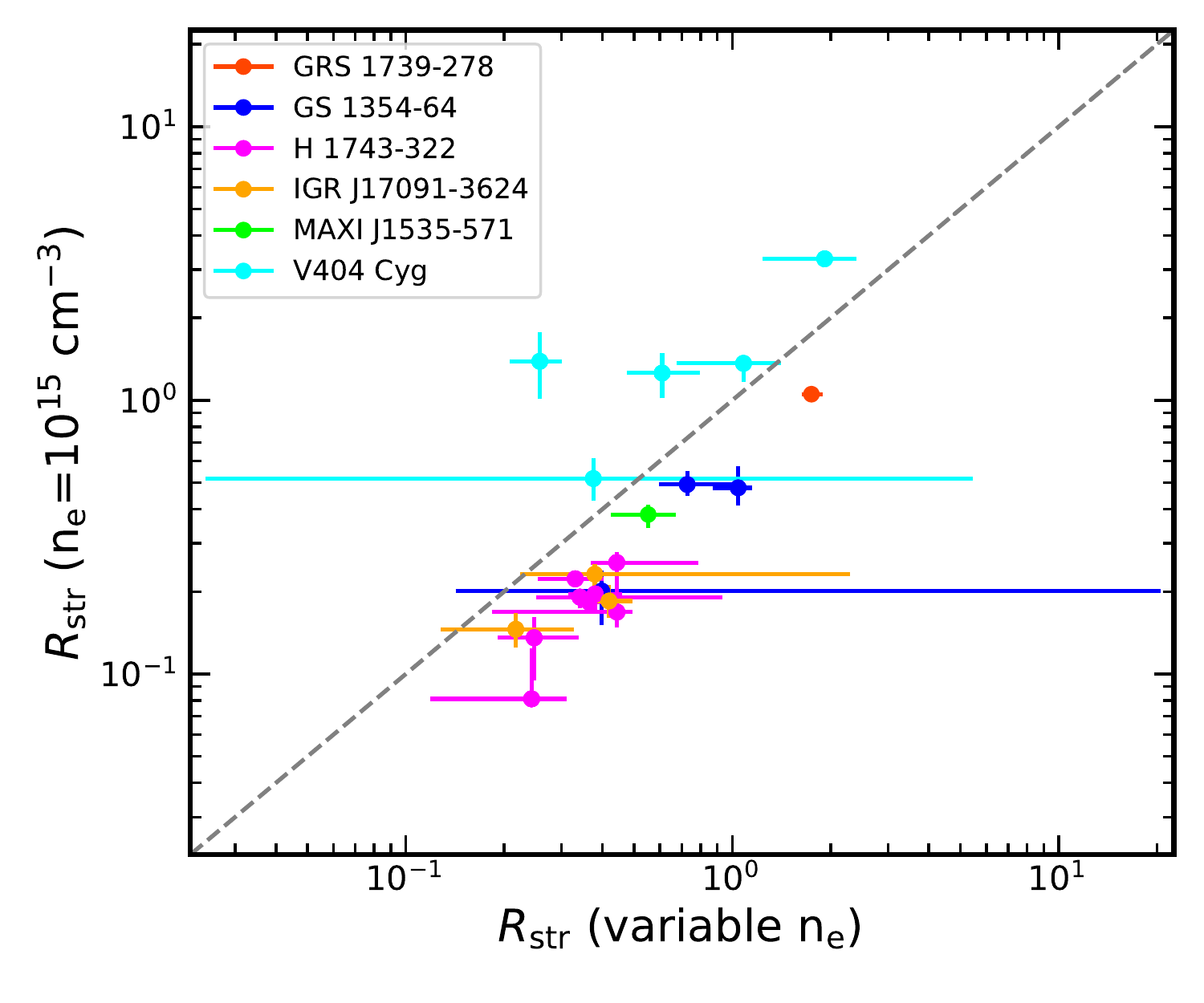}
    \caption{Comparison between the parameter measurements with the \texttt{reflionx\_HD} model that has a variable electron density and that with a fixed density at $\log(n_{\rm e}/{\rm cm}^{-3})=15$. The left panel is for the disk ionization parameter, the middle panel is for the iron abundance and the right panel is for the reflection strength (the energy flux ratio between the reflected and coronal emission in the 0.1--100~keV band). Colors are coded as in Fig.~\ref{ne_mdot2}.}
    \label{compare}
\end{figure*}

In this section, we discuss the impact of the high-density reflection model on spectral parameters compared to the traditional low density model. In Fig.~\ref{compare}, we show the measurements of the disk ionization parameter, the iron abundance and the reflection fraction. It is shown that when assuming a disk density of $\log(n_{\rm e}/{\rm cm}^{-3})=15$, the ionization parameter is systematically overestimated. This is likely because increasing the ionization parameter produces more soft X-ray emissions in the reflected spectrum \citep[see Fig.~3 of][]{Garcia2013}, which mimics the effect of high-density reflection \citep[see Fig.~4 of][]{Garcia2016}.

When fitting the relativistic reflection models to X-ray spectra of XRBs and AGNs, a super-solar iron abundance is commonly required \citep[e.g.][]{Jiang2018,Garcia2018ASPC}. Some studies have shown that high-density models could lower the measured iron abundance \citep[e.g.][]{Tomsick2018, Jiang2019gx339}. We do not see this effect in our sample (see the middle panel of Fig.~\ref{compare}) possibly because most of our observations do not need a super-solar iron abundance even in the case of $\log(n_{\rm e}/{\rm cm}^{-3})=15$.

We define the reflection strength parameter ($R_{\rm str}$) to be the energy flux ratio between the reflected and coronal emission in the 0.1--100~keV band. Fig.~\ref{compare} shows that, except for V404~Cyg, the model with variable $n_{\rm e}$ always returns a higher reflection strength. As evident in Fig.~\ref{example}, additional thermalized emission appearing in the soft bands contributes additional emission which may explain the correlated increase in reflection strength with density.

\begin{figure*}
    \centering
    \includegraphics[width=0.99\linewidth]{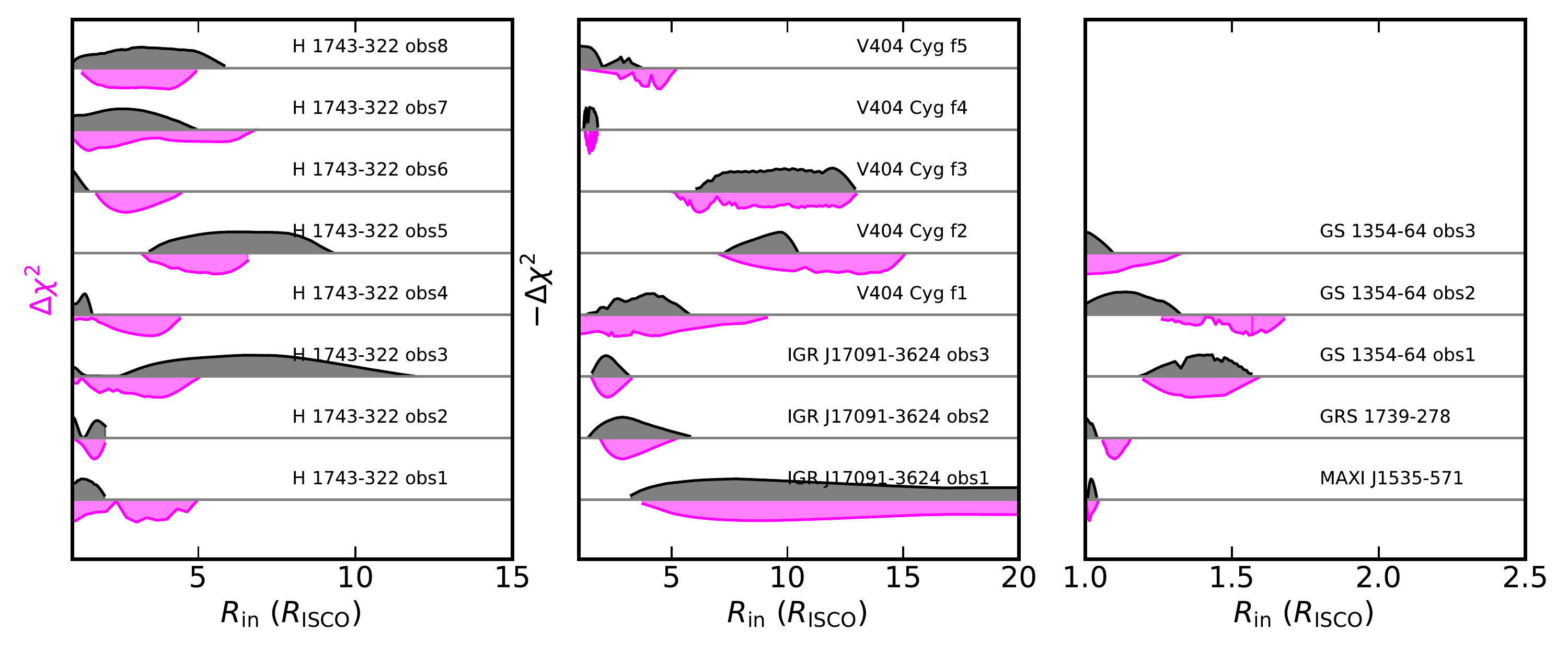}
    \caption{$\Delta\chi^2$ v.s. $R_{\rm in}$ for each observation obtained with \texttt{steppar} command in \texttt{XSPEC}. For the ISCO size, we assume $a_*=0.2$ for H~1743--322 \citep{Steiner2012}, $a_*=0$ for IGR~J17091--3624 (see Sec.~\ref{igrj17091}) and $a_*=0.998$ for the other sources. The horizontal grey lines represent the case $\Delta\chi^2=2.706$ (90\% confidence level for one parameter of interest). Data are offset in the vertical direction for visual clarity. The regions between $\Delta\chi^2$-$R_{\rm in}$ and $\Delta\chi^2=2.706$ are shaded with magenta ($n_{\rm e}$ free) and grey ($n_{\rm e}=10^{15}$) colors. The grey areas result after a mirror symmetry transformation with respect to the horizontal lines. For sources with multiple observations, the sequence follows Tab.~\ref{obs}}
    \label{delchi_rin}
\end{figure*}

It is of crucial importance to see if the assumption of the electron density would affect measurements of the inner disk radius. One of the reasons is that the black spin measurements with X-ray reflection spectroscopy rely on accurate determination of the ISCO size \citep[e.g.][]{Reynolds2014}. The constraints of $R_{\rm in}$ of each observation are shown in Fig.~\ref{delchi_rin} by plotting the distribution of $\Delta\chi^2$ as a function of the inner disk radius. It shows that allowing $n_{\rm e}$ to be free to vary provides consistent measurements on $R_{\rm in}$ compared to the case of fixed $n_{\rm e}$. This is expected since $R_{\rm in}$ is mainly constrained by the red wing of the broad iron line as a result of the gravitational redshift effect, which does not depend on the disk density.

\subsection{Disk densities} \label{dis_den}

\begin{figure*}
    \centering
    \includegraphics[width=0.49\linewidth]{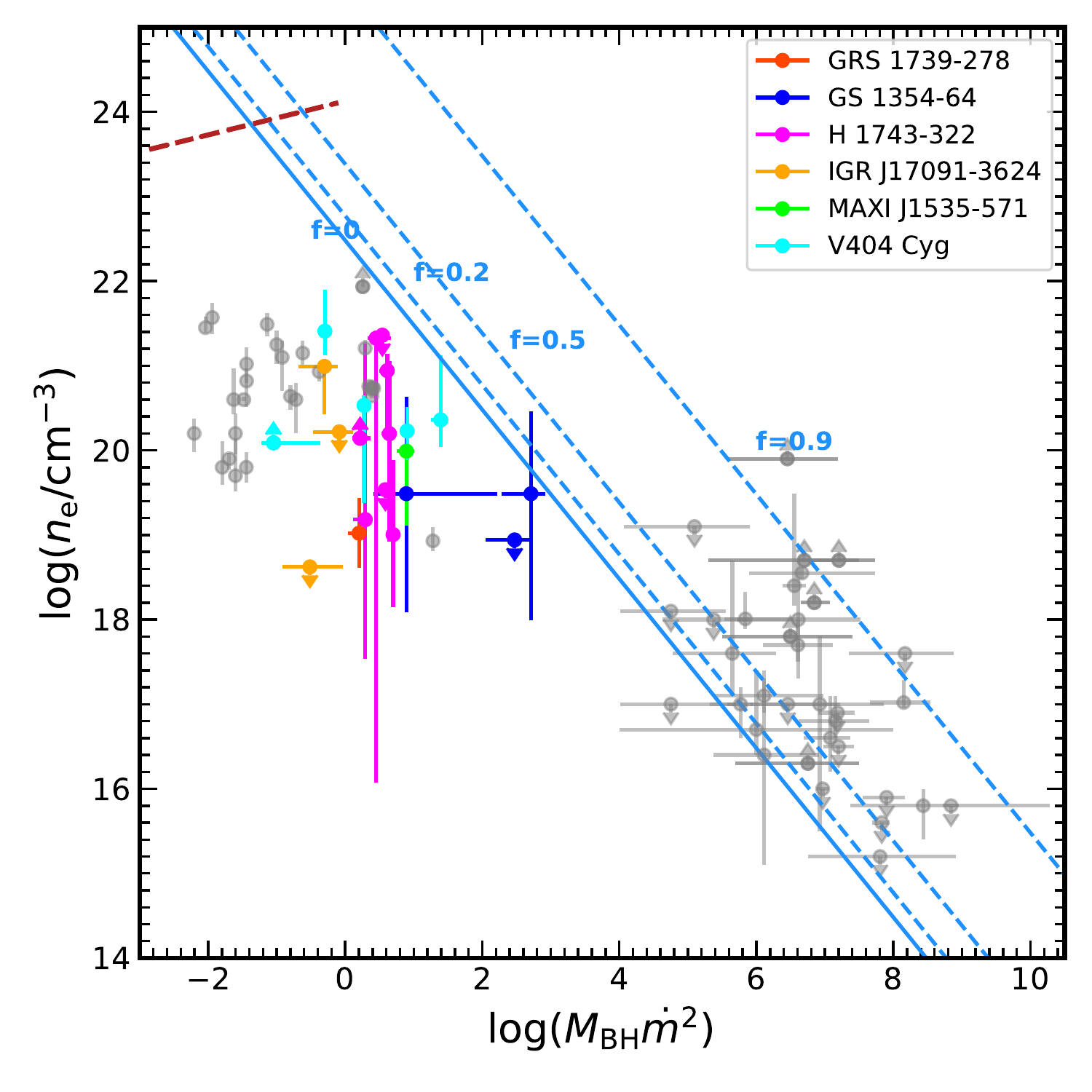}
    \includegraphics[width=0.49\linewidth]{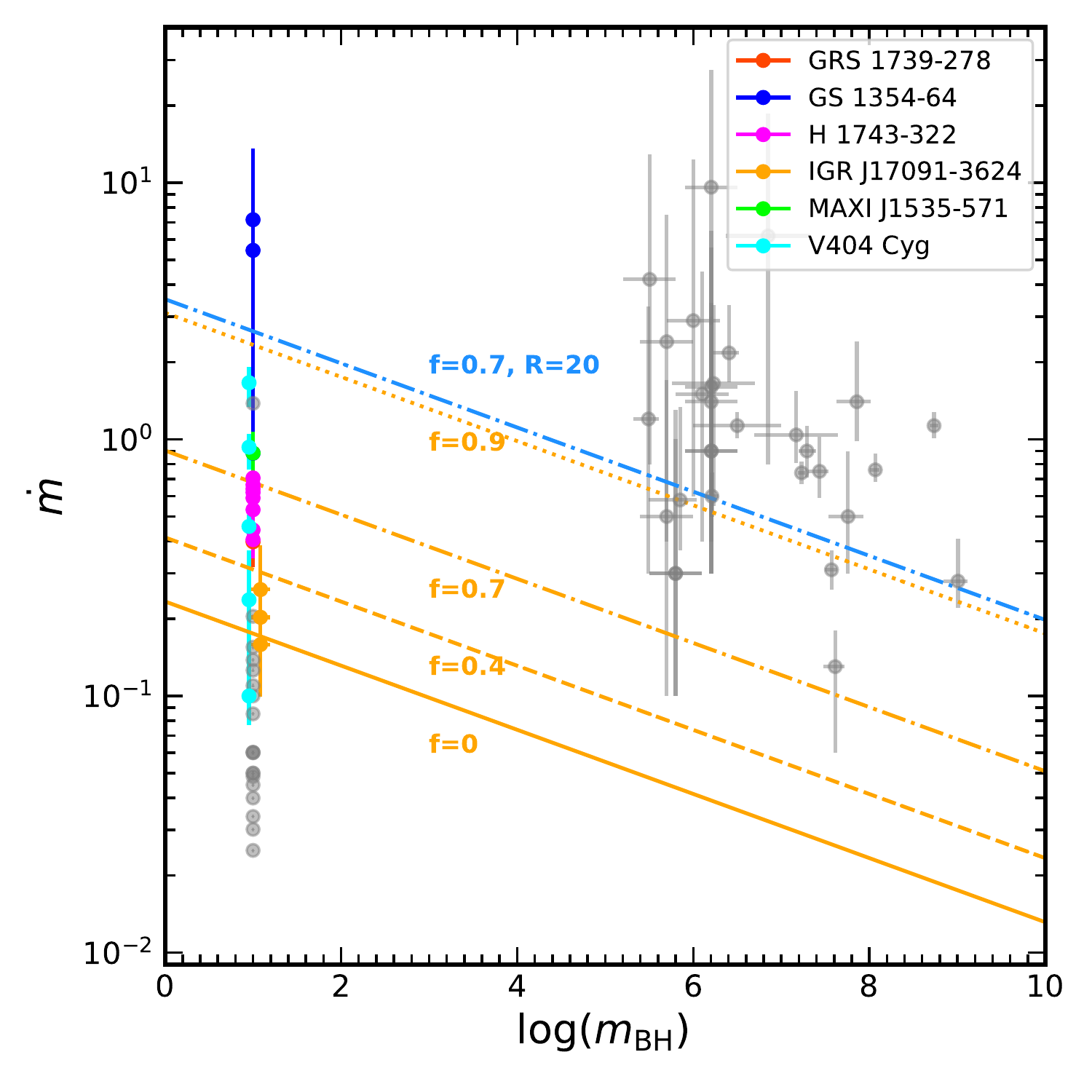}
    \caption{(Left) The variation of the disk electron density ($\log(n_{\rm e})$) with $\log(M_{\rm BH}\dot{m}^2)$, where $\dot{m}=\dot{M}c^2/L_{\rm Edd}=L_{\rm bol}/\eta L_{\rm Edd}$.  The lines in light blue represent solutions of a radiation pressure-dominated disk for different values of $f$ assuming $\xi^{\prime}=1$ and $R=R_{\rm s}$. The dashed dark-red line represents the solution for a gas pressure-dominated disk with $f=0$. (Right) The variation of the dimensionless mass accretion rate with the black hole mass. The orange lines show threshold mass accretion rates for a few values of $f$ and $R=5.7~R_{\rm s}$, below which the disk is dominated by gas pressure. The line in light blue shows the solution for $f=0.7$ and $R=20~R_{\rm s}$. The data in grey represent samples from \citealt{Jiang2019gx339} (GX~339--4), \citealt{Jiang2019AGN} (17 AGNs), \citealt{Tomsick2018} (Cygnus~X-1), \citealt{Jiang2020MNRAS.492.1947J} (GRS~1716--249), \citealt{Liu2022arXiv221109543L} (GX~339--4) and \citealt{Mallick2022} (13 low-mass AGNs). Our results are marked with other colors.}
    \label{ne_mdot2}
\end{figure*}

In the left panel of Fig.~\ref{ne_mdot2}, we present our measurements of the disk density ($\log(n_{\rm e})$) versus $\log(M_{\rm BH}\dot{m}^2)$. Previous studies in the literature of XRBs (the left cluster) and AGNs (the right cluster), which mainly cover the parameter space of $\log(M_{\rm BH}\dot{m}^2)<0$ and $\log(M_{\rm BH}\dot{m}^2)>4$, are plotted in grey. Our data fit into the gap $0<\log(M_{\rm BH}\dot{m}^2)<3$. Overall, we find that XRBs require a disk density significantly higher than AGNs. Moreover, 16 out of our 21 spectra show a disk density higher than $\log(n_{\rm e})=15$.

According to \cite{Svensson1994}, the density of radiation pressure-dominated standard disk \citep{Shakura1973} follows:
\begin{equation}
n_e=\frac{1}{\sigma_{\rm T}R_{\rm s}}\frac{256\sqrt{2}}{27}\alpha^{-1}R^{3/2}\dot{m}^{-2}[1-(R_{\rm in}/R)^{1/2}]^{-2}[\xi^{\prime}(1-f)]^{-3}
\label{ne_ra}
\end{equation}
where $\alpha=0.1$ is adopted for the viscosity parameter, $\sigma_{\rm T}=6.64\times 10^{-25}~{\rm cm}^{2}$ is the Thomson cross-section, $R_{\rm s}$ is the Schwarzschild radius, $R$ is the disc radius in units of $R_{\rm s}$, $\dot{m}$ is the dimensionless mass accretion rate defined as $\dot{m}=L_{\rm bol}/\eta L_{\rm Edd}$ ($\eta$ is the accretion efficiency that can reach 0.32 for $a_*=0.998$ and 0.057 for $a_*=0$, \citealt{Thorne1974}), $\xi^{\prime}$ is the conversion factor in the radiative diffusion equation that is chosen to be unity by \cite{Shakura1973} and $f$ is the fraction of power transported from the disc to the corona. The solutions for different values of $f$ are also plotted in the left panel of Fig.~\ref{ne_mdot2}. We can see that most of the AGN data (the lower right cluster) can be explained by these solutions, with a $f$ varying between 0.0 and 0.9 assuming $\xi^{\prime}=1$. However, except for a few observations with the highest accretion rates, most disk density measurements of black hole XRBs are below the theoretical predictions. 

We note that a large fraction (13/21) of our samples are in the range of $L_{\rm bol}/L_{\rm Edd}<10\%$. In this case, the gas pressure may play a significant role on the disk, especially when $f$ is large \citep{Svensson1994}. According to \cite{Svensson1994} the radius at which the radiation pressure ($P_{\rm rad}$) equals the gas pressure ($P_{\rm gas}$) is determined by:
\begin{equation}
\frac{R}{(1-R^{-1/2})^{16/21}}=40.5(\alpha M_{\rm BH}/10M_{\odot})^{2/21}\dot{m}^{16/21}[\xi^{\prime}(1-f)]^{6/7}   
\label{ra_ba}
\end{equation}
The left-hand side of this equation reaches minimum when $R\approx 5.7$. Therefore, given certain values for $M_{\rm BH}$, $f$ and $\dot{m}$, the left-hand side can be always larger than the right-hand side in which case the disk is dominated by gas pressure. In the right panel of Fig.~\ref{ne_mdot2}, we plot the threshold $\dot{m}$ as a function of $M_{\rm BH}$ for a few values of $f$. By fitting the AGN sample, \cite{Mallick2022} found $f\approx 0.7$. We can see that the AGN sample is well above the threshold line for $f=0.7$ and the radiation pressure should be important. However, this is not the case for our XRB sample, which indicates the important role of gas pressure. In the left panel of Fig.~\ref{ne_mdot2} we are also showing the solution for a gas pressure-dominated disk, which still predicts a disk density larger than our measurements by two orders of magnitude. One of the explanations could be that the reflection model only measures the density of the disk surface. The current reflection models assume a uniform density in the vertical direction, which may not be the case in reality. Developing reflection models that consider the vertical density structure of the disk would be important to understand this problem.

\begin{figure*}
    \centering
    \includegraphics[width=0.49\linewidth]{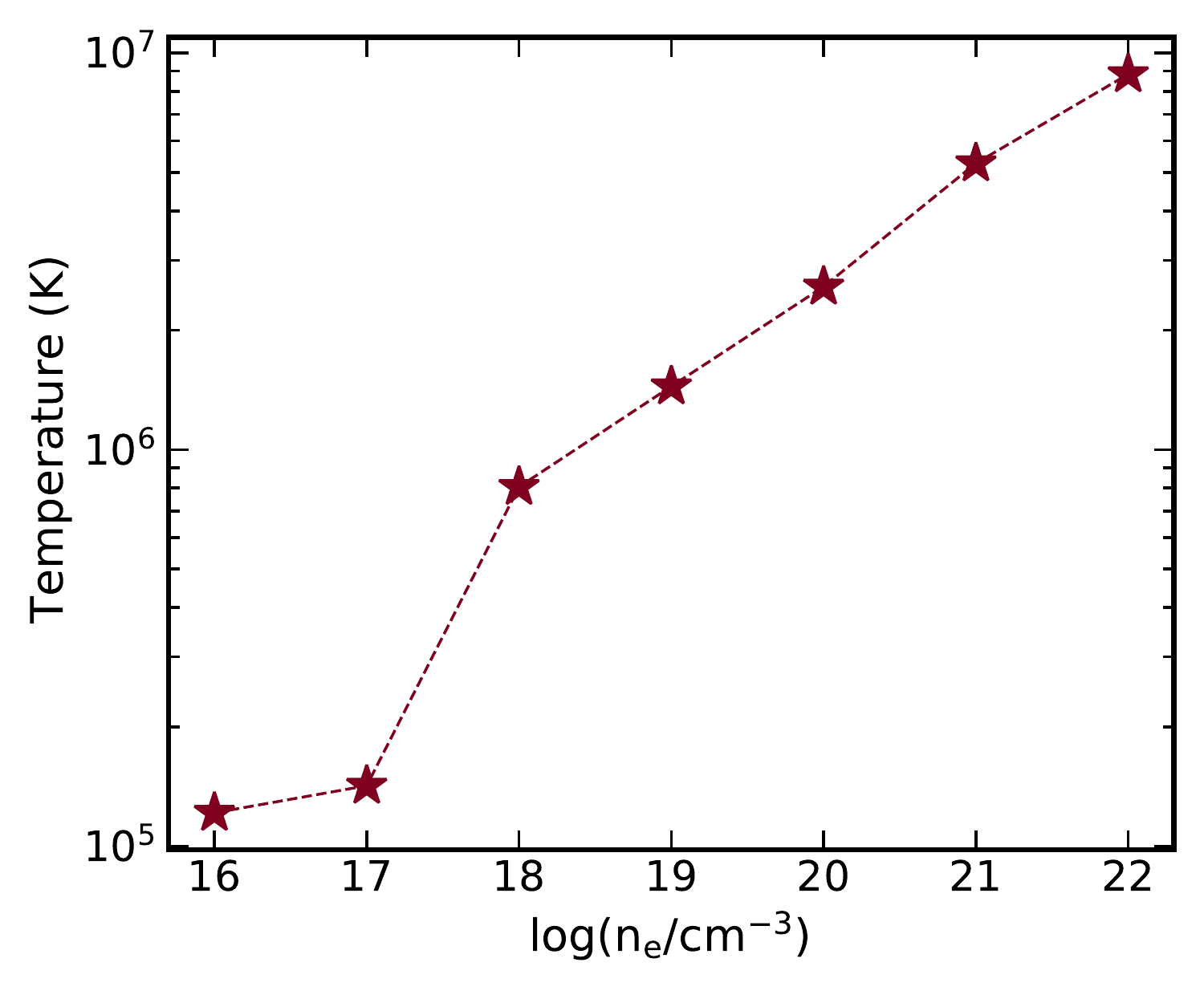}
    \includegraphics[width=0.49\linewidth]{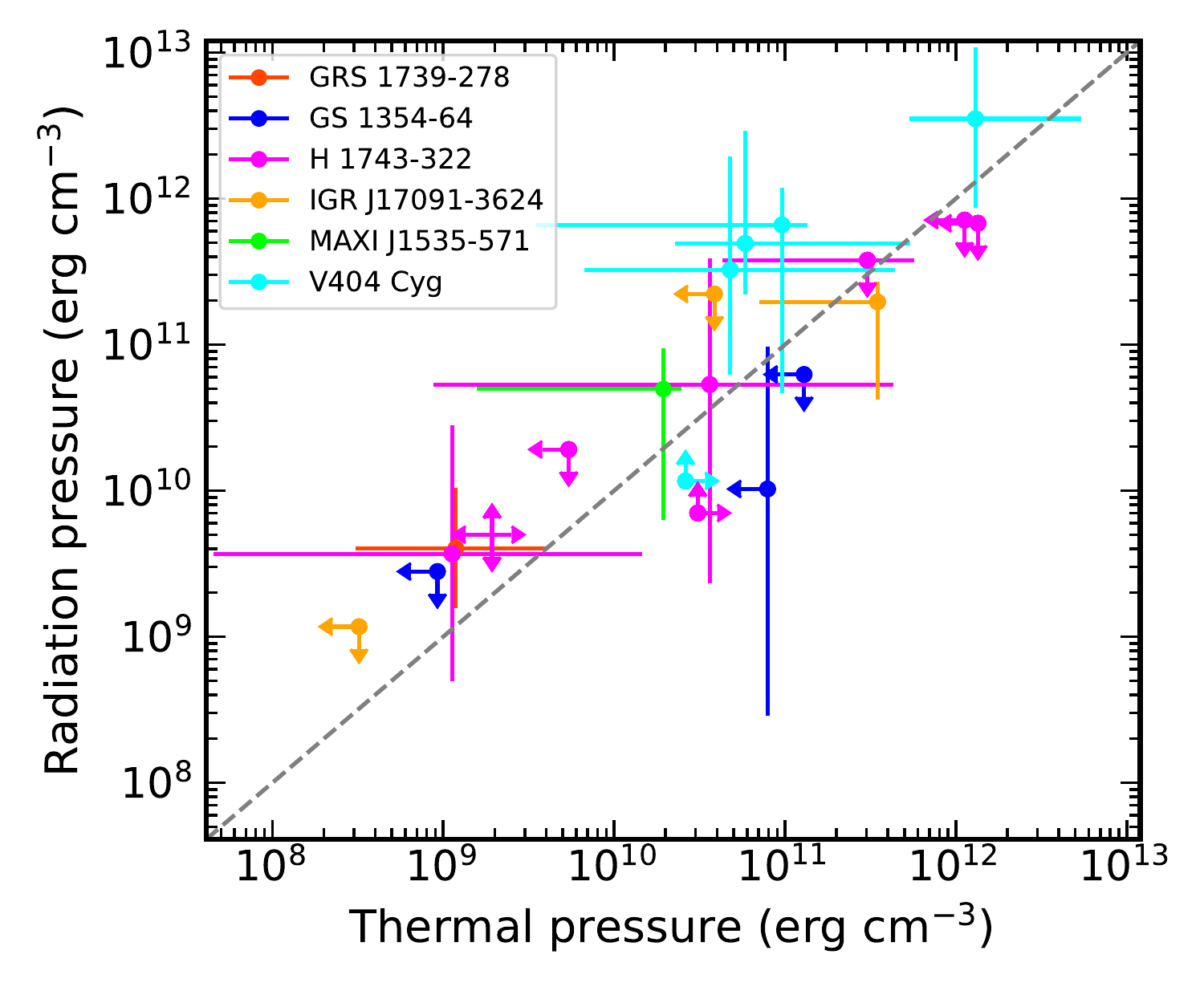}
    \caption{(Left) The temperature at the Thomson depth of $\tau=1$ of the corona-illuminated disk as a function of the density. The coronal emission component and the disk properties are assumed to be the mean values from our samples ($\Gamma=1.7$, $E_{\rm cut}$=100~keV, $\xi$=100 erg~cm~s$^{-1}$ and $A_{\rm Fe}$=1). The calculation is conducted with the \texttt{reflionx} code. (Right) The relation between the incident radiation pressure and the gas thermal pressure from this work. The grey line marks where $P_{\rm rad}=P_{\rm th}$.}
    \label{pressure_balance}
\end{figure*}

In \cite{Mallick2022}, the authors find a balance between the incident radiation pressure ($P_{\rm rad}=L/4\pi r^2c=\xi n_{\rm e}/4\pi c$ where $L$ is the corona luminosity) and the disk gas thermal pressure ($P_{\rm th}=n_{\rm e}k_{\rm B}T$). We explore this possibility with our XRBs sample. The radiation pressure can be directly calculated from the measured spectral parameters. To calculate the gas thermal pressure, we run the \texttt{reflionx} code to obtain the temperature at the Thomson depth $\tau=1$ for $\log(n_{\rm e})=16-22$ (see the left panel of Fig.~\ref{pressure_balance}). The inputs for the reflection calculation are chosen to be the average on our sample, i.e., $\Gamma=1.7$, $E_{\rm cut}$=100~keV, $\xi$=100 erg~cm~s$^{-1}$ and $A_{\rm Fe}$=1. The results are shown in the right panel of Fig.~\ref{pressure_balance}, where the dashed line shows $P_{\rm rad}=P_{\rm th}$. We can see that there is indeed a balance between the incident radiation pressure and the gas thermal pressure. Note that in this comparison we are neglecting the role of the magnetic field and the radiation pressure from the disk.

\subsection{Disk inner radius}
\label{dis_rin}

\begin{figure*}
    \centering
    \includegraphics[width=0.99\linewidth]{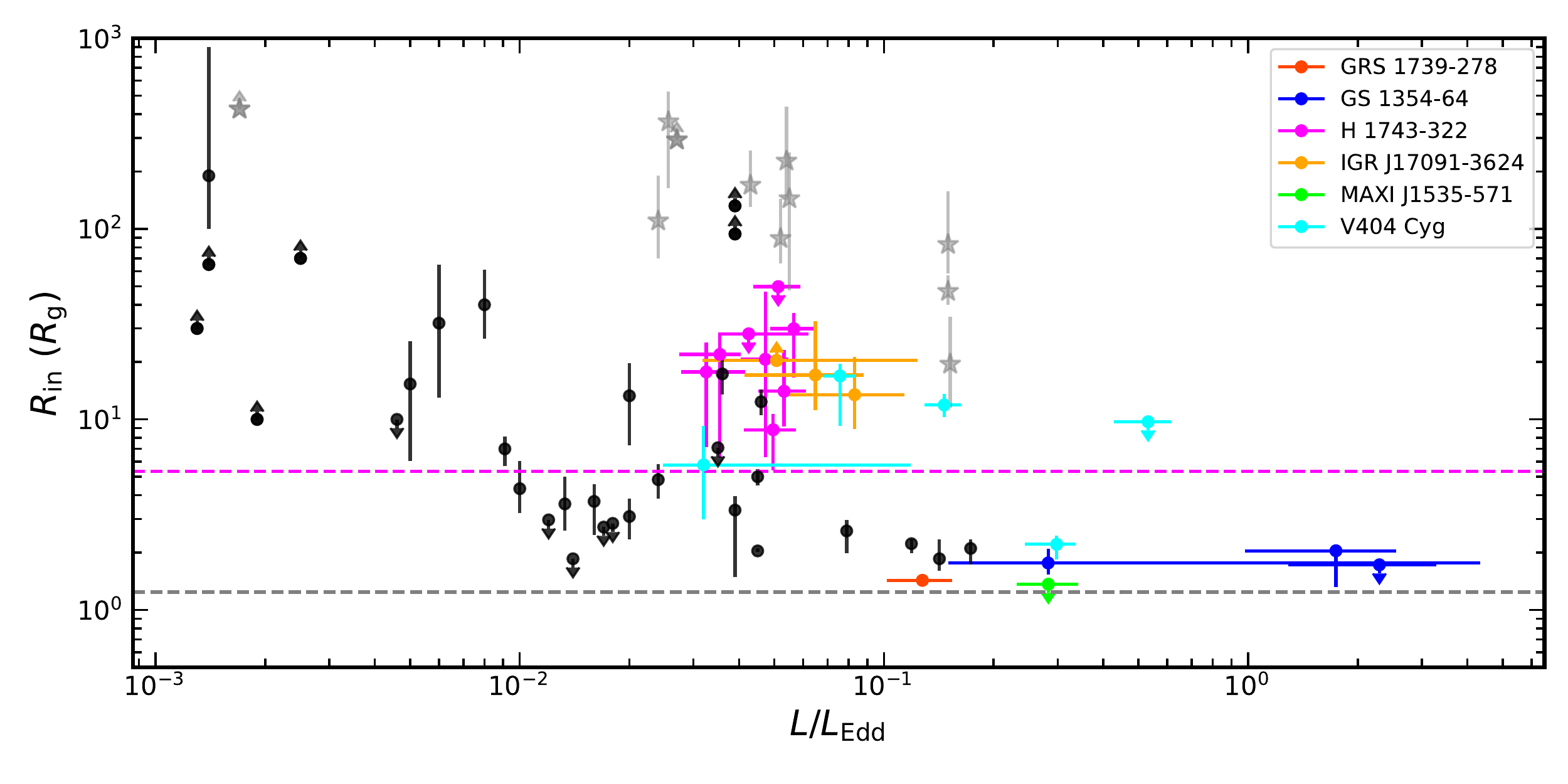}
    \caption{The evolution of the inner disk radius with the Eddington-scaled luminosity. Our results are color-coded as in Fig.~\ref{ne_mdot2}. The data in black and grey are from previous measurements of GX~339--4 \citep{Miller2006ApJ...653..525M, Reis2008,Tomsick2008ApJ...680..593T, Tomsick2009, Shidatsu2011, Kolehmainen2014MNRAS.437..316K, Petrucci2014, Plant2015, Garcia2015,Basak2016, Wang2018} where the grey stars mark the measurements with data in the timing mode of \textit{XMM-Newton} EPIC/pn. The grey and magenta dashed horizontal lines represent the ISCO radii for $a_*=$ 0.998 and 0.2 respectively.}
    \label{rin_edd}
\end{figure*}

The inner disk radius ($R_{\rm in}$) is an important parameter to understand the accretion process. The disk truncation scenario has been commonly used to explain the distinctive states of XRBs \citep[e.g.][]{Esin1997}. It is generally believed that the standard cold disk is truncated at large radius in the low hard state when the source luminosity is low \citep[e.g.][]{Tomsick2009} and it reaches the innermost stable circular orbit (ISCO) in the high luminosity soft state \citep[e.g.][]{Gierlinski2004, Steiner2010}. However, it is still under debate whether the disk is truncated in the bright hard state \citep[e.g.][]{Dzielak2019, Mahmoud2019}. We plot our reflection-based measurement of the inner disk radius versus the Eddington-scaled luminosity in Fig.~\ref{rin_edd} along with previous measurements in the hard state of GX~339--4, which is one of the best-studied sources on $R_{\rm in}$ \citep[see also Fig.~8 of][]{Wang2018}. For GX~339--4, the measurements of $R_{\rm in}$ from timing mode data of \textit{XMM-Newton} EPIC/pn are systematically larger than those from \textit{NuSTAR} or \textit{RXTE} data. This may be due to the complex pile-up effects that are hard to eliminate \citep[see the discussion in][]{Wang2018}.

Our measurements of $R_{\rm in}$ from six black hole XRBs in the hard state are broadly consistent with the trend of GX~339--4 and we extend this tendency to the regime where $L/L_{\rm Edd}>20\%$. We can see that a small $R_{\rm in}$ ($<10~R_{\rm g}$) is found when $L/L_{\rm Edd}>1\%$ even though all the data in Fig.~\ref{rin_edd} are from the hard state. For $L/L_{\rm Edd}>10\%$, almost all measurements are consistent with $R_{\rm in}$ being smaller than twice of the ISCO size (assuming $a_*=0.998$). These results support the idea that the inner disk radius can reach the ISCO in the bright hard state. Although it appears that the H~1743-322 disk is truncated at a large radius in the hard state, the black hole in this system has a low spin measured by the independent continuum-fitting method \citep[$a_*\sim 0.2$,][]{Steiner2012} and thus the ISCO size is large ($R_{\rm ISCO}=5.3~R_{\rm g}$, see the magenta dashed line in Fig.~\ref{rin_edd}). Things are more complicated for IGR~J17091--3624 since its black hole spin parameter is still uncertain (see Sec.~\ref{igrj17091}).

\subsection{Inclination angle}

In this analysis, we always fit the inclination as a free parameter. It allows us to compare the inclination angle measured from reflection modelling with that from other methods (often for the jet inclination or the binary inclination). For GS~1354--64, only an upper limit of 79$^{\circ}$ has been obtained for its binary inclination through the absence of X-ray eclipses \citep{Casares2009ApJS..181..238C}. There is a tight constraint on the binary inclination of V404~Cyg \citep[($67_{-1}^{+3}$)$^{\circ}$,][]{Khargharia2010} while the reflection analysis gives a lower value. 

Similar discrepancies also have been found in other sources, e.g. for Cygnus~X-1 the reflection spectra always require an inclination around 40$^{\circ}$ \citep[e.g.][]{Tomsick2014, Walton2016, Liu2019} while its binary inclination is $27.51_{-0.57}^{+0.77}$ \citep{Miller-Jones2021Sci}. We note that the relativistic reflection spectra are only sensitive to the inner part of the accretion disk while the inner disk inclination may not align with the orbital inclination, which leads to a warped disk \citep[e.g.][]{Pringle1996}. Simulations have shown that the misalignment between the black hole spin axis and orbital angular momentum can be common for black hole XRBs \citep[e.g.][]{Brandt1994,Fragos2010}. Observational indications of a warped disk have been found in a few sources, e.g. MAXI~J1535--571 \citep{Miller2018} and MAXI~J1820+070 \citep{Poutanen2022,Thomas2022}. The warped disk may have an impact on the relativistic reflection spectra \citep[e.g.][]{Abarr2021} but a detailed analysis along this aspect is out of the scope of this work.

For H~1743--322, \cite{Steiner2012} measured an inclination angle of $75^{\circ}\pm3^{\circ}$ for its large-scale ballistic jets, which are supposed to be aligned with the black hole spin axis. The inner disk region should also align with the spin axis due to the Bardeen-Petterson effect \citep{Bardeen1975ApJ...195L..65B} while the reflection model gives a lower measurement of $30^{\circ}-40^{\circ}$. We note that the inclination angle measurements with reflection analysis can be affected by systematic uncertainties, e.g. the lack of knowledge of the corona geometry or simplifications in the model calculations \citep[see discssuions in][]{Bambi2021}. In some cases, the systematic uncertainty can be as large as $\sim30^{\circ}$ \citep[e.g.][]{Garcia2018ApJ86425G,Connors2022}. This might explain the discrepancy we are seeing.


\subsection{Reflection strength}

Note that the reflection strength ($R_{\rm str}$) defined in this work is not the reflection fraction ($R_{\rm f}$) parameter in \cite{Dauser2016} which is the ratio between the corona intensity that shines on the disk and that reaches infinity. $R_{\rm f}$ is determined by the accretion geometry while $R_{\rm str}$ could be affected by inclination between the line-of-sight and the black hole spin axis \citep{Dauser2016}. We note that, in our sample, most sources are consistent with an inclination in the range of 20$^\circ$-30$^\circ$ (except for MAXI~J1535--571). Therefore, $R_{\rm str}$ can still provide insight into the disk-corona geometry. In the left panel of Fig.~\ref{Rs}, we are showing the relation between $R_{\rm str}$ and the Eddington-scaled X-ray luminosity from our analysis. There is no strong correlation between the two parameters with a Pearson correlation coefficient ($r$) of 0.32 ($1-p=75\%$). We also test the correlation between the photon index and the reflection strength (see the middle panel of Fig.~\ref{Rs}) but still find no strong correlation ($r=0.22, 1-p=66\%$).

\subsection{Photon index}

\begin{figure*}
    \centering
    \includegraphics[width=0.33\linewidth]{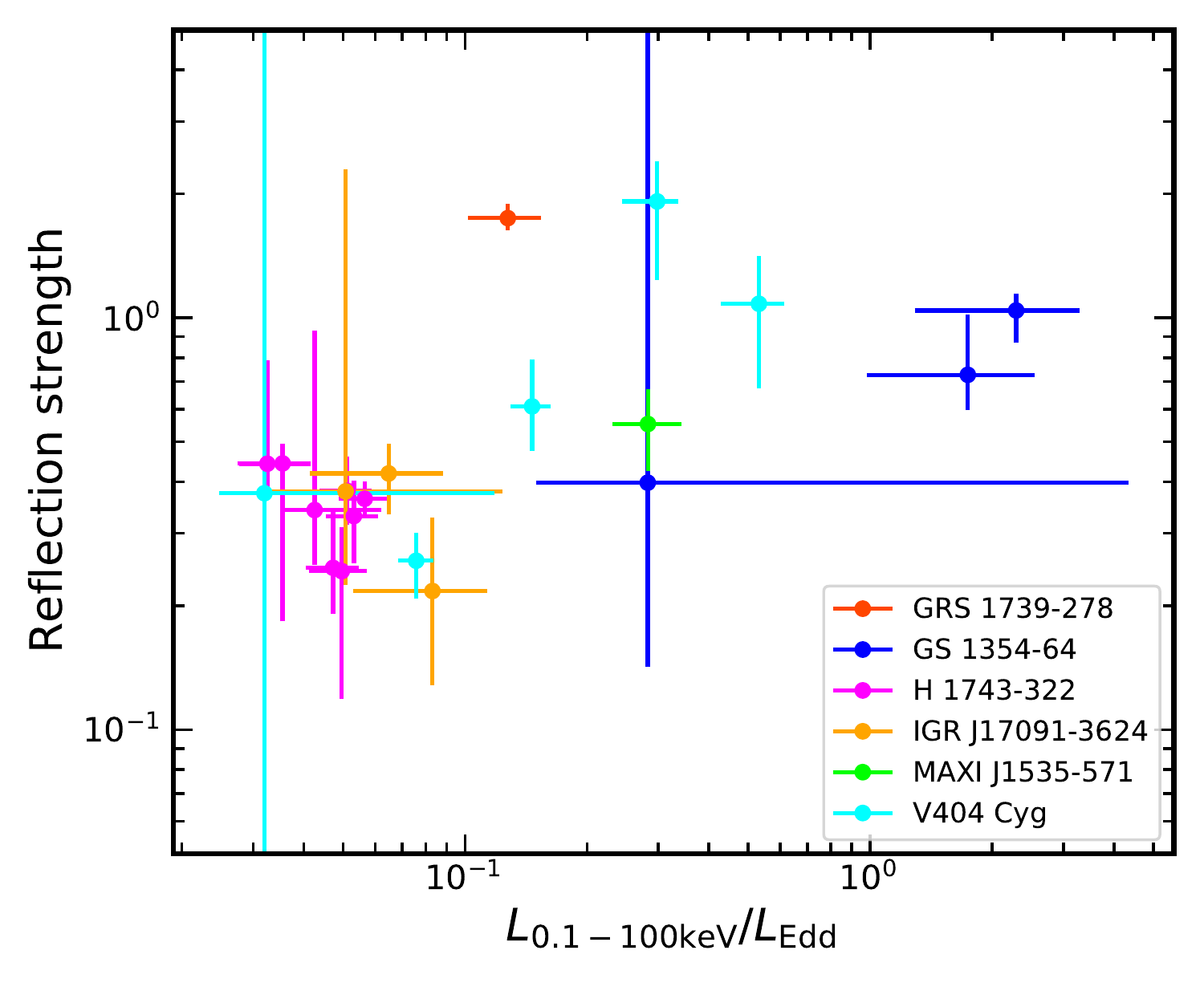}
    \includegraphics[width=0.33\linewidth]{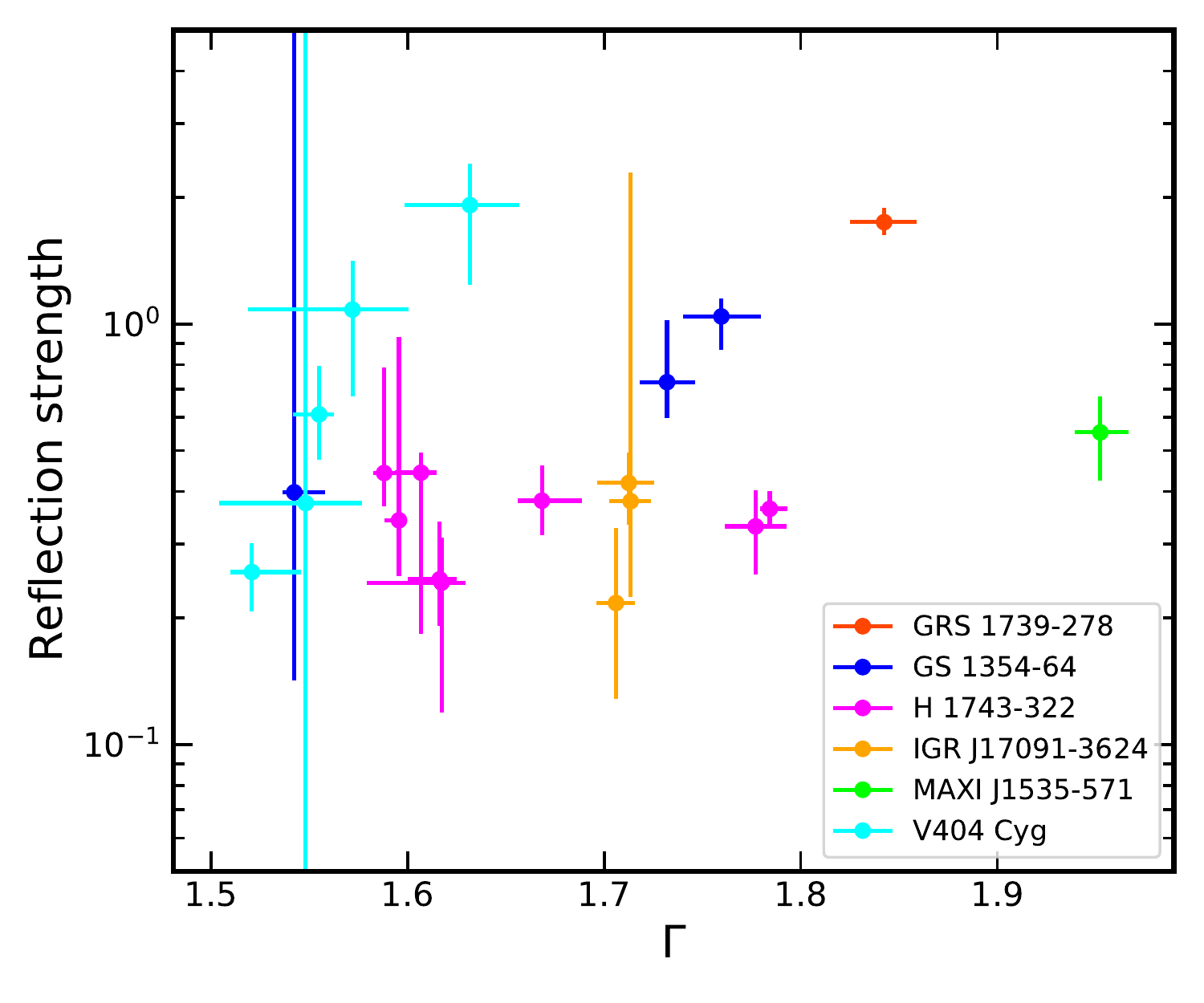}
    \includegraphics[width=0.33\linewidth]{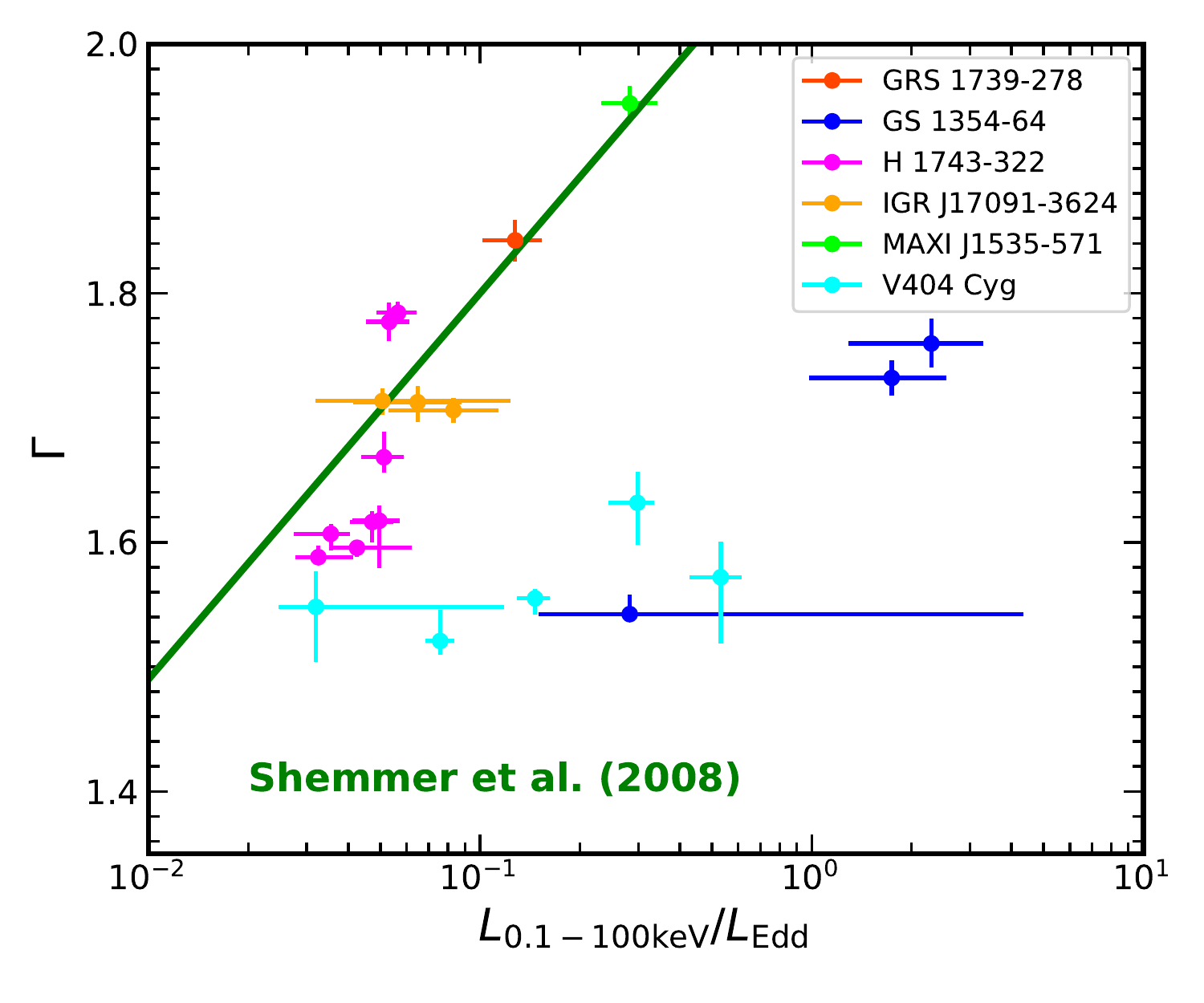}
    \caption{(Left) The relation between the reflection strength parameter ($R_{\rm str}$, defined as the flux ratio between the reflected component and the power-law component in 0.1--100 keV band) and the Eddington-scaled luminosity. (Middle) The relation between $R_{\rm str}$ and the photon index $\Gamma$. (Right) The variation between the photon index $\Gamma$ and Eddington-scaled luminosity. The line in green represents the statistical relation in AGNs found by \cite{Shemmer2008}.}
    \label{Rs}
\end{figure*}

It is well known in the literature that there is a positive statistical correlation between the Eddington ratio ($\lambda_{\rm Edd}=L_{\rm bol}/L_{\rm Edd}$) and X-ray photon index ($\Gamma$) at high accretion rates in AGNs \citep[e.g.][]{Shemmer2008, Risaliti2009, Brightman2013} and XRBs \citep[e.g.][]{Kubota2004, Wu2008ApJ...682..212W, You2023ApJ...945...65Y}. At low accretion rates \citep[e.g. $\lambda_{\rm Edd}<1\%$,][]{Yuan2007ApJ...658..282Y, Constantin2009ApJ...705.1336C}, a negative correlation is often found \citep[e.g.][]{Gu2009MNRAS.399..349G, Diaz2022arXiv221015376D}. In the regime of extremely low accretion rates (e.g. $\lambda_{\rm Edd}<10^{-4}$), $\Gamma$ has been found to be saturated \citep[e.g.][]{Corbel2006, Sobolewska2011MNRAS.417..280S, Gultekin2012, Plotkin2013}. This switch between correlation behaviors may suggest transition between different accretion modes \citep[e.g.][]{Cao2014ApJ...788...52C, Yang2015MNRAS.447.1692Y}. We test the correlation with our sample in the right panel of Fig.~\ref{Rs}. Our data are all in the range of $\lambda_{\rm Edd}>1\%$ and divided into two branches on the $\Gamma$-$\lambda_{\rm Edd}$ diagram. In the lower branch, the data of V404~Cyg are not from canonical outbursts but strong repeated flaring events that last less than 1~ks (see Sec.~\ref{v404}). Moreover, the flaring events might be related to the transient jet activities instead of changes of the mass accretion rate \citep{Walton2017}. These may explain why $\Gamma$ stays stable even though the X-ray luminosity changes by more than one order of magnitude. Besides V404~Cyg, the data of GS~1354--64 are also deviating from the trend of other sources. If considering only the upper branch, there is a strong positive correlation between the two parameters ($r=0.83, 1-p=99.96\%$). Moreover, we plot in Fig.~\ref{Rs} the statistical relation found in AGNs by \cite{Shemmer2008}, which is in good agreement with our upper branch. This indicates that the mass accretion rate changes the physical properties of the hot corona in a similar manner for XRBs and AGNs.

\subsection{The corona properties in the compactness-temperature plane}
\label{corona_tem}

The coronal properties can be studied on the compactness-temperature ($\ell-\Theta$) plane. The compactness is defined as:
\begin{equation}
\ell=\frac{L}{R}\frac{\sigma_{\rm T}}{m_{\rm e}c^3}
\end{equation}
where $L$ is the luminosity, $R$ is the source radius assuming a spherical geometry, $\sigma_{\rm T}$ the Thomson cross section and $m_{\rm e}$ the electron mass. By studying a sample of \nustar{} measurements AGNs and black hole XRBs on the, \cite{Fabian2015MNRAS.451.4375F} find that most of the measurements are clustered close to the limit set by electron-positron pair production \citep[see also][]{Ricci2018MNRAS.480.1819R}.  This supports the idea that pair production is an important ingredient in the corona. This process could lead to runaway pair production to share the energy and thus limits the coronal temperature below the pair line \citep[e.g.][]{Svensson1984,Zdziarski1985ApJ...289..514Z,Stern1995,Coppi1999}. Moreover, the fact that many sources are above the electron-electron coupling line means that energetic electrons do not have enough time to thermalize, which suggests the possibility of a magnetized and hybrid plasma that contains both thermal and non-thermal particles \citep[e.g.][]{Zdziarski1993ApJ...414L..93Z,Grove1998ApJ...500..899G}. The inclusion of a non-thermal component would tend to lower the temperature of the thermal component and produce a high-energy excess compared to the spectra of purely thermal Comptonization. This behavior matches well with the data of GRS~1739--278 which has the lowest observed coronal temperature in our sample. \cite{Fabian2017} investigated the hybrid plasma scenario with a sample of AGNs and found that a non-thermal fraction of 10\%-30\% could account for most of the objects.

\begin{figure}
    \centering
    \includegraphics[width=0.99\linewidth]{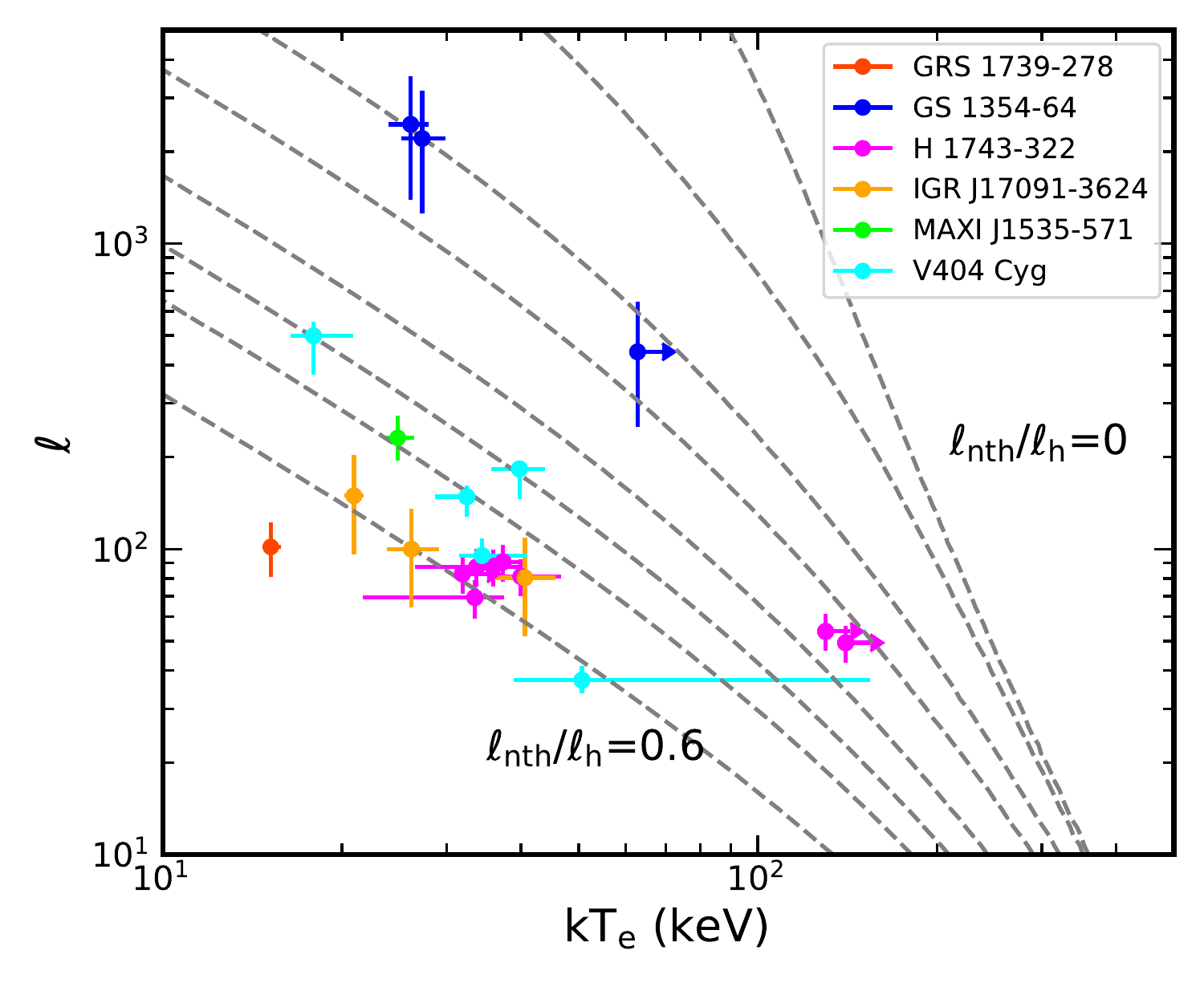}
    \caption{The relation between the coronal temperature and and the dimensionless compactness parameter from our analysis. The grey dashed lines are theoretical predictions of a hybrid plasma with $\ell_{\rm h}/\ell_{\rm s}=1$, where  $\ell_{\rm h}$ is the total heating parameter and $\ell_{\rm s}$ is the compactness of the soft photons. From right to left, the lines represent solutions with a non-thermal fraction ($\ell_{\rm nth}/\ell_{\rm h}$) of 0, 0.01, 0.05, 0.1, 0.2, 0.3, 0.4 and 0.6 respectively.}
    \label{compact}
\end{figure}

We plot the measurements of our sample on the compactness-temperature diagram along with the theoretical predictions of hybrid plasma to understand the physical properties of the corona in the hard state (Fig.~\ref{compact}). The calculations of the theoretical lines are done with the \texttt{eqpair}\footnote{The description of the model can be found here: \url{http://www.astro.yale.edu/coppi/eqpair/eqpap4.ps}. The model is available in \texttt{XSPEC}: \url{https://heasarc.gsfc.nasa.gov/xanadu/xspec/manual/XSmodelEqpair.html}.} model \citep{Coppi1999}. The non-thermal fraction is defined as the ratio between the compactness parameter of non-thermal and the total heating power ($\ell_{\rm nth}/\ell_{\rm h}$). For a fixed heating power, a higher non-thermal fraction results in a lower equilibrium temperature. We assume a corona size of 10~$R_{\rm g}$, which is a reasonable value given existing measurements on AGNs with X-ray reflection modelling or reverberation analysis \citep[e.g.][]{Fabian2009Natur.459..540F,Wilkins2011MNRAS.414.1269W,Emmanoulopoulos2014}.

Fig.~\ref{compact} shows that all of our measurements are below the theoretical line of purely thermal plasma. A large fraction (15/21) of our data are below the line for $\ell_{\rm nth}/\ell_{\rm h}=30\%$ which suggests an even higher non-thermal fraction. We note that the hybrid plasma model has been applied to a number of black hole XRBs \citep[e.g.][]{Gierlinski1999, Zdziarski2001, Droulans2010, Parker2015,Zdziarski2021} and in some cases a significant non-thermal fraction is indeed required in the hard state \citep{Wardzinski2002,Nowak2011,DelSanto2016}. 

For individual sources, we can see that GS~1354--64 and IGR~J17091--3624 follow the expected trend for a pair-dominated plasma that the coronal temperature drops when it gets radiatively more compact. This is even true for the flux-resolved data of V404~Cyg although the timescale of the flaring activities we analyze here is as short as $\sim$~ks (see Sec.~\ref{v404}). For H~1743--322, there are two outliers with higher temperatures compared to the other six observations with a similar level of radiation compactness. This may suggest a lower non-thermal fraction but it is also possible that the assumption of 10~$R_{\rm g}$ for the corona size is too small for the two observations. The latter may not be the main reason since the measurements of $R_{\rm in}$ for the two observations are consistent with 10~$R_{\rm g}$. We also note that the two outliers are from the so-called ``failed'' outbursts \citep[e.g.][]{Stiele2021} during which the source stays in the low hard state only.

We further investigate these H~1743--322 outliers by considering the correlation between X-ray and radio luminosity. This source is known to switch between two different radio-X-ray correlations: the `radio loud' and `radio quiet' branches \citep{Coriat2011MNRAS.414..677C}. We therefore explore whether our inferred high/low non-thermal fraction corresponds to the source being on the radio loud/quiet branch, which may be expected if, for example, the non-thermal fraction is related to the jet properties. We find no such correspondance. Only one of our outliers has a simultaneous radio observation, and it belongs to the radio quiet branch (1.28 GHz flux density of $0.88\pm0.05$~mJy; \citealt{Williams2020MNRAS.491L..29W}), as do several of the non-outliers.

\section{Conclusion}

In this work, we analyzed the broadband X-ray spectra of six stellar-mass black hole XRBs in the hard state with data from 
\nustar{} and \sw{}. The purpose is to test the effect of X-ray reflection by a high-density accretion disk. The main results are as follow:
\begin{itemize}
\item The model with the disk electron density fixed at $10^{15}$~cm$^{-3}$ systematically overestimates the ionization degree of the disk atmosphere. The measurement of the inner disk radius is not affected by the assumption of disk density.
\item The X-ray spectra of black hole XRBs require a higher disk density than that of AGNs. For the selected observations in this work, the accretion disk should be dominated by gas pressure. However, the measured densities are lower than either the prediction of gas or radiation pressure-dominated disk. The discrepancy might represent the vertical density structure of the disk.
\item From our analysis and previous measurements, we find that the inner disk radius can be close to the ISCO in the hard state.
\item We find that the reflection strength is not correlated to the Eddington ratio or photon index. There is a strong correlation between the Eddington ratio and photon index which is in good agreement with the statistical relation found in AGNs.
\item The coronal temperature is lower than the prediction of a purely thermal plasma and can be explained by a hybrid plasma with a non-thermal fraction around 30\%. 
\end{itemize}

{\bf Acknowledgments --} This work was supported by the National Natural Science Foundation of China (NSFC), Grant No.~12250610185 and Grant No.~11973019, the Natural Science Foundation of Shanghai, Grant No.~22ZR1403400, the Shanghai Municipal Education Commission, Grant No.~2019-01-07-00-07-E00035, and Fudan University, Grant No.~JIH1512604. JJ acknowledges support from the Leverhulme Trust, the Isaac Newton Trust and St Edmund's College. JAT acknowledges partial support from NASA under Astrophysics Data Analysis Program (ADAP) grant 80NSSC19K0586.



\bibliographystyle{apj}
\bibliography{bibliography}



\clearpage
\appendix

\section{Modeling with \texttt{relxillcp}}
\label{relxillcp}

Besides the \texttt{reflionx}-based models, the \texttt{relxill} \citep{Garcia2014} model is also commonly used to fit relativistic reflection spectra. To compare our results with existed analysis in the literature, we replace the reflection component with the model \texttt{relxillcp} (Model 2), which is a flavor of the package \texttt{relxill} v1.4.3. This model assumes a electron density of the disk $\log(n_{\rm e}/{\rm cm}^{-3})=$ 15\footnote{Note that the latest version of \texttt{relxillcp} allows a variable electron density in the range $\log(n_{\rm e}/{\rm cm}^{-3})$= 15-20 (see \url{http://www.sternwarte.uni-erlangen.de/~dauser/research/relxill/})}. Moreover, relativistic effects are properly taken into account by combining the rest-frame reflection code \texttt{xillver} \citep{Garcia2010, Garcia2013} and the relativistic broadening code \texttt{relline} \citep{Dauser2010,Dauser2013}. The reflection fraction is set to -1 so the model returns only the reflection component. The best-fit parameters are shown in Tab.~\ref{para_relxill}.

\section{Spectral analysis results for individual sources} 
\label{detail}

\subsection{GRS~1739--278}
\label{grs1739}

GRS~1739--278 is discovered in 1996 by \textit{Granat} \citep{Paul1996}. It is located at a distance of 6--8.5 kpc \citep{Greiner1996}. The \nustar{} observation in the analysis has been studied by \cite{Miller2015} without including the \sw{} data. The authors tested both \texttt{reflionx} and \texttt{relxill}-based reflection models and found a high black hole spin ($a_*=0.8\pm0.2$), an intermediate inclination (20$^{\circ}$--40$^{\circ}$) and an inner disk radius close to the ISCO. We obtain similar measurements with our reflection modelling (see Tab~\ref{para_n15} and Tab~\ref{para_relxill}). The very low temperature of the corona has already been revealed by \cite{Miller2015}. There is another \nustar{} observation in the archive (obsID: 80101050002) from September 2015 but the reflection features are too weak \citep{Furst2016grs1739}.

\subsection{GS~1354--64}
\label{gs1354}

GS~1354--64 is a Galactic black hole candidate discovered in 1987 \citep{Makino1987}. Only a lower limit ($\sim~8~M_{\odot}$) has been determined for its black hole mass and its distance is also not well-constrained \citep[25-61~kpc,][]{Casares2009ApJS..181..238C}. The first two observations in June and July in our analysis have been studied by \cite{El-Batal2016ApJ...826L..12E} without including the \sw{} data. The authors found a large truncation of the accretion disk for the June observation with a low density \texttt{relxill} model, which is consistent with our results in Tab.~\ref{para_relxill}. The high disk inclination and small inner disk radius measured from the July observation are also in agreement with previous studies \citep[e.g.][]{El-Batal2016ApJ...826L..12E, Xu2018ApJ...865..134X}.

\subsection{IGR~J17091--3624}
\label{igrj17091}

IGR~J17091--3624 is a Galactic black hole candidate at a distance of 11~kpc to 17~kpc (\citealt{Rodriguez2011}, but see \citealt{Altamirano2011ApJ...742L..17A} for debates on its distance.). The black hole spin parameter is still quite uncertain. Previous studies have reported both negative spin \citep{Rao2012, Wang2018MNRAS.478.4837W} and high spin \citep[$a_*>0.9$,][]{Reis2012ATel.4382....1R} measurements. The three \nustar{} and \sw{} observations in this analysis have been studied by \cite{Xu2017} with \texttt{relxill}-based low density reflection models. By fixing $a_*=0.998$, the authors found a truncated disk ($R_{\rm in}\sim 20R_{\rm g}$) and an intermediate disk inclination angle ($i\sim 30^{\circ}-40^{\circ}$), which are confirmed with our analysis (see Tab.~\ref{para_relxill}).

\subsection{V404~Cyg} 
\label{v404}

V404~Cyg is a dynamically confirmed black hole XRB located at $2.39\pm0.14$~kpc \citep{Miller-Jones2009}. The black hole mass should be $9.0_{-0.6}^{+0.2}~M_{\odot}$ \citep{Khargharia2010}. The two observations of V404~Cyg analyzed in this work show strong flaring variability \citep[see Fig.~2 of][]{Walton2017} with the count rate (per FPM) changing from 100 to more than 10$^4$ ct~s$^{-1}$. It was shown by \cite{Walton2017} that there are strong relativistic reflection features in the X-ray spectra and it is possible to constrain the black hole spin parameter if dividing the data according to flux levels and neglecting the data with deep absorption edge. We process the data in the same way as \cite{Walton2017} and obtain the spectra for five flux levels (see Tab.~1 of \cite{Walton2017}). 

To fit the spectra of V404~Cyg, it requires to include additional components: a neutral local absorption layer, an ionized absorption layer and a distant reflection component. The ionized absorption is modeled with a grid calculated with \texttt{XSTAR} \citep{Kallman2001, Kallman2004}. The full model of our Model 2 for this source reads as: \texttt{constant * tbabs$_{\rm gal}$ * (tbabs$_{\rm local}$ * xstar * (cflux * nthcomp + cflux * relxillcp) + cflux * xillvercp)}. The column density for the Galactic absorption is fixed at 10$^{22}$~cm$^{-2}$ and the absorption column locally to the source is a free parameter. The ionization state of the distant reflection component (\texttt{xillvercp}) is fixed at $\log(\xi)=0$ and the other parameters are linked to the relativistic reflection component.

\subsection{H~1743--322}
\label{h1743}

H~1743--322 was discovered in 1977 \citep{Kaluzienski1977}. It is located at $8.5\pm0.8$~kpc and the jet inclination is $75^{\circ}\pm 3^{\circ}$ \citep{Steiner2012}. The continuum fitting method indicates a black hole spin around $a_*\sim 0.2$ \citep{Steiner2012}. \cite{Molla2017} estimated the black hole mass to be $11.21_{-1.96}^{+1.65}~M_{\odot}$ using the two-component advective flow (TCAF) solution and the correlation between photon index and frequency of the quasi-periodic oscillation (QPO). 

There are eleven \nustar{} observation in the archive, three of which do not show reflection features. The other eight observations analyzed in this work are all in the hard state. The two observations from 2016 have been analyzed by \cite{Chand2020} and a truncated disk was found. However, our analysis with \texttt{relxillcp} model requires a disk extends to the ISCO. This difference is possibly due to the fact that we treat the inclination angle as a free parameter while in \cite{Chand2020} it is fixed at 75$^\circ$. The two 2018 observations have also been analyzed with reflection models assuming a fixed inclination angle \citep{Stiele2021} and again their measurements on $R_{\rm in}$ are larger than ours.

\subsection{MAXI~J1535--571}
\label{maxij1535}

MAXI~J1535--571 is a black hole candidate discovered in 2017 \citep{Negoro2017}. It is located at a distance of $4.1_{-0.5}^{+0.6}$~kpc \citep{Chauhan2019}. Previous reflection analysis indicates a high black hole spin and high inclination angle \citep[e.g.][]{Xu2018maxij1535, Miller2018}. Following \cite{Xu2018maxij1535}, we include a disk component and a distant reflection component to the full model to fit the \nustar{} data. The high inclination angle is well recovered with our fit (see Tab~\ref{para_relxill}).


\begin{sidewaystable}
    \centering
    \caption{Best-fit parameters with \texttt{reflionx} that has variable electron density}
    \label{para_reflionx}
    \renewcommand\arraystretch{1.8}
    \begin{tabular}{lcccccccccccc}
        \hline\hline
        Source & Date & nH & $R_{\rm in}$ ($R_{\rm g}$) & $\Gamma$ & $\log(n_{\rm e})$ & $\xi$ & kT$_{\rm e}$ (keV) & Incl (deg) & $A_{\rm Fe}$ (solar) & $\log(F_{\rm ref})$ & $\log(F_{\rm po})$ & $\chi^2/\nu$ \\
\hline
GRS 1739-278 & 20140326 & $3.15_{-0.05}^{+0.12}$ & $1.43_{-0.07}^{+0.08}$ & $1.842_{-0.017}^{+0.016}$ & $19.0_{-0.4}^{+0.4}$ & $145.0_{-10}^{+10}$ & $15.2_{-0.5}^{+0.6}$ & $24.2_{-4}^{+2.5}$ & $0.500_{-P}^{+0.011}$ & $-7.810_{-0.023}^{+0.03}$ & $-8.053_{-0.018}^{+0.015}$ & 3431.9/3238 \\
GS 1354-64 & 20150613 & $1.18_{-0.07}^{+0.07}$ & $1.76_{-0.23}^{+0.3}$ & $1.542_{-0.006}^{+0.016}$ & $19.5_{-1.4}^{+1.1}$ & $5.0_{-P}^{+50}$ & $500_{-430}^{+P}$ & $25_{-P}^{+10}$ & $5.0_{-1.6}^{+P}$ & $-9.4_{-0.5}^{+1.7}$ & $-8.960_{-0.021}^{+0.07}$ & 2030.5/1872 \\
GS 1354-64 & 20150711 & $1.21_{-0.07}^{+0.07}$ & $2.04_{-0.7}^{+0.16}$ & $1.732_{-0.014}^{+0.014}$ & $17.7_{-P}^{+1.2}$ & $115_{-5}^{+6}$ & $27.3_{-2.1}^{+2.5}$ & $20_{-P}^{+9}$ & $0.50_{-P}^{+0.13}$ & $-8.40_{-0.08}^{+0.15}$ & $-8.262_{-0.015}^{+0.015}$ & 3115.3/2918 \\
GS 1354-64 & 20150806 & $0.44_{-0.19}^{+0.24}$ & $1.3_{-P}^{+0.4}$ & $1.760_{-0.019}^{+0.02}$ & $19.5_{-1.5}^{+1}$ & $130_{-11}^{+30}$ & $26.1_{-2.2}^{+1.9}$ & $22_{-P}^{+8}$ & $0.50_{-P}^{+0.06}$ & $-8.197_{-0.08}^{+0.029}$ & $-8.22_{-0.02}^{+0.03}$ & 2898.9/2821 \\
IGR J17091 & 20160307 & $1.79_{-0.07}^{+0.07}$ & $55_{-34}^{+P}$ & $1.713_{-0.011}^{+0.01}$ & $18.0_{-P}^{+0.64}$ & $34_{-13}^{+71}$ & $41_{-5}^{+5}$ & $41_{-18}^{+P}$ & $0.50_{-P}^{+0.07}$ & $-9.06_{-0.23}^{+0.8}$ & $-8.641_{-0.012}^{+0.012}$ & 2821.7/2751 \\
IGR J17091 & 20160312 & $2.05_{-0.15}^{+0.08}$ & $17_{-6}^{+16}$ & $1.712_{-0.016}^{+0.013}$ & $20.99_{-0.6}^{+0.02}$ & $75_{-22}^{+28}$ & $26.2_{-2.4}^{+3.0}$ & $22.0_{-P}^{+9.0}$ & $0.58_{-P}^{+0.15}$ & $-8.92_{-0.1}^{+0.07}$ & $-8.547_{-0.012}^{+0.009}$ & 2368.3/2409 \\
IGR J17091 & 20160314 & $1.57_{-0.07}^{+0.19}$ & $14_{-5}^{+8}$ & $1.71_{-0.01}^{+0.01}$ & $18.0_{-P}^{+2.3}$ & $190_{-40}^{+320}$ & $21.0_{-0.8}^{+0.8}$ & $17.3_{-P}^{+7.7}$ & $0.50_{-P}^{+0.08}$ & $-9.04_{-0.23}^{+0.18}$ & $-8.372_{-0.011}^{+0.011}$ & 2305.0/2321 \\
H 1743-322 & 20140918 & $1.99_{-0.04}^{+0.27}$ & $6_{-P}^{+22}$ & $1.5956_{-0.007}^{+0.0029}$ & $21.33_{-5}^{+0.09}$ & $33_{-8}^{+87}$ & $34_{-12}^{+4}$ & $5_{-P}^{+21}$ & $0.65_{-0.07}^{+0.12}$ & $-8.82_{-0.13}^{+0.4}$ & $-8.357_{-0.025}^{+0.004}$ & 3039.2/3006 \\
H 1743-322 & 20140923 & $1.46_{-0.5}^{+0.11}$ & $8.8_{-3}^{+1.9}$ & $1.617_{-0.04}^{+0.012}$ & $18.3_{-P}^{+1.3}$ & $180_{-36}^{+30}$ & $33.7_{-7}^{+4}$ & $28_{-4}^{+4}$ & $0.5_{-P}^{+0.81}$ & $-8.87_{-0.3}^{+0.11}$ & $-8.2562_{-0.0029}^{+0.022}$ & 3324.7/3187 \\
H 1743-322 & 20141009 & $2.03_{-0.23}^{+0.21}$ & $21_{-14}^{+26}$ & $1.616_{-0.016}^{+0.009}$ & $20.7_{-P}^{+0.68}$ & $53_{-21}^{+64}$ & $49_{-17}^{+P}$ & $17_{-P}^{+6}$ & $1.1_{-0.3}^{+1.0}$ & $-8.89_{-0.11}^{+0.14}$ & $-8.279_{-0.009}^{+0.009}$ & 2498.7/2605 \\
H 1743-322 & 20150703 & $2.41_{-0.3}^{+0.17}$ & $18_{-11}^{+8}$ & $1.588_{-0.006}^{+0.01}$ & $22.0_{-1.9}^{+P}$ & $23.0_{-4}^{+3}$ & $500_{-360}^{+P}$ & $5_{-P}^{+16}$ & $0.71_{-0.14}^{+0.17}$ & $-8.86_{-0.08}^{+0.25}$ & $-8.505_{-0.007}^{+0.005}$ & 2387.6/2416 \\
H 1743-322 & 20160313 & $2.11_{-0.12}^{+0.11}$ & $30_{-13}^{+6}$ & $1.784_{-0.005}^{+0.009}$ & $19.0_{-0.9}^{+0.9}$ & $137_{-10}^{+12}$ & $37_{-3}^{+3}$ & $27_{-3}^{+3}$ & $0.6_{-P}^{+0.1}$ & $-8.68_{-0.04}^{+0.04}$ & $-8.2400_{-0.004}^{+0.0023}$ & 3212.4/3088 \\
H 1743-322 & 20160315 & $1.93_{-0.18}^{+0.17}$ & $14_{-5}^{+9}$ & $1.777_{-0.016}^{+0.016}$ & $20.2_{-1.3}^{+0.9}$ & $128_{-18}^{+25}$ & $36_{-4}^{+5}$ & $31_{-3}^{+3}$ & $0.77_{-0.15}^{+0.17}$ & $-8.74_{-0.11}^{+0.09}$ & $-8.256_{-0.005}^{+0.005}$ & 3146.8/3060 \\
H 1743-322 & 20180919 & $2.59_{-0.15}^{+0.16}$ & $8_{-P}^{+42}$ & $1.668_{-0.012}^{+0.021}$ & $20.9_{-0.7}^{+0.2}$ & $101_{-64}^{+14}$ & $40_{-4}^{+7}$ & $5_{-P}^{+19}$ & $0.69_{-0.1}^{+0.11}$ & $-8.71_{-0.08}^{+0.08}$ & $-8.288_{-0.007}^{+0.005}$ & 2870.0/2814 \\
H 1743-322 & 20180926 & $2.17_{-0.14}^{+0.11}$ & $22_{-16}^{+7}$ & $1.607_{-0.013}^{+0.008}$ & $19.2_{-1.6}^{+2.1}$ & $124_{-38}^{+29}$ & $500_{-370}^{+P}$ & $19_{-P}^{+6}$ & $0.73_{-0.12}^{+0.15}$ & $-8.82_{-0.4}^{+0.05}$ & $-8.467_{-0.005}^{+0.008}$ & 3200.1/3019 \\
MAXI J1535 & 20170907 & $7.7_{-0.2}^{+0.5}$ & $1.32_{-P}^{+0.05}$ & $1.952_{-0.013}^{+0.014}$ & $19.99_{-0.9}^{+0.08}$ & $190_{-20}^{+170}$ & $24.8_{-1.3}^{+1.6}$ & $78.8_{-2.4}^{+0.8}$ & $0.64_{-0.1}^{+0.08}$ & $-7.46_{-0.11}^{+0.09}$ & $-7.20_{-0.01}^{+0.01}$ & 3481.1/3131 \\
V404 Cyg & 20150624 & $1.7_{-0.5}^{+0.6}$ & $5.8_{-2.8}^{+4}$ & $1.548_{-0.04}^{+0.029}$ & $21.7_{-1.6}^{+P}$ & $123_{-88}^{+122}$ & $51_{-12}^{+104}$ & $27_{-P}^{+10}$ & $0.69_{-0.09}^{+0.5}$ & $-8.0_{-1.2}^{+1.2}$ & $-7.571_{-0.015}^{+0.04}$ & 2027.1/2057 \\
V404 Cyg & 20150624 & $3.0_{-0.5}^{+0.6}$ & $16.9_{-8}^{+2.7}$ & $1.521_{-0.011}^{+0.025}$ & $21.4_{-0.3}^{+0.5}$ & $520_{-300}^{+120}$ & $34_{-3}^{+7}$ & $5_{-P}^{+9}$ & $0.521_{-0.019}^{+0.07}$ & $-7.75_{-0.09}^{+0.05}$ & $-7.163_{-0.022}^{+0.05}$ & 2578.3/2545 \\
V404 Cyg & 20150624 & $1.6_{-0.9}^{+0.3}$ & $11.9_{-1.6}^{+1.7}$ & $1.555_{-0.013}^{+0.007}$ & $20.53_{-1.2}^{+0.12}$ & $731_{-18.0}^{+537}$ & $32.4_{-4}^{+1.1}$ & $18_{-P}^{+4}$ & $0.50_{-P}^{+0.05}$ & $-7.19_{-0.09}^{+0.11}$ & $-6.970_{-0.05}^{+0.027}$ & 2632.8/2627 \\
V404 Cyg & 20150624 & $2.4_{-1.5}^{+0.3}$ & $1.84_{-0.24}^{+0.23}$ & $1.648_{-0.023}^{+0.027}$ & $20.2_{-0.3}^{+0.3}$ & $610_{-25}^{+130}$ & $40_{-4}^{+4}$ & $58.9_{-1.1}^{+6}$ & $0.500_{-P}^{+0.023}$ & $-6.62_{-0.09}^{+0.09}$ & $-6.879_{-0.09}^{+0.011}$ & 2284.1/2296 \\
V404 Cyg & 20150624 & $1.9_{-0.4}^{+0.6}$ & $6_{-P}^{+4}$ & $1.572_{-0.05}^{+0.028}$ & $20.4_{-0.3}^{+0.8}$ & $810_{-160}^{+380}$ & $17.9_{-1.5}^{+3.0}$ & $23_{-P}^{+7}$ & $0.50_{-P}^{+0.08}$ & $-6.41_{-0.15}^{+0.11}$ & $-6.44_{-0.12}^{+0.04}$ & 2209.8/2156 \\
\hline\hline
\end{tabular} \\
\textit{Note}: The flux of the power-law and reflection component are estimated in the 0.1--100 keV band. The symbol $P$ denotes the lower or higher limits. Note that for V404~Cyg, the column density (nH) of the Galactic absorption is fixed at $1\times 10^{22}$~cm$^{-2}$ and what shown in the table are for local absorption in the system (see Sec.~\ref{v404}).
\end{sidewaystable}


\begin{sidewaystable}
    \centering
    \caption{Best-fit parameters with \texttt{reflionx} that has $\log(n_{\rm e})=15$}
    \label{para_n15}
    \renewcommand\arraystretch{1.8}
    \begin{tabular}{lcccccccccccc}
        \hline\hline
        Source & Date & nH & $R_{\rm in}$ ($R_{\rm g}$) & $\Gamma$ & $\xi$ & kT$_{\rm e}$ (keV) & Incl (deg) & $A_{\rm Fe}$ (solar) & $\log(F_{\rm ref})$ & $\log(F_{\rm po})$ & $\chi/\nu$ \\
\hline
GRS 1739-278 & 20140326 & $3.02_{-0.04}^{+0.05}$ & $1.30_{-P}^{+0.06}$ & $1.839_{-0.012}^{+0.02}$ & $190_{-17}^{+21}$ & $15.41_{-0.23}^{+0.7}$ & $28.2_{-1.2}^{+0.9}$ & $0.50_{-P}^{+0.01}$ & $-8.045_{-0.019}^{+0.016}$ & $-8.067_{-0.007}^{+0.015}$ & 3492.5/3239 \\
GS 1354-64 & 20150613 & $1.22_{-0.04}^{+0.08}$ & $1.83_{-0.26}^{+0.09}$ & $1.544_{-0.01}^{+0.011}$ & $140_{-P}^{+60}$ & $500_{-430}^{+P}$ & $20_{-P}^{+15}$ & $5.0_{-2.5}^{+P}$ & $-9.66_{-0.12}^{+0.07}$ & $-8.963_{-0.022}^{+0.021}$ & 2030.3/1873 \\
GS 1354-64 & 20150711 & $1.19_{-0.06}^{+0.06}$ & $1.54_{-P}^{+0.18}$ & $1.742_{-0.015}^{+0.022}$ & $120_{-6}^{+13}$ & $28.2_{-2.3}^{+4}$ & $40_{-5}^{+14}$ & $0.50_{-P}^{+0.08}$ & $-8.57_{-0.04}^{+0.05}$ & $-8.263_{-0.022}^{+0.013}$ & 3116.8/2919 \\
GS 1354-64 & 20150806 & $0.51_{-0.17}^{+0.16}$ & $1.30_{-P}^{+0.12}$ & $1.762_{-0.03}^{+0.014}$ & $163_{-15}^{+106}$ & $26.1_{-2.7}^{+1.9}$ & $33_{-4}^{+3}$ & $0.50_{-P}^{+0.09}$ & $-8.521_{-0.04}^{+0.028}$ & $-8.20_{-0.05}^{+0.08}$ & 2911.9/2822 \\
IGR J17091 & 20160307 & $1.76_{-0.06}^{+0.04}$ & $47_{-29}^{+P}$ & $1.707_{-0.01}^{+0.012}$ & $90_{-45}^{+20}$ & $38_{-4}^{+5}$ & $43_{-12}^{+P}$ & $0.50_{-P}^{+0.05}$ & $-9.27_{-0.08}^{+0.03}$ & $-8.638_{-0.012}^{+0.012}$ & 2824.3/2752 \\
IGR J17091 & 20160312 & $1.63_{-0.05}^{+0.05}$ & $17_{-10}^{+22}$ & $1.698_{-0.01}^{+0.011}$ & $175_{-28}^{+51}$ & $24.1_{-1.5}^{+1.8}$ & $28_{-7}^{+10}$ & $0.50_{-P}^{+0.08}$ & $-9.29_{-0.06}^{+0.06}$ & $-8.553_{-0.011}^{+0.01}$ & 2372.0/2410 \\
IGR J17091 & 20160314 & $1.55_{-0.06}^{+0.06}$ & $13_{-4}^{+8}$ & $1.70_{-0.01}^{+0.01}$ & $330_{-110}^{+190}$ & $20.9_{-0.7}^{+0.8}$ & $18_{-13}^{+6}$ & $0.50_{-P}^{+0.07}$ & $-9.21_{-0.07}^{+0.06}$ & $-8.372_{-0.011}^{+0.011}$ & 2305.7/2322 \\
H 1743-322 & 20140918 & $1.74_{-0.18}^{+0.12}$ & $5.3_{-P}^{+6}$ & $1.601_{-0.003}^{+0.012}$ & $180_{-50}^{+19}$ & $34.0_{-12}^{+2.1}$ & $22_{-P}^{+12}$ & $0.51_{-P}^{+0.05}$ & $-9.087_{-0.04}^{+0.011}$ & $-8.3677_{-0.0015}^{+0.0018}$ & 3039.5/3007 \\
H 1743-322 & 20140923 & $1.06_{-0.16}^{+0.3}$ & $5.4_{-P}^{+5}$ & $1.584_{-0.014}^{+0.03}$ & $440_{-190}^{+150}$ & $26.6_{-1.8}^{+7}$ & $32_{-3}^{+4}$ & $1.08_{-P}^{+0.18}$ & $-9.33_{-0.02}^{+0.19}$ & $-8.2416_{-0.016}^{+0.0018}$ & 3325.4/3188 \\
H 1743-322 & 20141009 & $2.17_{-0.27}^{+0.22}$ & $40_{-24}^{+13}$ & $1.637_{-0.019}^{+0.012}$ & $114_{-32}^{+18}$ & $99_{-50}^{+P}$ & $29_{-3}^{+18}$ & $0.57_{-P}^{+0.06}$ & $-9.16_{-0.15}^{+0.07}$ & $-8.288_{-0.006}^{+0.01}$ & 2499.9/2606 \\
H 1743-322 & 20150703 & $1.72_{-0.13}^{+0.22}$ & $7_{-P}^{+4}$ & $1.583_{-0.012}^{+0.011}$ & $310_{-270}^{+140}$ & $500_{-390}^{+P}$ & $33_{-3}^{+6}$ & $0.50_{-P}^{+0.23}$ & $-9.12_{-0.12}^{+0.03}$ & $-8.528_{-0.005}^{+0.019}$ & 2389.9/2417 \\
H 1743-322 & 20160313 & $2.11_{-0.14}^{+0.06}$ & $38_{-20}^{+10}$ & $1.788_{-0.011}^{+0.011}$ & $155_{-17}^{+28}$ & $39_{-4}^{+4}$ & $33_{-4}^{+8}$ & $0.56_{-P}^{+0.11}$ & $-8.976_{-0.026}^{+0.03}$ & $-8.238_{-0.003}^{+0.003}$ & 3221.8/3089 \\
H 1743-322 & 20160315 & $2.31_{-0.06}^{+0.05}$ & $5.3_{-P}^{+16}$ & $1.822_{-0.006}^{+0.004}$ & $124.1_{-4}^{+2.5}$ & $48_{-3}^{+6}$ & $80_{-7}^{+P}$ & $0.65_{-0.08}^{+0.05}$ & $-8.910_{-0.021}^{+0.027}$ & $-8.2572_{-0.0028}^{+0.0028}$ & 3151.4/3061 \\
H 1743-322 & 20180919 & $2.48_{-0.28}^{+0.2}$ & $14_{-P}^{+14}$ & $1.679_{-0.02}^{+0.019}$ & $140_{-23}^{+78}$ & $44_{-7}^{+4}$ & $31_{-6}^{+20}$ & $0.51_{-P}^{+0.15}$ & $-9.00_{-0.07}^{+0.04}$ & $-8.293_{-0.004}^{+0.006}$ & 2876.7/2815 \\
H 1743-322 & 20180926 & $2.08_{-0.07}^{+0.11}$ & $17_{-P}^{+16}$ & $1.605_{-0.009}^{+0.009}$ & $150_{-29}^{+43}$ & $500_{-360}^{+P}$ & $24.7_{-7}^{+6}$ & $0.66_{-0.1}^{+0.12}$ & $-9.24_{-0.06}^{+0.04}$ & $-8.467_{-0.005}^{+0.005}$ & 3203.5/3020 \\
MAXI J1535 & 20170907 & $7.5_{-0.6}^{+0.4}$ & $1.326_{-0.025}^{+0.027}$ & $1.945_{-0.021}^{+0.014}$ & $245_{-55}^{+250}$ & $24.2_{-1.7}^{+1.3}$ & $78.4_{-1.0}^{+1.7}$ & $0.62_{-0.08}^{+0.06}$ & $-7.62_{-0.05}^{+0.03}$ & $-7.201_{-0.013}^{+0.009}$ & 3483.5/3132 \\
V404 Cyg & 20150624 & $0.56_{-P}^{+0.16}$ & $5.3_{-2.0}^{+2.1}$ & $1.57_{-0.05}^{+0.06}$ & $490_{-60}^{+220}$ & $80_{-35}^{+114}$ & $29_{-P}^{+5}$ & $0.89_{-0.27}^{+0.14}$ & $-7.88_{-0.08}^{+0.07}$ & $-7.590_{-0.004}^{+0.028}$ & 2028.1/2058 \\
V404 Cyg & 20150624 & $2.0_{-0.5}^{+0.8}$ & $12.7_{-2.7}^{+0.8}$ & $1.559_{-0.009}^{+0.015}$ & $2900_{-130}^{+185}$ & $58_{-9}^{+5}$ & $5_{-P}^{+7}$ & $0.52_{-P}^{+0.04}$ & $-7.31_{-0.09}^{+0.02}$ & $-7.46_{-0.1}^{+0.11}$ & 2589.0/2546 \\
V404 Cyg & 20150624 & $0.3_{-P}^{+0.4}$ & $12.0_{-1.3}^{+2.0}$ & $1.577_{-0.012}^{+0.005}$ & $2570_{-50}^{+320}$ & $42.2_{-1.6}^{+0.8}$ & $11_{-6}^{+8}$ & $0.50_{-P}^{+0.03}$ & $-7.051_{-0.008}^{+0.06}$ & $-7.15_{-0.09}^{+0.04}$ & 2640.5/2628 \\
V404 Cyg & 20150624 & $0.82_{-0.07}^{+0.23}$ & $1.921_{-0.018}^{+0.04}$ & $1.649_{-0.004}^{+0.027}$ & $1000_{-130}^{+200}$ & $43.8_{-3}^{+0.5}$ & $58.7_{-11}^{+0.3}$ & $0.500_{-P}^{+0.027}$ & $-6.60_{-0.03}^{+0.03}$ & $-7.12_{-0.01}^{+0.01}$ & 2288.2/2297 \\
V404 Cyg & 20150624 & $<0.2$ & $1.9_{-P}^{+1.3}$ & $1.660_{-0.016}^{+0.007}$ & $1330_{-145}^{+380}$ & $24.2_{-0.9}^{+1.8}$ & $22.0_{-2.5}^{+9}$ & $0.500_{-P}^{+0.007}$ & $-6.398_{-0.06}^{+0.016}$ & $-6.533_{-0.028}^{+0.023}$ & 2212.8/2157 \\
\hline\hline
\end{tabular} \\
\textit{Note}: The flux of the power-law and reflection component are estimated in the 0.1--100 keV band. The symbol $P$ denotes the lower or higher limits. Note that for V404~Cyg, the column density (nH) of the Galactic absorption is fixed at $1\times 10^{22}$~cm$^{-2}$ and what shown in the table are for local absorption in the system (see Sec.~\ref{v404}).
\end{sidewaystable}


\begin{sidewaystable}
    \centering
    \caption{Best-fit parameters with \texttt{relxillcp} that has $n_{\rm e}=10^{15}$~cm$^{-3}$}
    \label{para_relxill}
    \renewcommand\arraystretch{1.8}
    \begin{tabular}{lcccccccccccc}
        \hline\hline
        Source & Date & nH & $R_{\rm in}$ ($R_{\rm g}$) & $\Gamma$ & $\log(\xi)$ & kT$_{\rm e}$ (keV) & Incl (deg) & $A_{\rm Fe}$ (solar) & $\log(F_{\rm ref})$ & $\log(F_{\rm po})$ & $\chi^2/\nu$ \\
\hline
GRS 1739-278 & 20140326 & $2.87_{-0.05}^{+0.05}$ & $1.30_{-P}^{+0.03}$ & $1.61_{-0.03}^{+0.02}$ & $3.544_{-0.04}^{+0.028}$ & $20.0_{-1.4}^{+1.8}$ & $25.9_{-2.6}^{+2.4}$ & $2.3_{-0.4}^{+0.5}$ & $-7.897_{-0.015}^{+0.01}$ & $-8.71_{-0.06}^{+0.05}$ & 3409.5/3239 \\
GS 1354-64 & 20150613 & $1.20_{-0.07}^{+0.08}$ & $27_{-10}^{+26}$ & $1.525_{-0.009}^{+0.01}$ & $2.71_{-0.11}^{+0.24}$ & $41_{-8}^{+27}$ & $5_{-P}^{+13}$ & $0.7_{-P}^{+0.5}$ & $-9.97_{-0.16}^{+0.09}$ & $-8.933_{-0.018}^{+0.017}$ & 2049.0/1876 \\
GS 1354-64 & 20150711 & $1.07_{-0.06}^{+0.06}$ & $1.60_{-0.07}^{+0.1}$ & $1.705_{-0.012}^{+0.013}$ & $2.49_{-0.07}^{+0.06}$ & $29_{-3}^{+4}$ & $65_{-7}^{+3}$ & $0.50_{-P}^{+0.05}$ & $-8.66_{-0.04}^{+0.04}$ & $-8.251_{-0.016}^{+0.016}$ & 3117.3/2919 \\
GS 1354-64 & 20150806 & $1.00_{-0.07}^{+0.1}$ & $1.42_{-0.04}^{+0.21}$ & $1.751_{-0.009}^{+0.01}$ & $2.71_{-0.09}^{+0.04}$ & $26.6_{-2.3}^{+2.1}$ & $73.4_{-3.0}^{+1.2}$ & $0.50_{-P}^{+0.04}$ & $-8.55_{-0.05}^{+0.03}$ & $-8.19_{-0.01}^{+0.01}$ & 2901.7/2822 \\
IGR J17091 & 20160307 & $1.70_{-0.06}^{+0.06}$ & $19_{-8}^{+22}$ & $1.699_{-0.011}^{+0.01}$ & $2.04_{-0.21}^{+0.22}$ & $60_{-13}^{+18}$ & $34_{-6}^{+6}$ & $0.52_{-P}^{+0.16}$ & $-9.42_{-0.06}^{+0.05}$ & $-8.621_{-0.01}^{+0.011}$ & 2819.2/2752 \\
IGR J17091 & 20160312 & $1.56_{-0.05}^{+0.05}$ & $19_{-11}^{+16}$ & $1.684_{-0.011}^{+0.01}$ & $2.83_{-0.1}^{+0.11}$ & $28_{-3}^{+3}$ & $37_{-6}^{+9}$ & $0.52_{-P}^{+0.22}$ & $-9.33_{-0.1}^{+0.06}$ & $-8.552_{-0.012}^{+0.014}$ & 2367.5/2410 \\
IGR J17091 & 20160314 & $1.53_{-0.06}^{+0.05}$ & $18_{-8}^{+10}$ & $1.703_{-0.012}^{+0.008}$ & $2.94_{-0.12}^{+0.15}$ & $24.2_{-2.2}^{+1.2}$ & $28_{-6}^{+7}$ & $0.50_{-P}^{+0.17}$ & $-9.18_{-0.1}^{+0.04}$ & $-8.377_{-0.008}^{+0.015}$ & 2294.6/2322 \\
H 1743-322 & 20140918 & $1.44_{-0.09}^{+0.09}$ & $5.3_{-P}^{+5}$ & $1.559_{-0.01}^{+0.008}$ & $3.10_{-0.16}^{+0.18}$ & $31_{-3}^{+3}$ & $36_{-7}^{+3}$ & $1.5_{-0.6}^{+0.6}$ & $-9.35_{-0.08}^{+0.07}$ & $-8.350_{-0.004}^{+0.007}$ & 3042.7/3008 \\
H 1743-322 & 20140923 & $1.49_{-0.09}^{+0.11}$ & $10.0_{-2.0}^{+2.7}$ & $1.607_{-0.006}^{+0.007}$ & $2.98_{-0.08}^{+0.07}$ & $31.0_{-2.2}^{+3.0}$ & $30.8_{-2.1}^{+1.8}$ & $0.69_{-0.09}^{+0.09}$ & $-9.13_{-0.05}^{+0.05}$ & $-8.257_{-0.005}^{+0.004}$ & 3320.4/3189 \\
H 1743-322 & 20141009 & $1.8_{-0.2}^{+0.3}$ & $5.3_{-P}^{+3.3}$ & $1.611_{-0.015}^{+0.02}$ & $2.86_{-0.13}^{+0.16}$ & $41_{-8}^{+25}$ & $34_{-6}^{+5}$ & $0.82_{-0.29}^{+0.5}$ & $-9.18_{-0.1}^{+0.14}$ & $-8.296_{-0.015}^{+0.008}$ & 2503.7/2607 \\
H 1743-322 & 20150703 & $1.85_{-0.12}^{+0.25}$ & $5.3_{-P}^{+8}$ & $1.580_{-0.011}^{+0.008}$ & $2.72_{-0.17}^{+0.21}$ & $400_{-200}^{+P}$ & $34_{-8}^{+4}$ & $1.1_{-0.3}^{+1.0}$ & $-9.47_{-0.06}^{+0.08}$ & $-8.498_{-0.008}^{+0.005}$ & 2389.5/2417 \\
H 1743-322 & 20160313 & $1.56_{-0.09}^{+0.08}$ & $5.3_{-P}^{+1.3}$ & $1.725_{-0.007}^{+0.007}$ & $3.17_{-0.08}^{+0.08}$ & $30.4_{-2.3}^{+2.6}$ & $33.5_{-3.0}^{+2.2}$ & $1.23_{-0.28}^{+0.6}$ & $-9.16_{-0.05}^{+0.04}$ & $-8.244_{-0.005}^{+0.004}$ & 3212.6/3089 \\
H 1743-322 & 20160315 & $1.30_{-0.08}^{+0.09}$ & $5.3_{-P}^{+1.7}$ & $1.714_{-0.007}^{+0.01}$ & $3.46_{-0.2}^{+0.17}$ & $28.2_{-1.8}^{+2.2}$ & $33_{-8}^{+3}$ & $2.7_{-1.2}^{+0.9}$ & $-9.17_{-0.09}^{+0.07}$ & $-8.271_{-0.009}^{+0.01}$ & 3140.6/3061 \\
H 1743-322 & 20180919 & $1.92_{-0.12}^{+0.12}$ & $5.3_{-P}^{+0.7}$ & $1.608_{-0.013}^{+0.012}$ & $3.21_{-0.16}^{+0.22}$ & $30.2_{-2.4}^{+4}$ & $38.8_{-6}^{+1.3}$ & $1.9_{-0.8}^{+0.5}$ & $-9.20_{-0.07}^{+0.07}$ & $-8.290_{-0.012}^{+0.008}$ & 2878.5/2815 \\
H 1743-322 & 20180926 & $1.97_{-0.1}^{+0.09}$ & $15_{-P}^{+15}$ & $1.582_{-0.004}^{+0.008}$ & $3.02_{-0.14}^{+0.09}$ & $400_{-280}^{+P}$ & $23_{-P}^{+6}$ & $0.98_{-0.13}^{+0.27}$ & $-9.41_{-0.07}^{+0.05}$ & $-8.458_{-0.006}^{+0.005}$ & 3199.4/3020 \\
MAXI J1535 & 20170907 & $6.9_{-0.6}^{+0.7}$ & $4_{-P}^{+5}$ & $1.806_{-0.009}^{+0.009}$ & $3.66_{-0.04}^{+0.03}$ & $22.9_{-1.1}^{+1.4}$ & $62.6_{-2.4}^{+10}$ & $0.87_{-0.08}^{+0.11}$ & $-7.29_{-0.05}^{+0.07}$ & $-7.41_{-0.09}^{+0.05}$ & 3458.6/3132 \\
V404 Cyg & 20150624 & $0.6_{-0.3}^{+0.4}$ & $2.9_{-P}^{+1.9}$ & $1.53_{-0.03}^{+0.03}$ & $3.0_{-0.2}^{+0.3}$ & $62_{-30}^{+300}$ & $35_{-15}^{+20}$ & $1.7_{-0.8}^{+0.7}$ & $-7.94_{-0.25}^{+0.05}$ & $-7.56_{-0.03}^{+0.04}$ & 2030.6/2058 \\
V404 Cyg & 20150624 & $<0.07$ & $5.7_{-0.3}^{+0.3}$ & $1.5542_{-0.008}^{+0.0018}$ & $3.076_{-0.005}^{+0.009}$ & $50.0_{-1.1}^{+9}$ & $27_{-3}^{+3}$ & $1.003_{-0.021}^{+0.04}$ & $-7.468_{-0.014}^{+0.013}$ & $-7.220_{-0.014}^{+0.005}$ & 2611.6/2546 \\
V404 Cyg & 20150624 & $<0.03$ & $7.1_{-0.9}^{+1.2}$ & $1.5486_{-0.0027}^{+0.0025}$ & $3.076_{-0.014}^{+0.005}$ & $36.4_{-0.9}^{+0.7}$ & $32.7_{-0.7}^{+0.8}$ & $0.997_{-0.009}^{+0.018}$ & $-7.149_{-0.006}^{+0.007}$ & $-6.9573_{-0.015}^{+0.0011}$ & 2720.0/2628 \\
V404 Cyg & 20150624 & $<0.2$ & $6.6_{-0.5}^{+1.5}$ & $1.567_{-0.009}^{+0.019}$ & $3.116_{-0.04}^{+0.024}$ & $29.5_{-1.8}^{+4}$ & $34.4_{-1.0}^{+0.8}$ & $1.00_{-0.03}^{+0.11}$ & $-6.809_{-0.017}^{+0.025}$ & $-6.751_{-0.009}^{+0.022}$ & 2328.6/2297 \\
V404 Cyg & 20150624 & $<0.1$ & $7.7_{-1.0}^{+1.1}$ & $1.646_{-0.013}^{+0.007}$ & $2.936_{-0.03}^{+0.017}$ & $21.7_{-0.3}^{+1.7}$ & $35.1_{-0.6}^{+0.5}$ & $0.966_{-0.07}^{+0.019}$ & $-6.544_{-0.008}^{+0.02}$ & $-6.345_{-0.008}^{+0.008}$ & 2222.3/2157 \\

\hline\hline
\end{tabular} \\
\textit{Note}: The flux of the power-law and reflection component are estimated in the 0.1--100 keV band. The symbol $P$ denotes the lower or higher limits. Note that for V404~Cyg, the column density (nH) of the Galactic absorption is fixed at $1\times 10^{22}$~cm$^{-2}$ and what shown in the table are for local absorption in the system (see Sec.~\ref{v404}).
\end{sidewaystable}

\begin{figure*}
    \centering
    \includegraphics[width=0.24\linewidth]{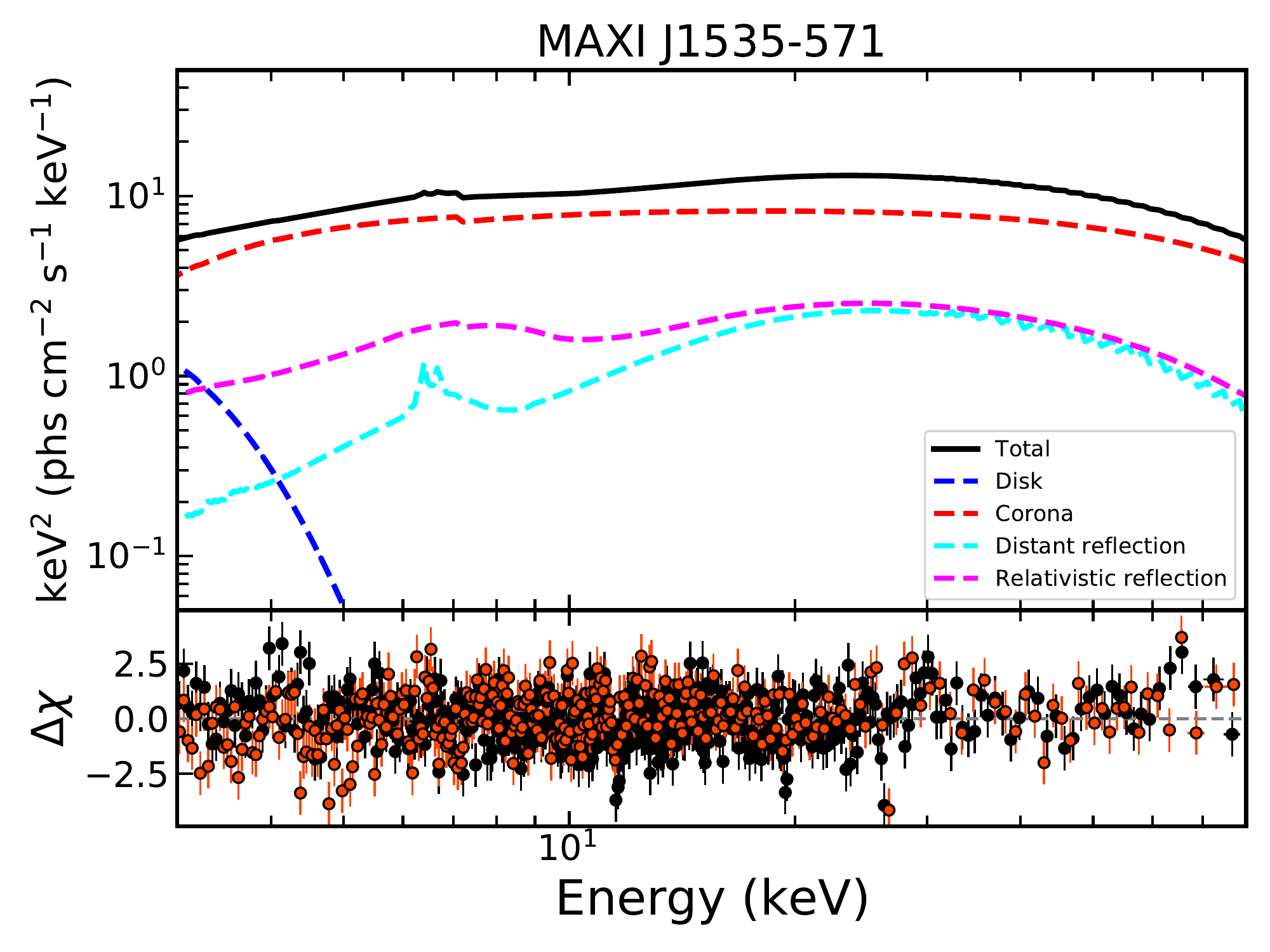}
    \includegraphics[width=0.24\linewidth]{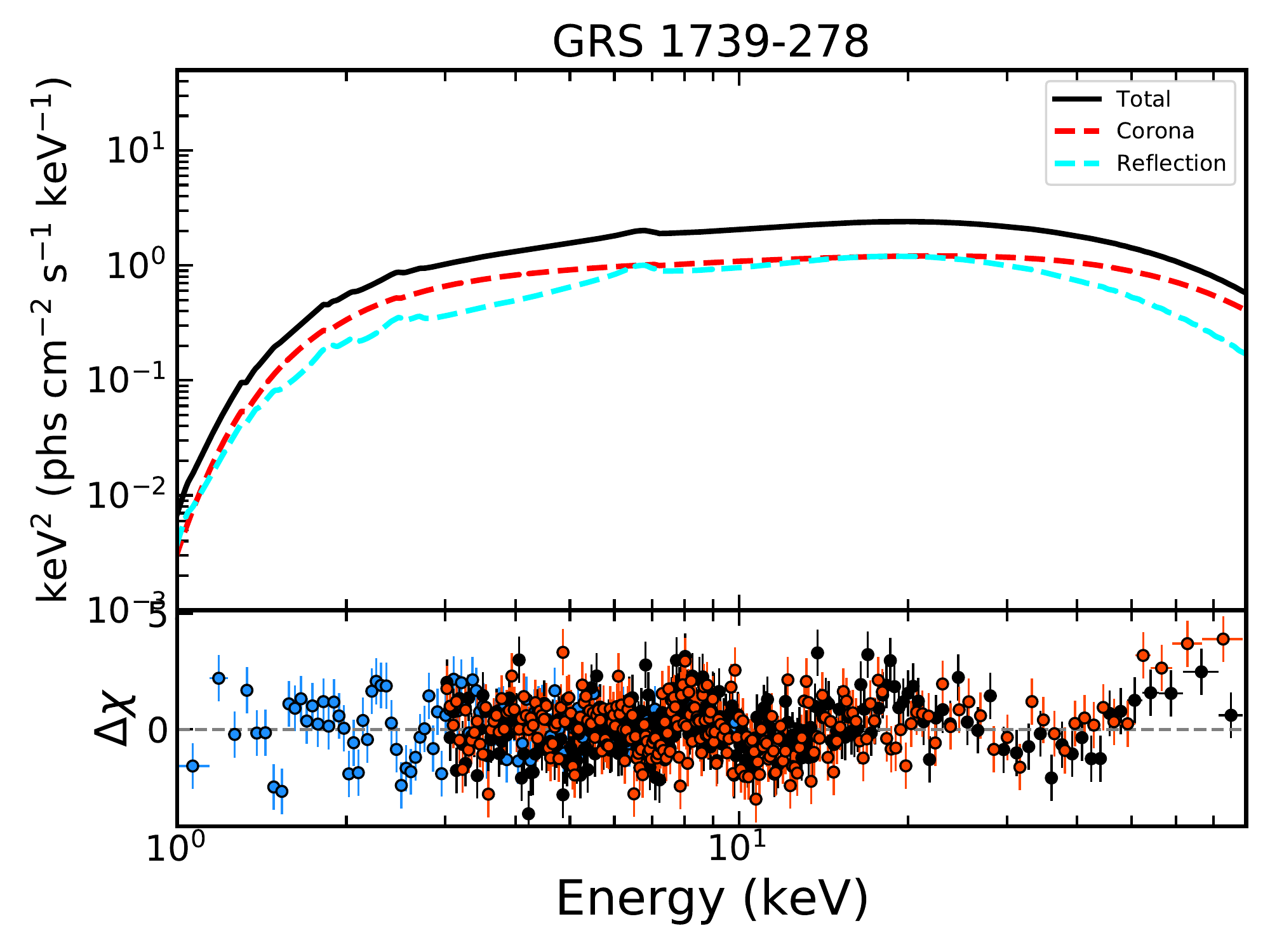}
    \includegraphics[width=0.24\linewidth]{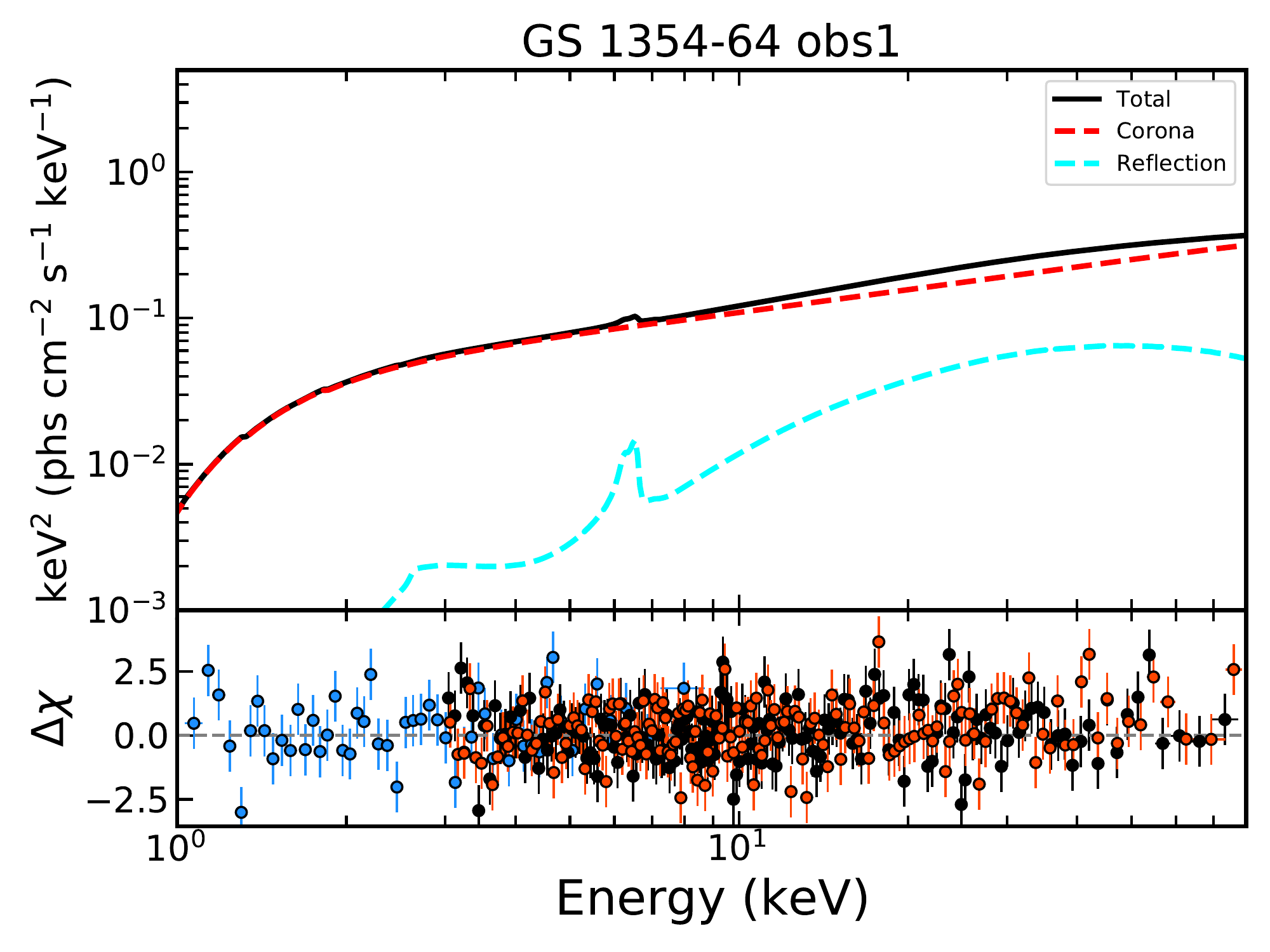}
    \includegraphics[width=0.24\linewidth]{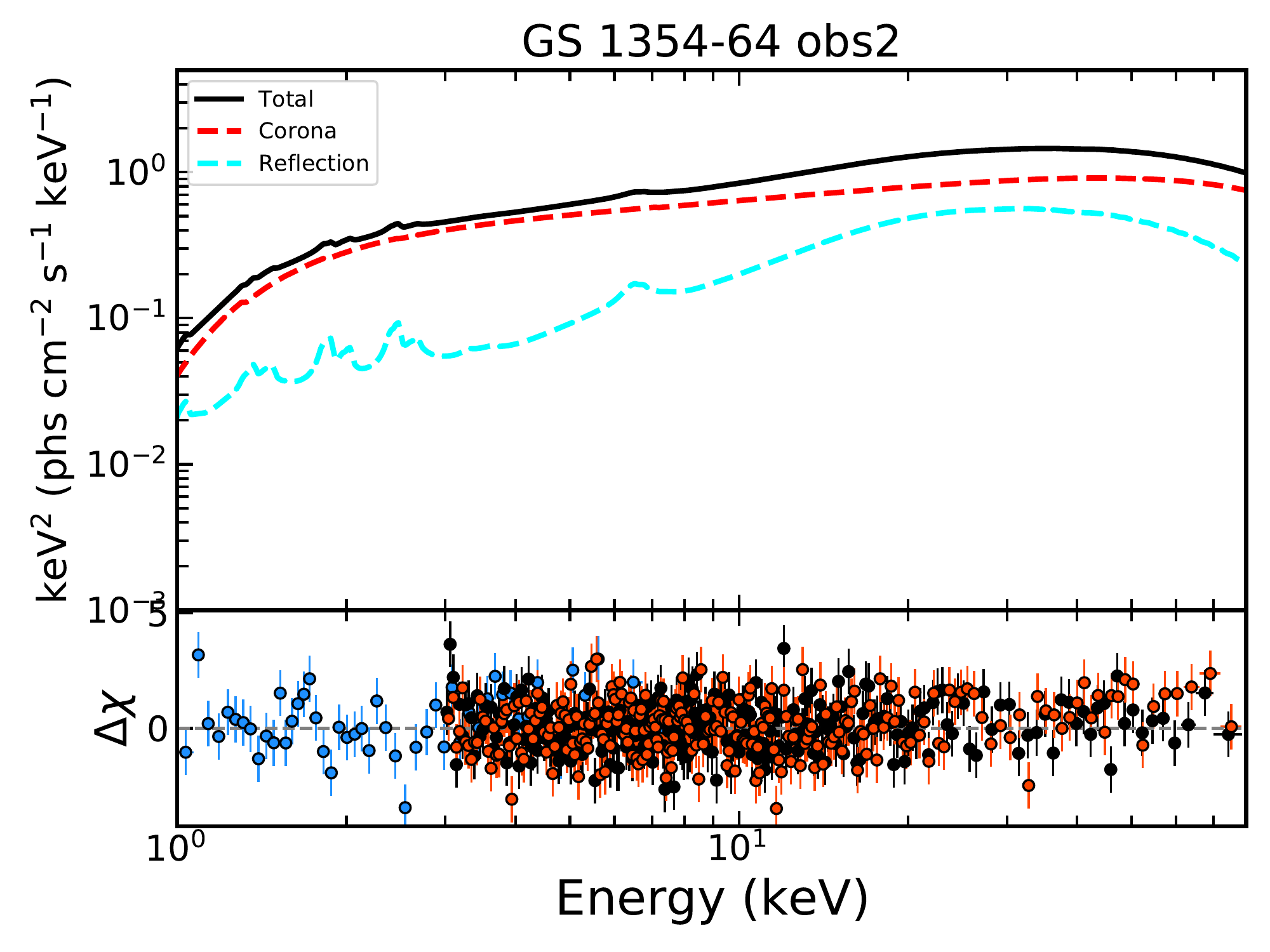}\\
    \includegraphics[width=0.24\linewidth]{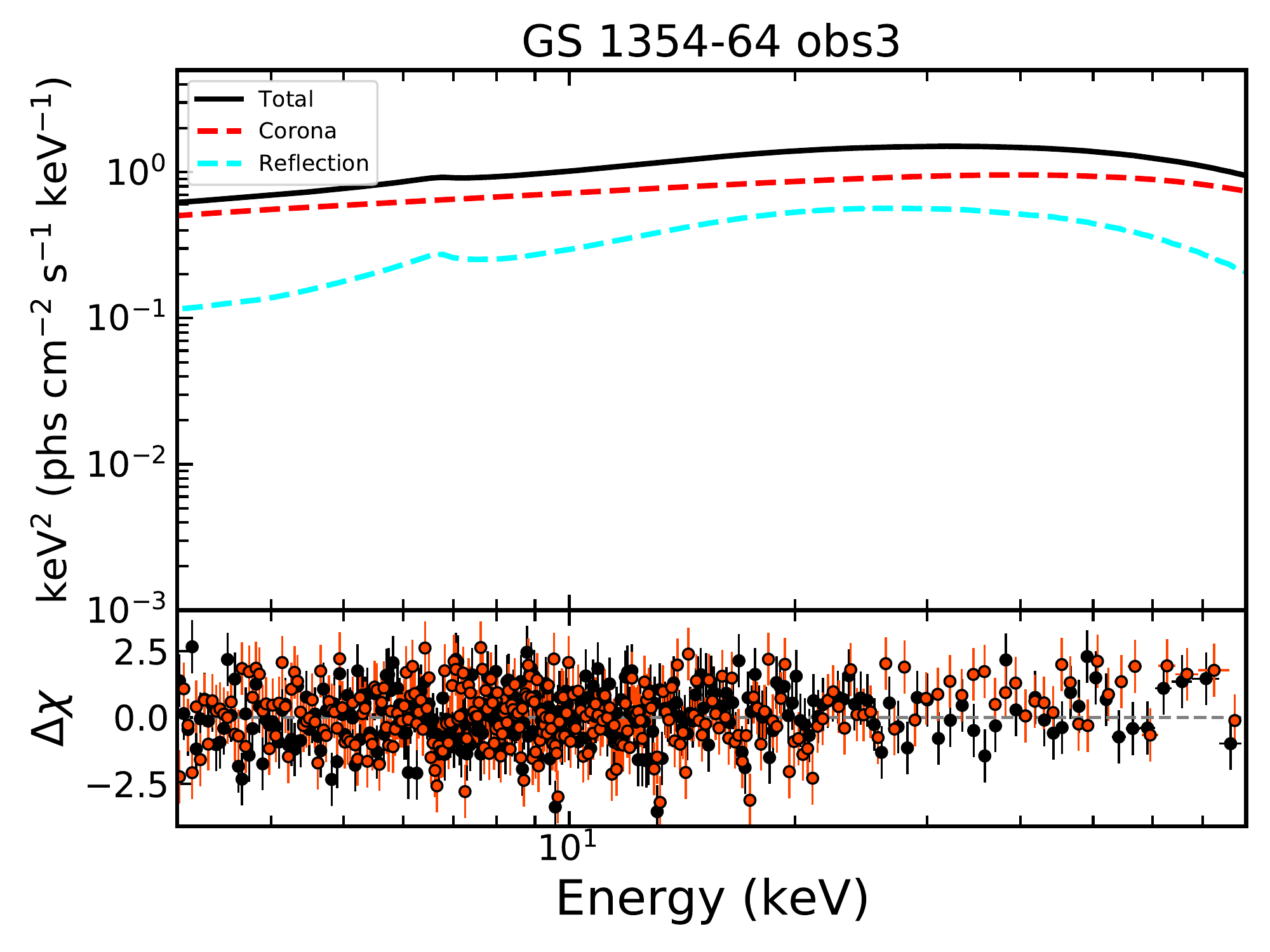}
    \includegraphics[width=0.24\linewidth]{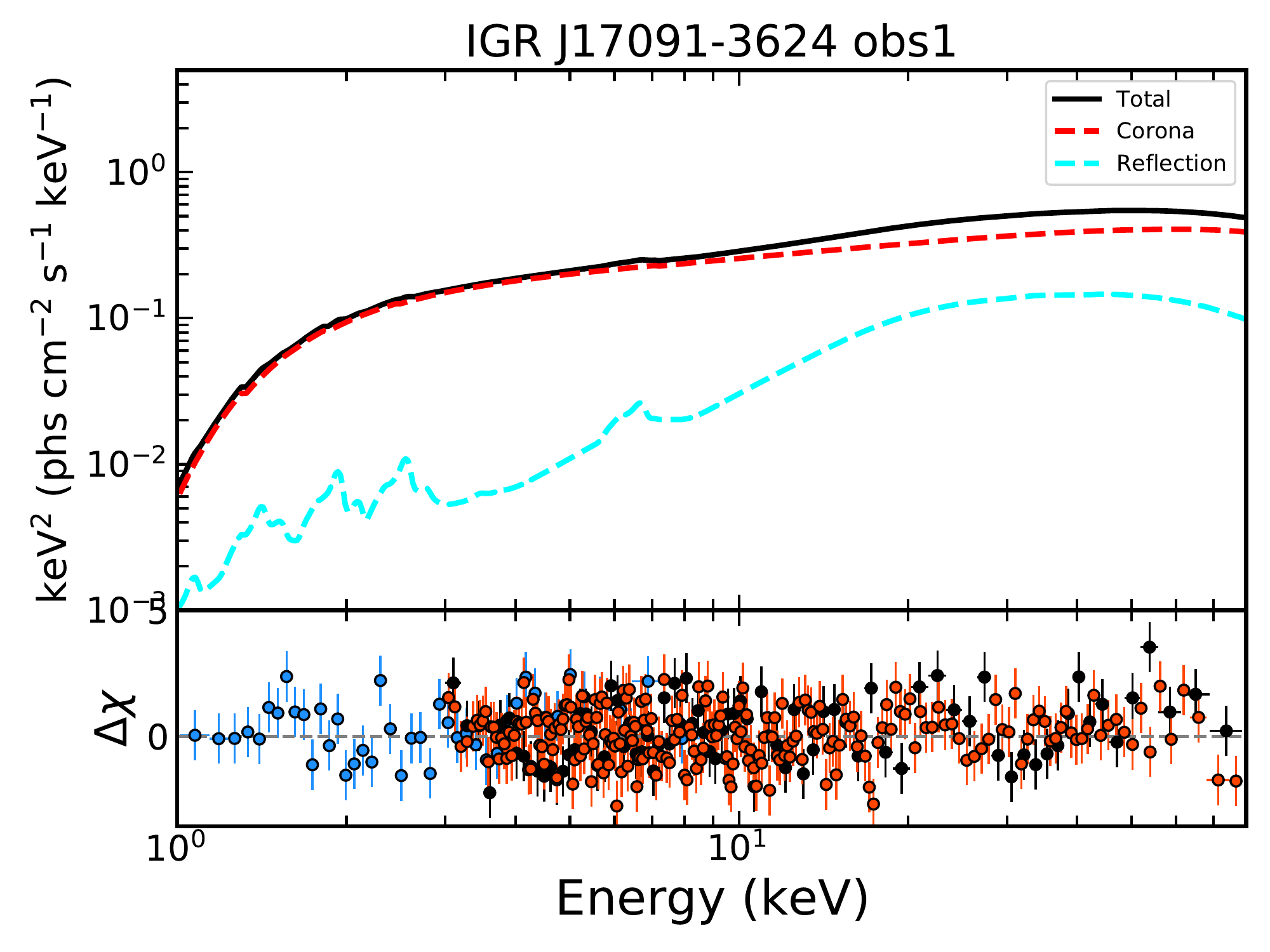}
    \includegraphics[width=0.24\linewidth]{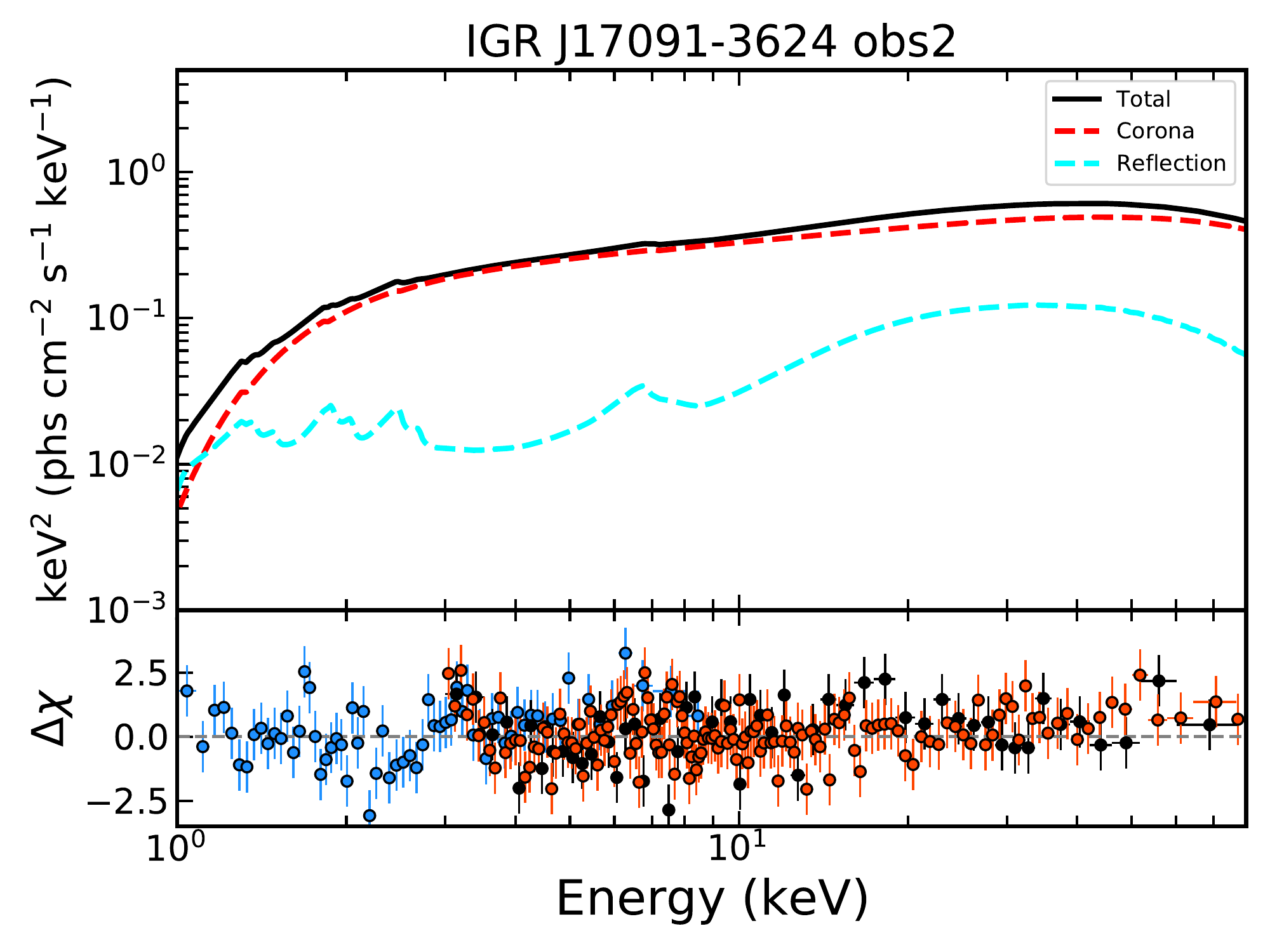}
    \includegraphics[width=0.24\linewidth]{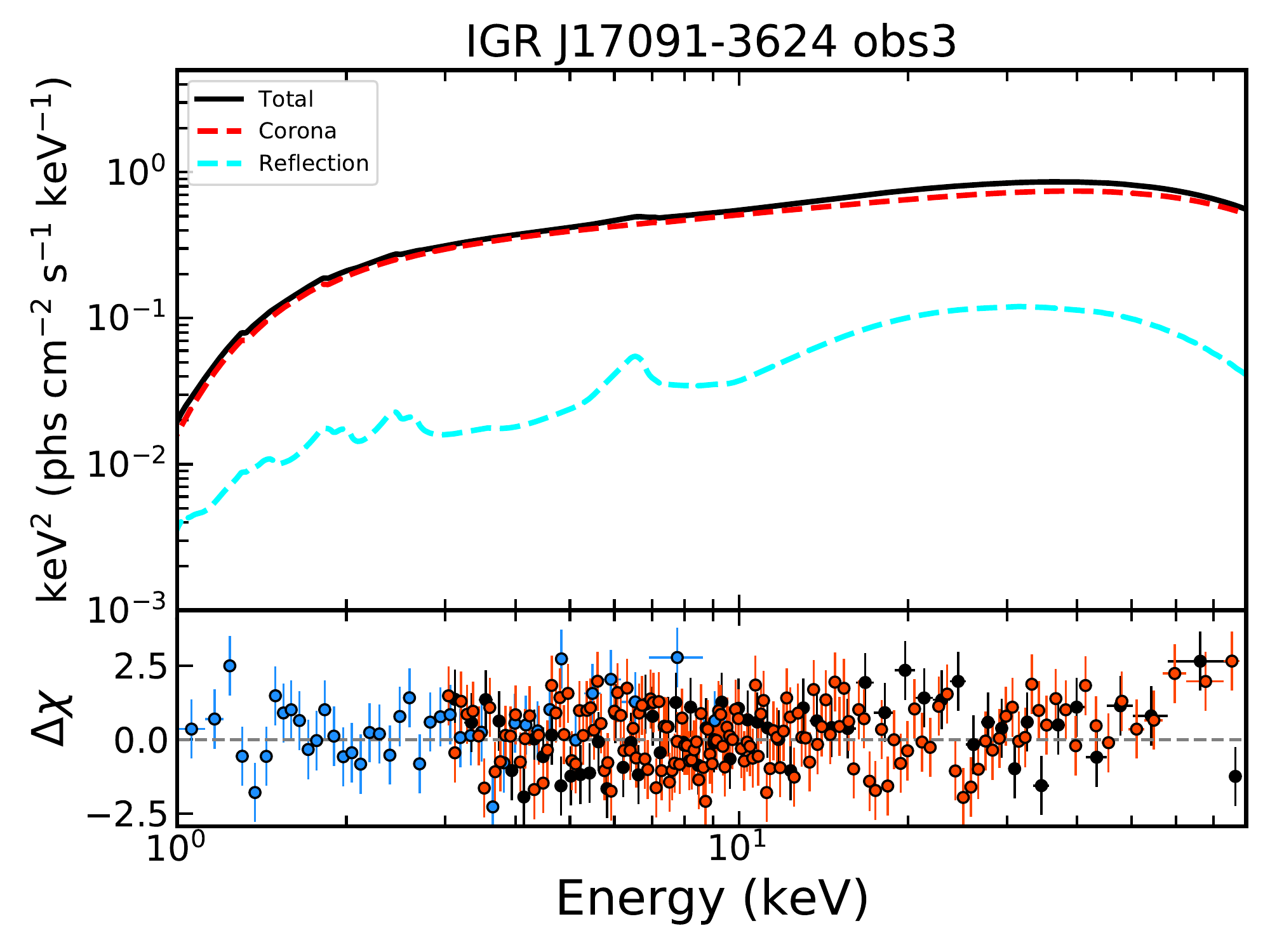}\\
    \includegraphics[width=0.24\linewidth]{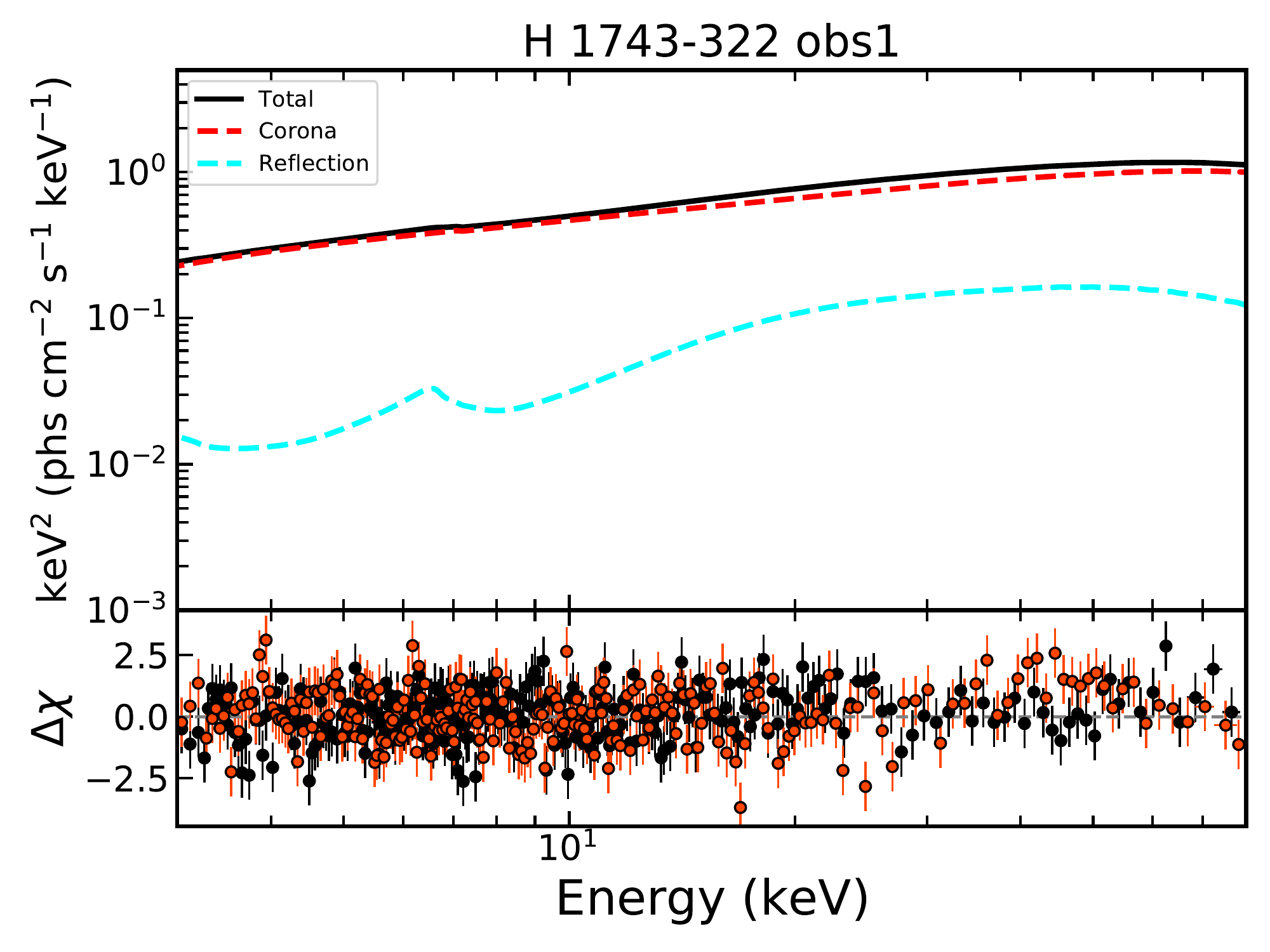}
    \includegraphics[width=0.24\linewidth]{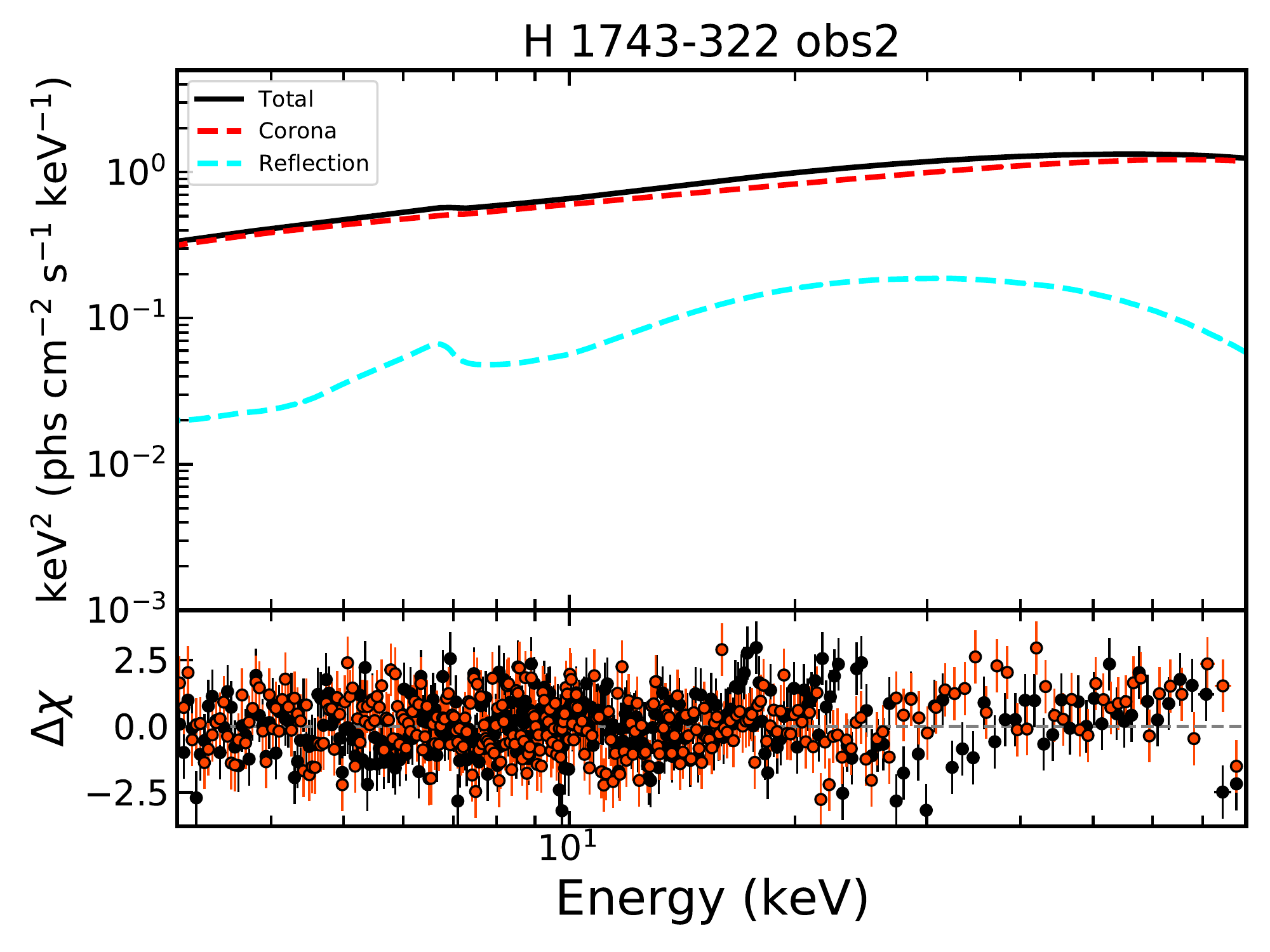}
    \includegraphics[width=0.24\linewidth]{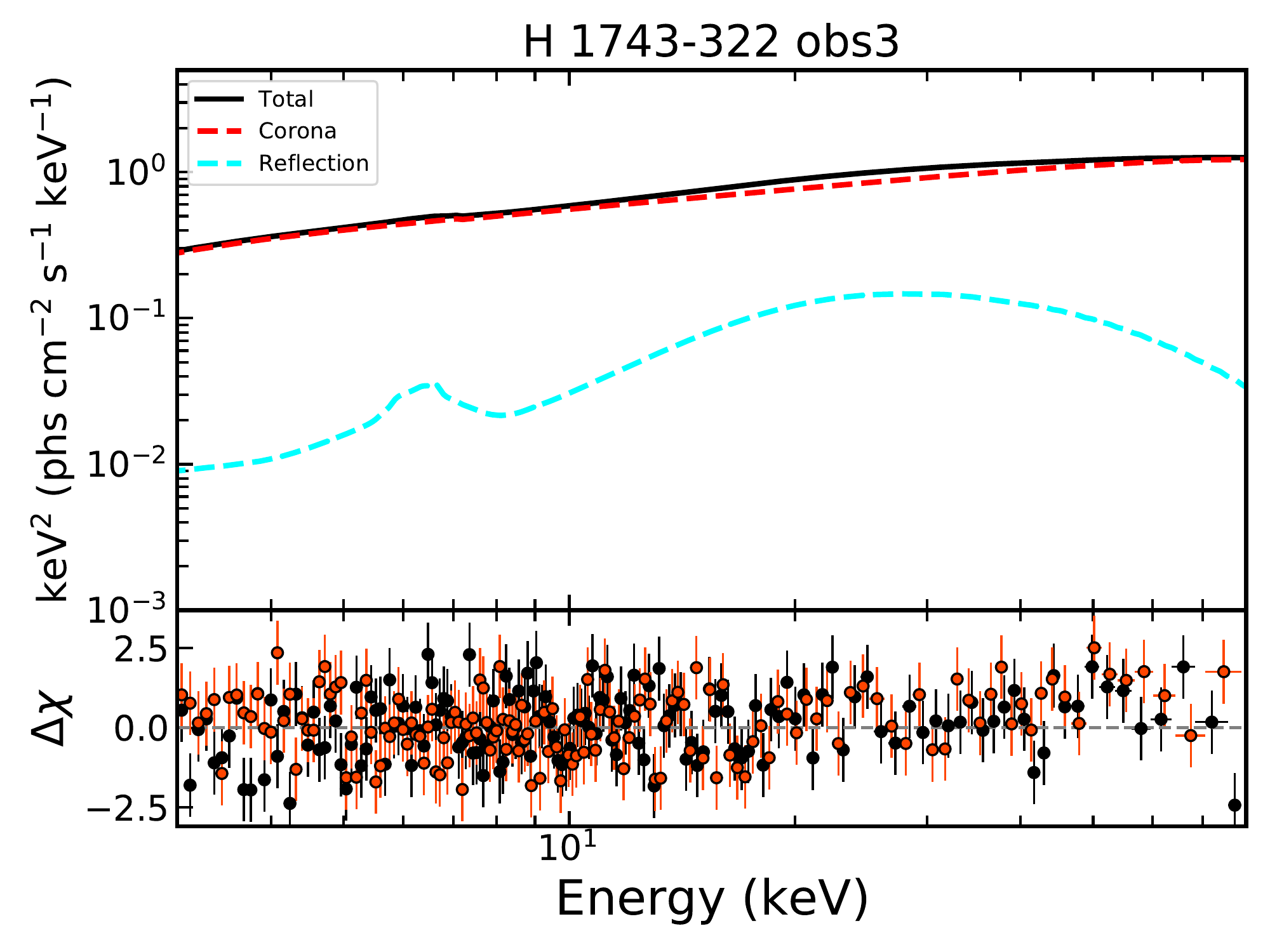}
    \includegraphics[width=0.24\linewidth]{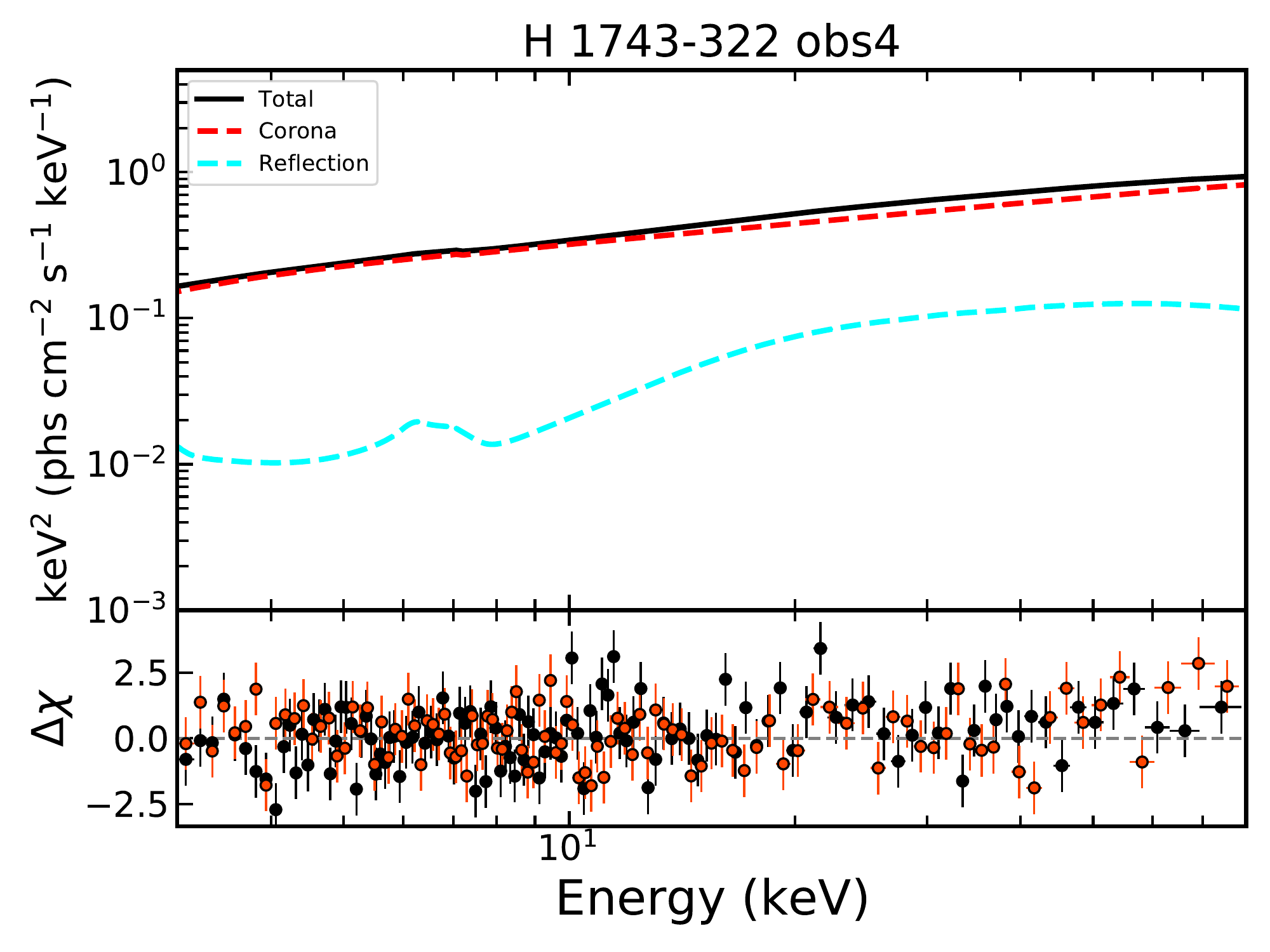}\\
    \includegraphics[width=0.24\linewidth]{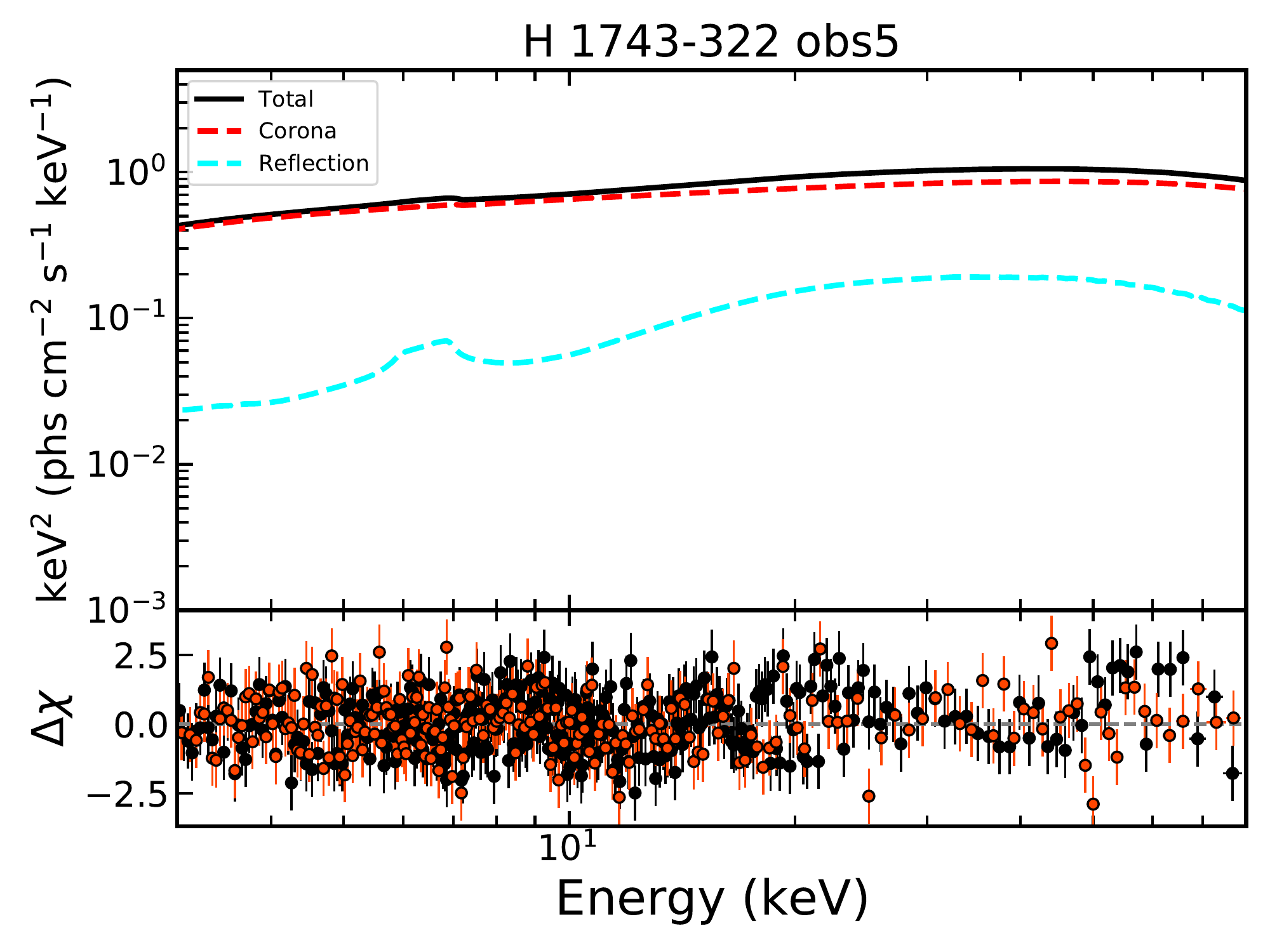}
    \includegraphics[width=0.24\linewidth]{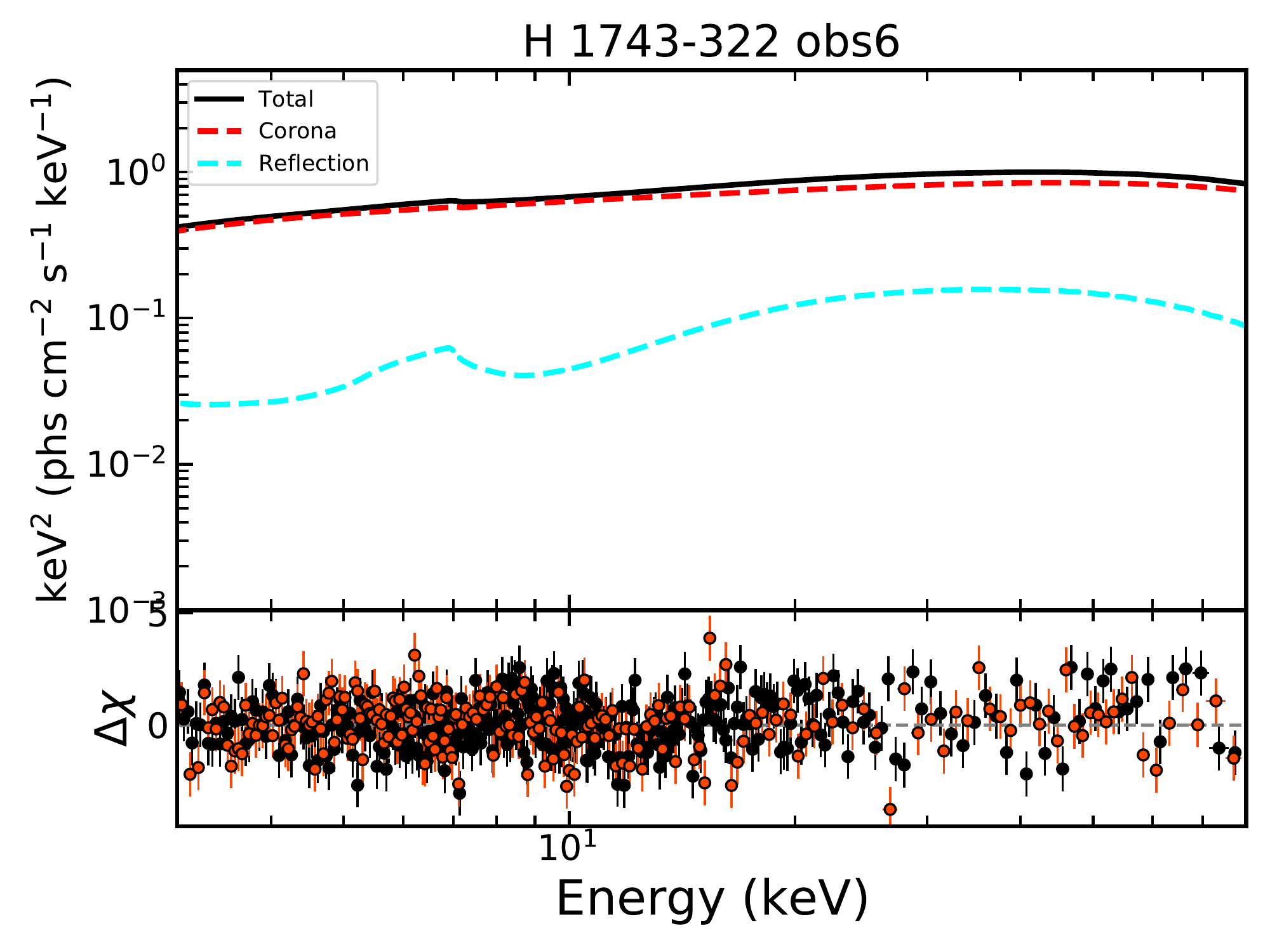}
    \includegraphics[width=0.24\linewidth]{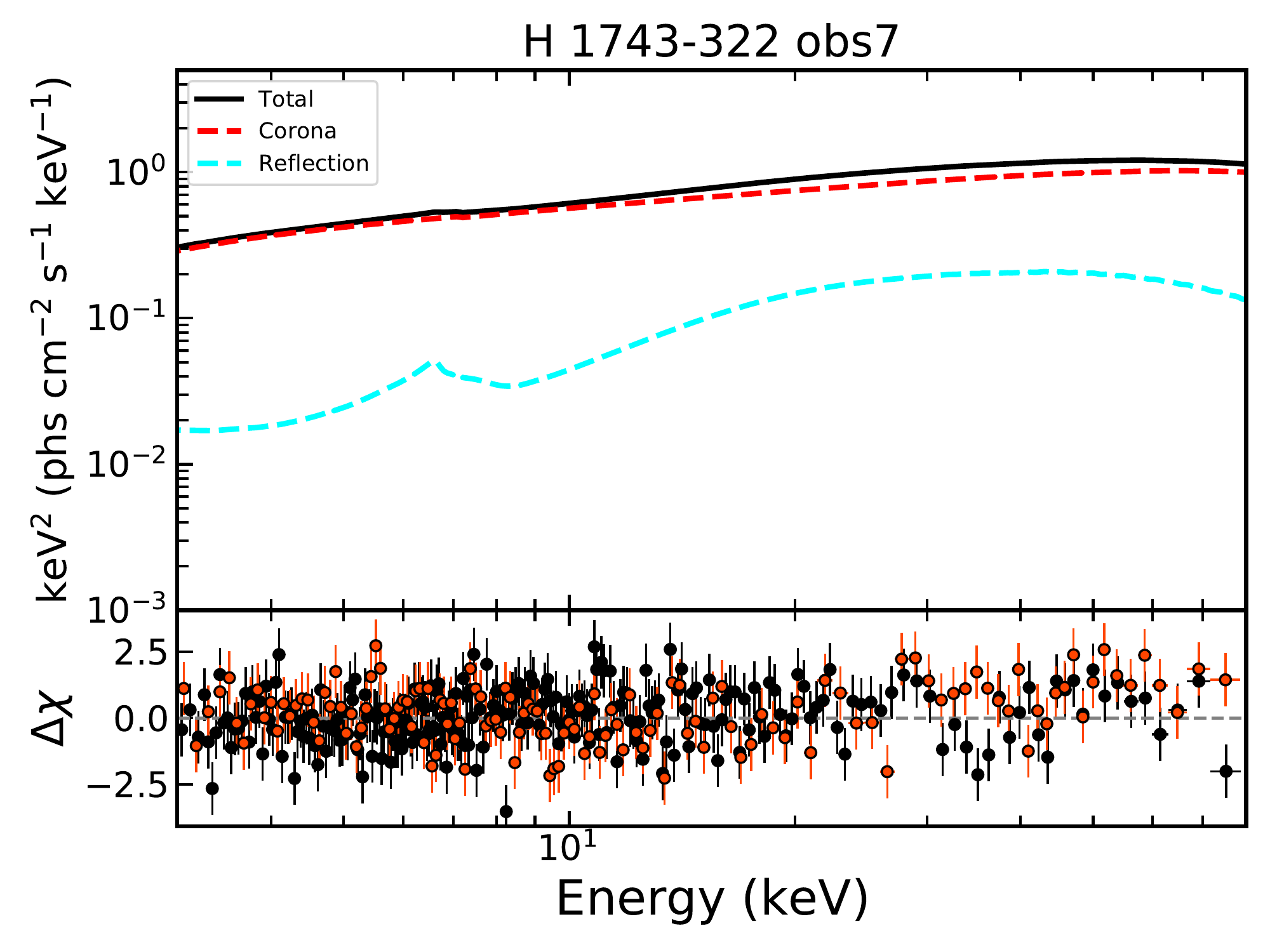}
    \includegraphics[width=0.24\linewidth]{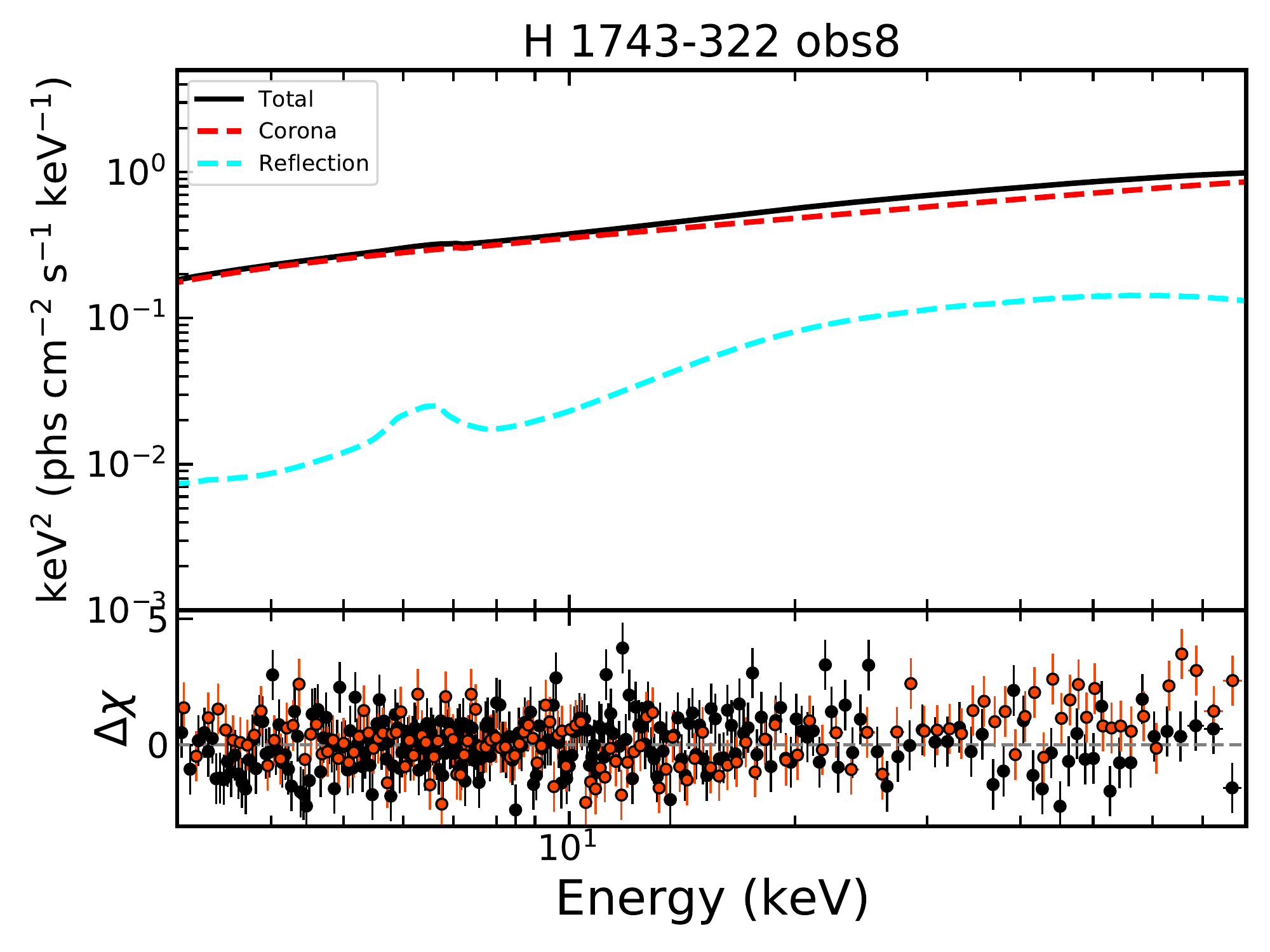}\\
    \includegraphics[width=0.24\linewidth]{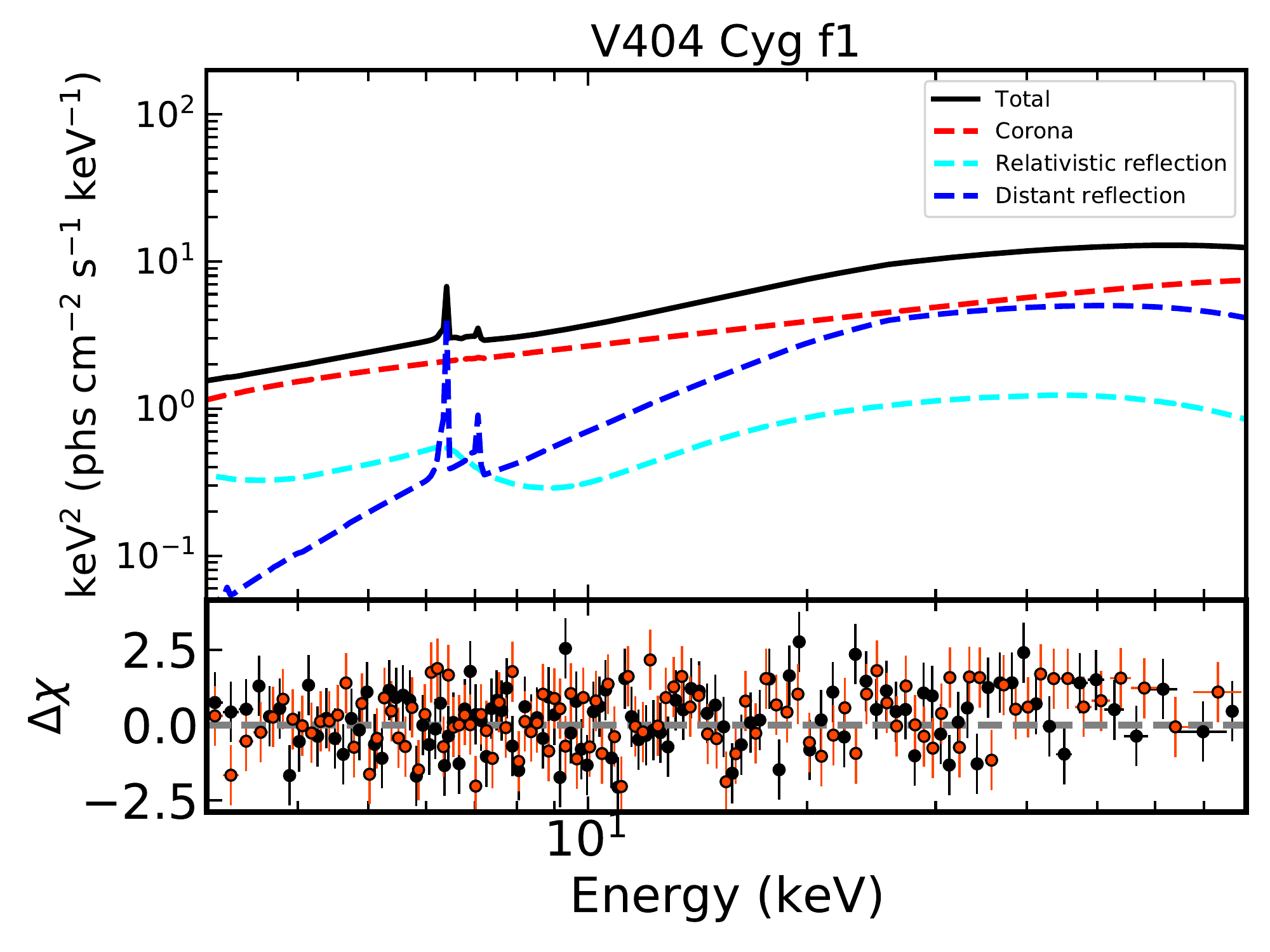}
    \includegraphics[width=0.24\linewidth]{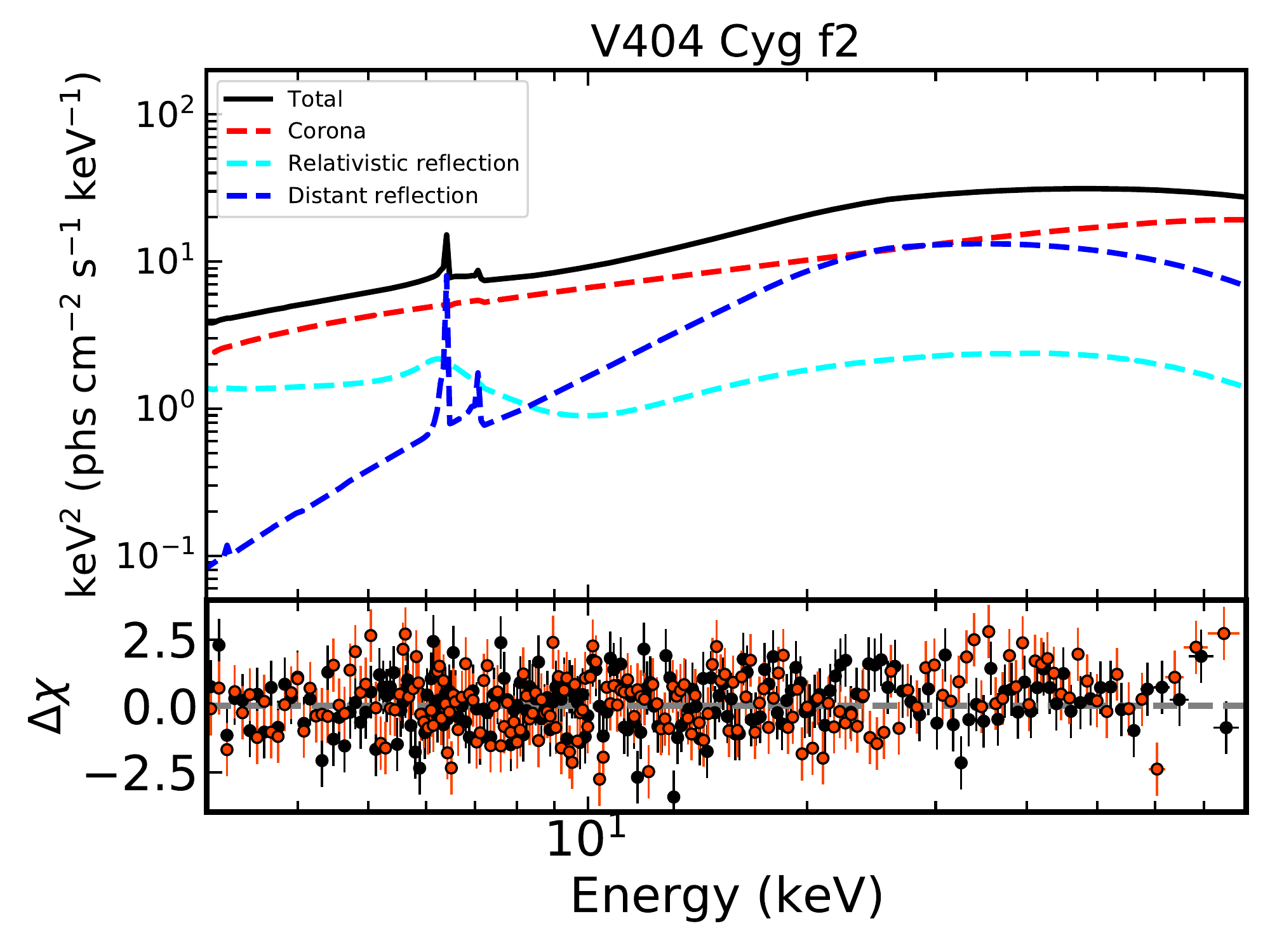}
    \includegraphics[width=0.24\linewidth]{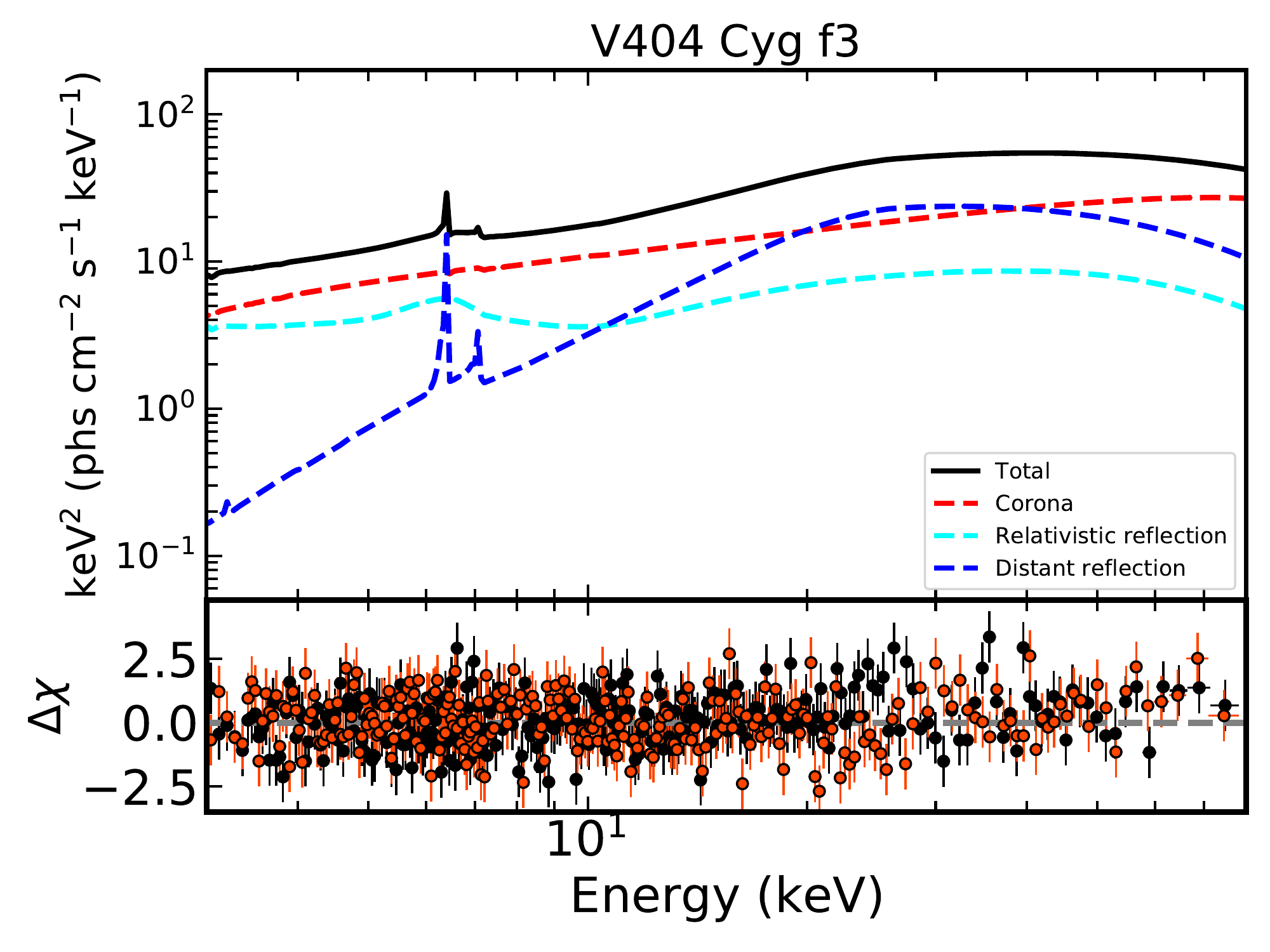}
    \includegraphics[width=0.24\linewidth]{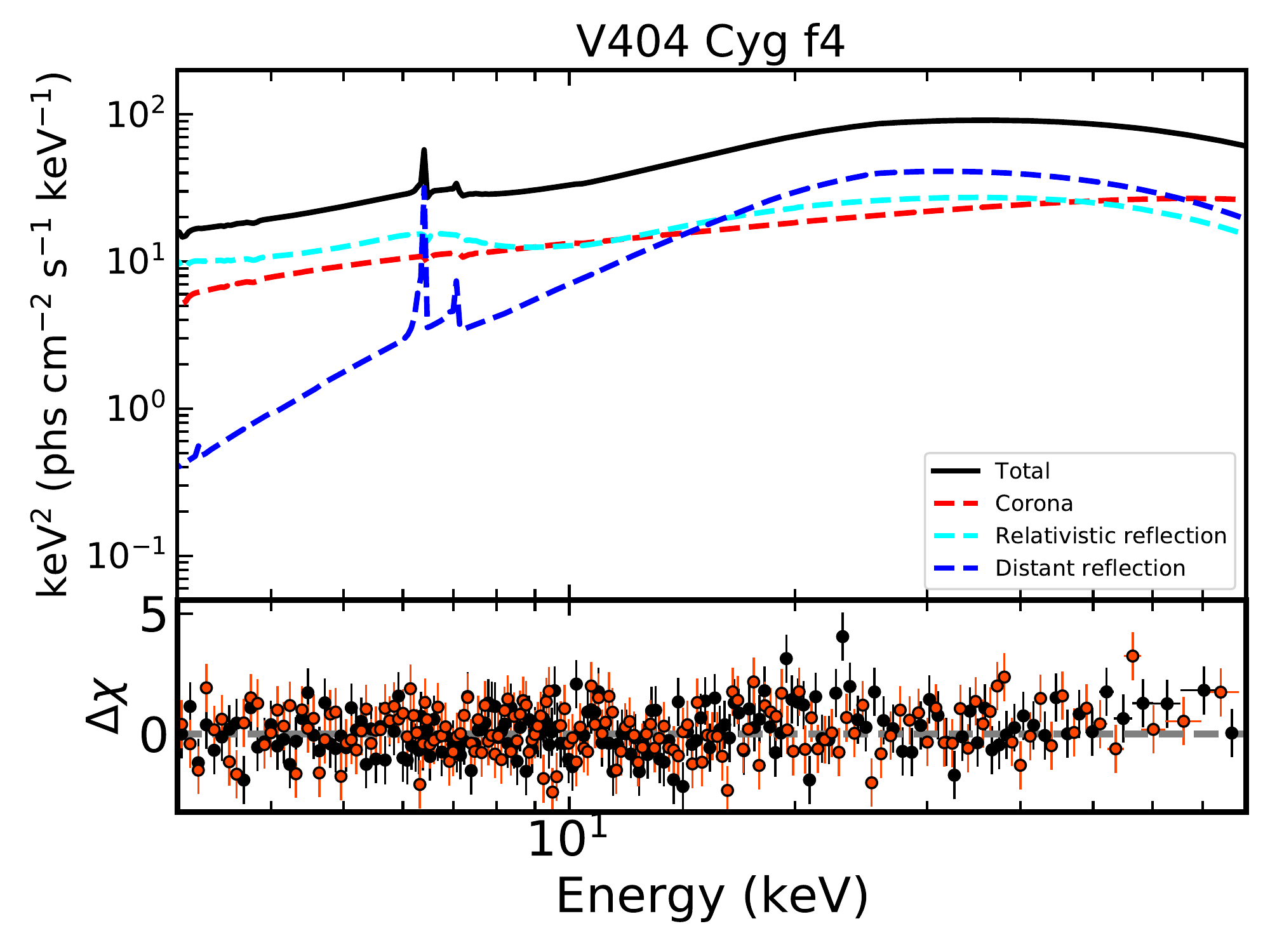}\\
    \includegraphics[width=0.24\linewidth]{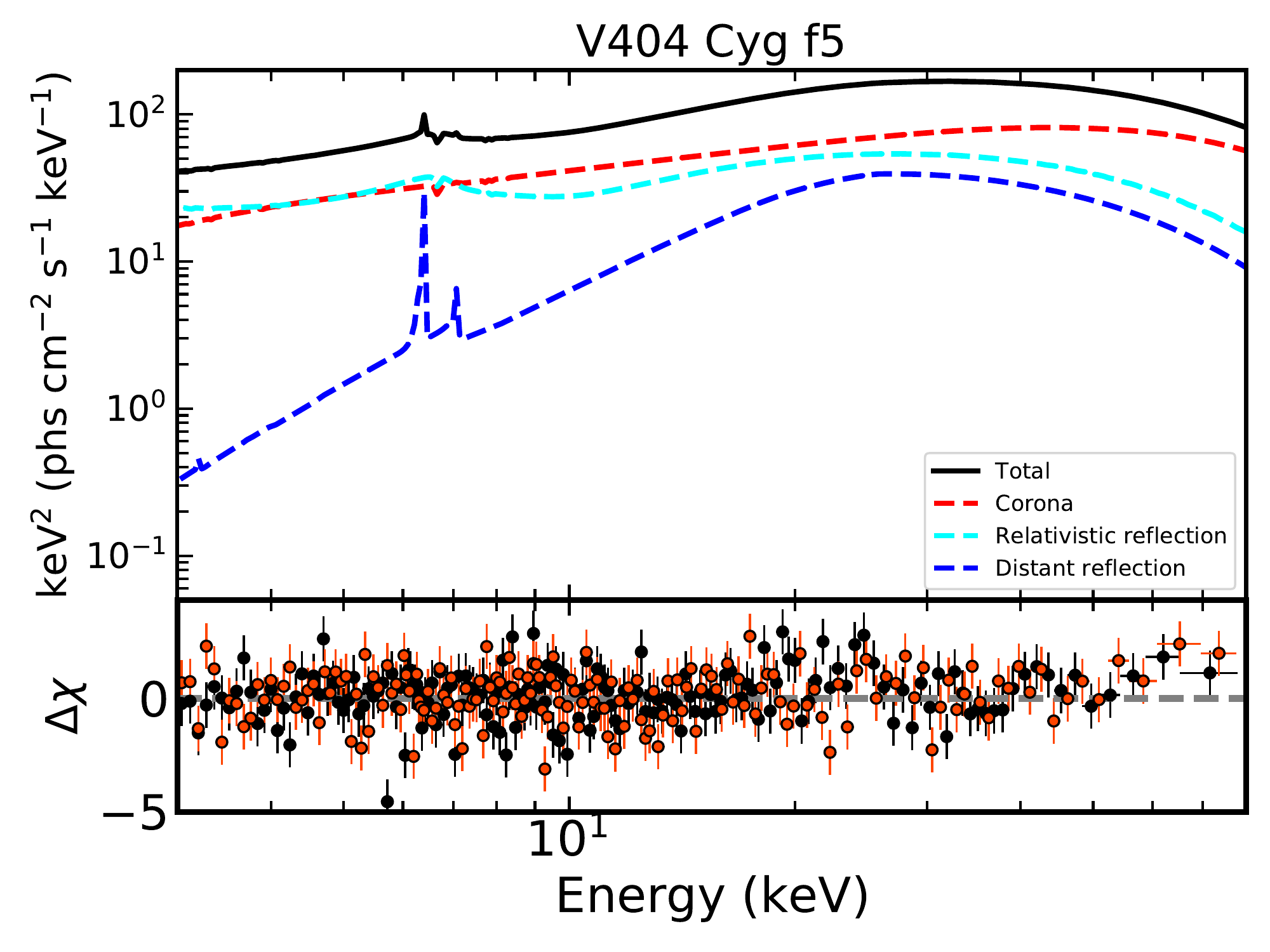}\\
    \caption{The best-fit model components (top panels) and the corresponding residual plots (bottom panels) for observations analyzed in this work.}
    \label{eemod_de}
\end{figure*}

\end{document}